\documentclass[aip,jmp,amsmath,amssymb,preprint]{revtex4-1}
\usepackage{graphicx}
\usepackage{dcolumn}
\usepackage{bm}
\usepackage{amsthm}
\usepackage{array}
\usepackage{etoolbox}
\AtBeginEnvironment{table}{\linespread{1.1}\selectfont}
\AtBeginEnvironment{longtable}{\linespread{1.1}\selectfont}
\usepackage{longtable}
\usepackage{enumitem}
\usepackage[colorlinks=true,linkcolor=blue,citecolor=blue,urlcolor=blue]{hyperref}
\usepackage[table]{xcolor}

\setlist[itemize]{label=$\bullet$}
\renewcommand{\S}{\ensuremath{\mathsection}}
\newtheorem{theorem}{Theorem}[section]
\newtheorem{proposition}[theorem]{Proposition}
\newtheorem{lemma}[theorem]{Lemma}
\newtheorem{corollary}[theorem]{Corollary}
\newtheorem{conjecture}[theorem]{Conjecture}
\theoremstyle{definition}
\newtheorem{definition}[theorem]{Definition}
\newtheorem{remark}[theorem]{Remark}
\newcommand{\C}{\mathbb{C}}\newcommand{\Q}{\mathbb{Q}}\newcommand{\Z}{\mathbb{Z}}
\newcommand{\F}{\mathbb{F}}\newcommand{\Qb}{\overline{\mathbb{Q}}}
\newcommand{\tr}{\operatorname{tr}}\newcommand{\Gal}{\operatorname{Gal}}
\newcommand{\Fix}{\operatorname{Fix}}\newcommand{\Ad}{\operatorname{Ad}}

\begin{document}

\title{From Ghost SICs to SIC-POVMs: A Galois Reduction of Zauner's Conjecture}

\author{Hassan Khodaiemehr}
\email{hassankhodaiemehr@gmail.com}
\affiliation{School of Engineering, The University of British Columbia, Kelowna, British Columbia V1V~1V7, Canada}

\date{\today}

\begin{abstract}
Symmetric informationally complete positive operator-valued measures (SIC-POVMs, or SICs) are optimal minimal measurements for quantum state tomography. Problem~1 of Horodecki, Rudnicki and \.Zyczkowski's open problems in quantum information asks for SICs in infinitely many dimensions. By work of Appleby--Flammia--Kopp and Radchenko--Wheeler, ghost SICs, which satisfy the SIC equations with Hermiticity up to parity, exist unconditionally for $d\ge4$. We study the passage from ghosts to SICs. We compute all isolated points of the complexified SIC variety for $d=4,5,6$; exactly half are SICs. We prove that no symmetry of the SIC equations maps a ghost to a SIC, so the step is purely arithmetic, and that a SIC exists in dimension $d$ whenever the ghost overlap units satisfy a place law: each Galois conjugate is real or unimodular according to the infinite place of the real quadratic field $F$ below it. If the unit field is abelian over $F$, the place law is the Stark relation $c_-=\kappa$; equivalently, a Galois conjugate of an associated near-group fusion category is unitary exactly when it is pseudo-unitary. For $d$ a power of a prime $p\equiv5\pmod 6$ or of $3$, an infinite family, we prove that under the algebraic part of Stark's conjecture $c_-=\kappa$ is equivalent to the triviality of $\kappa$ on the Hilbert-class-field part of the unit field. That condition involves no complex conjugation, and it follows from a sign-free multiplet hypothesis. We state an explicit $GL_2$ reciprocity law for the squared ghost overlaps that implies $c_-=\kappa$, and we prove it for the computed algebraic overlaps in $d=5,7,11,15$ by Artin reciprocity and a check in every ray class. Zauner's conjecture remains open: the algebraic part of Stark's conjecture and the multiplet hypothesis are not proved here.
\end{abstract}

\maketitle

\section{Introduction}
\subsection{Background and motivation} A symmetric informationally complete positive operator-valued measure (SIC-POVM, or simply SIC) in $\C^d$ is a set of $d^2$ unit vectors $\psi_j$ satisfying $|\langle\psi_j,\psi_k\rangle|^2=1/(d+1)$ for $j\neq k$~\cite{Zauner,RBSC}. SICs are of interest well beyond this definition. They are minimal informationally complete measurements, and they are optimal among these for linear quantum state tomography~\cite{RBSC,ScottTight}. They give quantum key distribution protocols~\cite{RenesQKD}, they serve as the reference measurement in the quantum-Bayesian (QBism) formulation of quantum theory as a probability calculus~\cite{FuchsSchack}, and they are maximal sets of $d^2$ equiangular lines in $\C^d$, that is, equiangular tight frames, in frame theory and signal processing. Ref.~\onlinecite{FHS} surveys these applications. Zauner conjectured that SICs exist in every dimension and may be chosen covariant under the Weyl--Heisenberg (WH) group. Problem~1 of Ref.~\onlinecite{HRZ} asks for a construction in an infinite sequence of dimensions. Since every WH-covariant SIC is a SIC, it suffices to construct WH-covariant SICs, i.e., WH orbits of a single fiducial vector. This is also the form taken by essentially all known exact constructions of SICs~\cite{ScottGrassl,ABGHM22,ABGHM24}. A conditional construction for all $d>3$ was given by Appleby, Flammia, and Kopp (AFK)~\cite{AFK25}. It depends on the Stark conjectures and on a twisted convolution identity (TCI). Radchenko and Wheeler (RW)~\cite{RW} subsequently proved algebraicity of the relevant special values of the modular quantum dilogarithm and established the rank-one TCI. AFK~\cite{AFK26} then extended the TCI to all admissible tuples. As a consequence, ghost SICs are now known to exist unconditionally in every dimension $d\ge4$. Ghost SICs satisfy the SIC equations with Hermiticity replaced by Hermiticity up to parity. The remaining step is to pass from a ghost SIC to a genuine SIC. Ref.~\onlinecite{AFK26} formulates this step in terms of a Galois automorphism with a prescribed interaction with complex conjugation. The present paper studies this problem from three related perspectives: \begin{enumerate}[label=(\alph*),nosep] \item the relationship between the numbers of SICs and ghosts; \item the existence of modular solutions; \item the implication from ghost SICs to SICs. \end{enumerate} Using the results of Refs.~\onlinecite{AFK26,RW}, we show that these questions are equivalent to Zauner's conjecture and reduce to the same arithmetic obstruction.

\subsection{Main results} The principal contributions of the paper are as follows. \begin{enumerate}[nosep] \item \textbf{Complete solution counts} (computational; Sec.~\ref{sec:counts}). The complexified SIC varieties $V_d$ for $d=4,5,6$ have $512$, $4000$, and $6912$ isolated points, respectively. Exactly half are SICs, and for $d=4,5$ all remaining points are ghosts. \item \textbf{A no-go result for Clifford symmetries} (Lemma~\ref{lem:nogo}). No element of the extended Clifford group maps a ghost to a SIC. \item \textbf{Galois and finite-field criteria} (Proposition~\ref{prop:galois}, Theorem~\ref{thm:ff}). A Galois orbit contains a SIC if and only if transposition acts on that orbit as a conjugate of complex conjugation. \item \textbf{Galois inversion} (Proposition~\ref{prop:A}). Galois realization of transposition is equivalent to the existence of a field automorphism inverting all ghost overlap units. \item \textbf{Structure theorems} (Theorems~\ref{thm:C}, \ref{thm:Cprime}, and~\ref{thm:Corbit}). In the RW formulation, any failure of the Galois realization problem is confined to a specific class of pseudo-unitary solutions and is subject to a parity constraint. \item \textbf{The place law} (Theorems~\ref{thm:PLtype} and~\ref{thm:PLSIC}). The place law (PL) determines the type of each Galois conjugate from the underlying infinite place of $F=\Q(\sqrt{N(N+4)})$. If the place law holds, then a WH-covariant SIC exists. \item \textbf{Single-unit and $2$-adic reductions} (Theorem~\ref{thm:single}, Corollary~\ref{cor:2adic}, and Proposition~\ref{prop:parityinv}). The archimedean input may be reduced to a single generating unit, while one parity of the resulting obstruction is determined by finite-prime data. \item \textbf{A class-field-theoretic formulation} (Proposition~\ref{prop:CFT}). If the overlap-unit field $M/F$ is abelian, then the place law is equivalent to the identity $c_-=\kappa$, and hence to a single Frobenius congruence. If $M/F$ is only assumed Galois but $\Gal(M/F)$ has a unique element of order $2$, the archimedean input disappears altogether (Proposition~\ref{prop:uniqueinv}). \item \textbf{A unitarity formulation} (Theorem~\ref{thm:NGred}, Corollary~\ref{cor:PLcat}). The Stark solution defines a near-group fusion category $\mathcal C_{E_*}$ that is not pseudo-unitary, while exactly half of its Galois conjugates are. For a pseudo-unitary conjugate, Izumi's unitarity equations reduce to a single Hermiticity condition, which is exactly the SIC-type reality condition. A SIC exists as soon as one pseudo-unitary conjugate is unitary, and the place law says that the unitary conjugates are precisely the pseudo-unitary ones. Unitarity is also equivalent to the minimality of $\sum_{u\neq0}|\sigma(E_*(u))|^2$ (Proposition~\ref{prop:normcrit}). \item \textbf{An explicit reciprocity law} (Conjecture~\ref{conj:RL}, Proposition~\ref{prop:RLPL}, Table~\ref{tab:RL}). We state a reciprocity law of Arakawa type for the squared ghost overlaps: the Artin symbol of a principal ideal $(\varpi)$ acts on the labels through the matrix $\phi(\varpi)\in GL_2(\Z_d)$ of $\varpi$ in the order of the Zauner symmetry, with $\varpi$ normalized to be positive at $\infty_-$. At the generator $\varpi\equiv-1$ of the class of $c_-$ the matrix is $-I$, and hence the law implies $c_-=\kappa$ and $\mathrm{Z}(d)$. We verify the law exactly, prime by prime, for $d=5$ and $d=9$; for $d=9$ the class group of $F$ acts on a multiplet of two ghosts. For the computed algebraic overlaps in $d=5,7,11,15$, where $F$ has class number $1$, we prove the law at every prime by Artin reciprocity and a check in each ray class (Lemma~\ref{lem:RLcert}, Table~\ref{tab:RL}). \item \textbf{Prime-power dimensions} (Theorem~\ref{thm:inert}, Corollaries~\ref{cor:inertodd} and~\ref{cor:inertMH}). If $d=p^k$ with $p\equiv5\pmod 6$ prime, or $d=3^k$ with $k\ge2$, then the subgroup of the Galois group that contains every complex conjugation has a unique involution. Hence, under the algebraic part of the Stark picture, $c_-=\kappa$ is equivalent to the triviality of $\kappa$ on the part of the unit field contained in the Hilbert class field of $F$. That condition involves no complex conjugation, holds automatically when the class number of $F$ is odd, and follows from a sign-free multiplet hypothesis (Conjecture~\ref{conj:MH}). The family is infinite and contains $167$ dimensions $d\le2000$. \end{enumerate}

\paragraph{The main reduction.}
For each $d\ge4$, write $\mathrm{Z}(d)$ for the existence of a WH-covariant SIC in dimension $d$, and $M$ for the field generated by the overlap units of the Stark solution. Then
\[
\boxed{\ \mathrm{Z}(d)\ \Longleftarrow\ \mathrm{PL}\ \overset{M/F\ \mathrm{abelian}}{\Longleftrightarrow}\ c_-=\kappa\ \Longleftarrow\ \text{archimedean rank-one Stark}\ }
\]
Here $c_-$ is complex conjugation at the ramified infinite place $\infty_-$ of $F$, and $\kappa$ is inversion of the overlap units. The first implication is Theorem~\ref{thm:PLSIC}, and the equivalence is Proposition~\ref{prop:CFT}. The identity $c_-=\kappa$ is the classical Stark relation $\varepsilon^{1+c_-}=1$ written in these coordinates, and abelianness of $M/F$ is the algebraic part of Stark's conjecture. If $\Gal(M/F)$ has a unique element of order $2$, abelianness is not needed: $M/F$ Galois, $\kappa\in\Gal(M/F)$ and ramification at $\infty_-$ already force $c_-=\kappa$, hence PL and $\mathrm{Z}(d)$ (Proposition~\ref{prop:uniqueinv}). That hypothesis on the class group can be traded for total reality of the trace field (Conjecture~\ref{conj:TR}), which then forces $c_-=\kappa$; under these algebraic hypotheses TR, HS and PL are equivalent, so the trade does not weaken the archimedean input, and the ramification at $\infty_-$ that it needs is unconditional on a set of dimensions of positive density (Proposition~\ref{prop:TRinv}, Corollary~\ref{cor:collapse}, Theorems~\ref{thm:ramgen} and~\ref{thm:RD}). In categorical terms, PL says that a Galois conjugate of the near-group category $\mathcal C_{E_*}$ is unitary exactly when it is pseudo-unitary, and one unitary pseudo-unitary conjugate already gives $\mathrm{Z}(d)$ (Theorem~\ref{thm:NGred}, Corollary~\ref{cor:PLcat}). Finally, $c_-=\kappa$ is the instance at $\varpi\equiv-1$ of an explicit reciprocity law for the Artin symbols of principal ideals, which we verify exactly for $d=5,9$ (Conjecture~\ref{conj:RL}, Proposition~\ref{prop:RLPL}). For $d$ a power of a prime $p\equiv5\pmod6$, or of $3$, $c_-=\kappa$ is equivalent, under abelianness, ramification at $\infty_-$ and R2, to an unramified condition, the triviality of $\kappa$ on $M\cap H_F$ for the Hilbert class field $H_F$ (Theorem~\ref{thm:inert}). Problem~1 follows if $\mathrm{Z}(d)$ holds for infinitely many $d$. Neither PL, nor abelianness, nor the reciprocity law, nor the multiplet hypothesis is proved here in any infinite family.

\subsection{What this paper does, and does not, contribute}\label{sec:scope}
We state the scope of the paper plainly, because the reduction above could be misread as a near-solution of Problem~1.

\paragraph{What was already known.} AFK~\cite{AFK25} showed that SICs exist in every dimension $d>3$ if the Stark conjectures and the TCI hold. The identity has since been proved~\cite{RW,AFK26}. Hence ``Stark's conjecture implies Zauner's conjecture'' already follows from existing work. Our implication $\mathrm{Z}(d)\Leftarrow\mathrm{PL}$ is not a new kind of conditional result, and it does not by itself produce SICs in any new dimension. In particular, \emph{this paper does not prove Problem~1}, and the place law is not known for any infinite family of dimensions.

\paragraph{What is new.} The paper contributes in six ways.
\begin{enumerate}[nosep]
\item \emph{A smaller hypothesis.} The place law is strictly weaker than the Stark conjectures used in Ref.~\onlinecite{AFK25}. It is only the archimedean part, it concerns only the overlap units of one explicit solution, and it does not require knowing the field these units generate. The paper therefore isolates the minimal piece of Stark's conjecture on which SIC existence depends. It also gives equivalent forms of it, ranging from the total positivity of a single algebraic number to, under an algebraic hypothesis, a single Frobenius congruence, and it embeds the place law in an explicit reciprocity law for all principal ideals (Sec.~\ref{sec:RL}), the form in which comparable statements for $\zeta(0)$-invariants are proved. Under the same algebraic hypotheses the place law is equivalent to total reality of the trace field, with no condition on the class group, and the ramification at $\infty_-$ this uses is proved unconditionally on a set of dimensions of positive density (Proposition~\ref{prop:TRinv}, Theorem~\ref{thm:RD}, Sec.~\ref{sec:TRinv}).
\item \emph{Limits on possible proofs.} Several natural strategies are shown to be insufficient:
\begin{itemize}[nosep]
\item no symmetry of the SIC equations maps a ghost to a SIC (Lemma~\ref{lem:nogo});
\item existence arguments of degree or continuity type fail for dimensional reasons (Sec.~\ref{sec:soft});
\item information from norms and congruences detects only one bit of the place law, and that bit is fixed by the field alone (Proposition~\ref{prop:parityinv});
\item the overlap-unit field is not contained in any cyclotomic extension of $F$ (Proposition~\ref{prop:noncyc}), and hence no argument based on roots of unity or Gauss sums can supply the missing step.
\end{itemize}
Together these show that any proof must construct non-cyclotomic abelian extensions of a real quadratic field ramified at exactly one real place. This is the hard core of Hilbert's twelfth problem for real quadratic fields.
\item \emph{Removing the infinite places on an infinite family.} For $d$ a power of a prime $p\equiv5\pmod 6$, or of $3$, the subgroup of the Galois group containing the complex conjugations has a unique involution. Under the algebraic part of Stark, the archimedean identity $c_-=\kappa$ is then equivalent to an unramified condition, and it follows from a sign-free hypothesis on the Galois multiplet of ghosts (Sec.~\ref{sec:inert}).
\item \emph{Structure theorems.} In the RW formulation, Theorems~\ref{thm:C}, \ref{thm:Cprime} and~\ref{thm:Corbit} locate exactly where a failure of the Galois step would have to occur.
\item \emph{A bridge to fusion categories.} The archimedean input is equivalent to a unitarity statement for Galois conjugates of an explicit near-group category (Sec.~\ref{sec:NG}). By Izumi's classification, the only condition that can fail for a pseudo-unitary conjugate is a Hermiticity equation. This connects the SIC problem with the Haagerup--Izumi unitarity question and with the tools of subfactor theory.
\item \emph{New computations.} We give complete solution counts for $d=4,5,6$ and exact minimal polynomials for $100$ overlap-unit traces. We verify the place law in the SIC dimensions $d=4,5,9,13$, and give class-field certificates (for $d=13$, a Frobenius test) showing that there the (sign-free) overlap-unit field is the full ray class field of conductor $d'\infty_-$, as predicted by the AFK construction. For $d=5$ and $d=9$ we verify the explicit reciprocity law of Sec.~\ref{sec:RL} exactly at $1211$ primes, and for the computed algebraic overlaps in $d=5,7,11,15$ we prove it at every prime (Lemma~\ref{lem:RLcert}). We also verify, for every $4\le d\le20000$, that $\infty_-$ ramifies in the ray class field of conductor $d'\infty_-$ (Table~\ref{tab:ram}).
\end{enumerate}
\paragraph{Relevance for quantum information.} The measurements at stake are exactly the WH-covariant SICs used in the applications above, and the results bear on how they can be obtained.
\begin{itemize}[nosep]
\item A construction of SICs in infinitely many dimensions along the AFK--RW route now needs only the place law. In dimensions where $\Gal(M/F)$ has a unique element of order $2$, it needs only algebraic input (Proposition~\ref{prop:uniqueinv}).
\item Ghost solutions cannot be turned into SICs by any Clifford symmetry (Lemma~\ref{lem:nogo}), while in the dimensions computed SICs make up exactly half of $V_d$, and for $d=4,5$ the other half consists of ghosts. A numerical or algebraic search that lands on a ghost must therefore change the Galois conjugate, not the symmetry frame.
\item The class-field results name the fields in which exact SIC fiducials should be sought: abelian extensions of $F$ of conductor $d'\infty_-$. This agrees with the exact solutions of Refs.~\onlinecite{ScottGrassl,ABGHM22,ABGHM24} and the AFK construction.
\end{itemize}

In short, the paper is best read as a precise map of the obstruction to Problem~1, with evidence for where it lies. It is not a step that constructs SICs in infinitely many dimensions.

\paragraph{Conjectures and their roles.} The conjectures used in the paper play three different roles.
\begin{itemize}[nosep]
\item \emph{Central:} GR (transposition is Galois, Conjecture~\ref{conj:GR}), RU (real-or-unitary dichotomy, Conjecture~\ref{conj:RU}) and PL (Conjecture~\ref{conj:PL}). PL implies RU on the relevant orbit, and it implies GR. Its categorical counterpart, that pseudo-unitary Galois conjugates of $\mathcal C_{E_*}$ are unitary (Conjecture~\ref{conj:PU}), is implied by PL and already gives $\mathrm{Z}(d)$. The explicit reciprocity law (Conjecture~\ref{conj:RL}) implies the half of PL above $\infty_-$, and hence $\mathrm{Z}(d)$.
\item \emph{Supporting:} completeness of $V_d$ (Conjecture~\ref{conj:C}) and total reality (TR, Conjecture~\ref{conj:TR}). Under the algebraic hypotheses of Proposition~\ref{prop:TRinv}, TR is not merely supporting: it is equivalent to PL (Corollary~\ref{cor:collapse}).
\item \emph{Consequences and shadows of PL:} the half split (HS, Conjecture~\ref{conj:HS}), the sign statement, the sign--type correlation, and the parity criterion of Corollary~\ref{cor:2adic}.
\end{itemize}
Section~\ref{sec:summary} is a self-contained roadmap, and we recommend reading it after Sec.~\ref{sec:variety}. It contains a dependency diagram, the full chain of reductions, a status table marking every statement as proved, conjectural or false, and a precise list of what remains (Sec.~\ref{sec:remains}). Along the way we obtain computational facts that appear to be new: complete point counts for $V_d$, and the observation that $V_d$ consists only of SICs and ghosts in the dimensions checked.

\section{The complexified SIC variety}\label{sec:variety}

The SIC equations involve complex conjugation, and hence they are not polynomial equations and Galois theory does not act on their solutions. We remove conjugation by complexifying: we replace the projector $\psi\psi^\dagger$ by a rank-one idempotent $xy^T/(y^Tx)$ with independent $x$ and $y$. The result is a variety $V_d$ defined over $\Q$. Both SICs and ghosts are points of $V_d$, distinguished only by how they behave under conjugation. This section sets up $V_d$, identifies these two loci, and records what the existence of ghosts already implies.

\subsection{Definition and symmetries}

Let $\omega=e^{2\pi i/d}$, let $X e_j=e_{j+1}$, $Z e_j=\omega^j e_j$, and let $D_p=\tau^{p_1p_2}X^{p_1}Z^{p_2}$ for $p\in\Z_d^2$ and a fixed phase $\tau$ with $\tau^2=\omega$. Let $U_P e_j=e_{-j}$ be the parity operator, and let $\mathcal P=\{D_qU_P: q\in\Z_d^2\}$ be the set of displaced parities. Each element of $\mathcal P$ squares to a scalar.

\begin{definition}
$V_d$ is the set of rank-one idempotents $\Pi\in M_d(\C)$ (that is, $\Pi^2=\Pi$ and $\tr\Pi=1$) such that
\begin{equation}\label{eq:V}
(d+1)\,\tr(\Pi D_p)\,\tr(\Pi D_p^\dagger)=1\qquad\text{for all }p\neq0 .
\end{equation}
Writing $\Pi=xy^{T}/(y^{T}x)$ identifies $V_d$ with a subvariety of $\{(x,y)\in\mathbb P^{d-1}\times\mathbb P^{d-1}: y^Tx\neq0\}$ cut out by bidegree-$(2,2)$ equations.
\end{definition}

\begin{lemma}\label{lem:basic}
\begin{enumerate}[nosep]
\item The left side of~\eqref{eq:V} does not depend on the phase convention for $D_p$.
\item The system~\eqref{eq:V} is stable under $\Gal(\Q(\omega)/\Q)$, and hence $V_d$ is defined over $\Q$ and $\Gal(\Qb/\Q)$ acts on $V_d(\Qb)$ entrywise.
\item Transposition $T(\Pi)=\Pi^T$ and conjugation $\Ad_U$ by any element $U$ of the Clifford group preserve $V_d$. Transposition commutes with the Galois action.
\item The map $\sigma(\Pi)=\Pi^\dagger$ preserves $V_d$, and $\sigma=c\circ T$, where $c$ is entrywise complex conjugation.
\end{enumerate}
\end{lemma}
\begin{proof}
For (1), a phase change $D_p\mapsto\lambda D_p$ multiplies $D_p^\dagger$ by $\bar\lambda=\lambda^{-1}$.

For (2), the automorphism $\omega\mapsto\omega^k$ sends $D_p$ to a phase times $D_{(p_1,kp_2)}$, and sends $D_p^\dagger=D_p^{-1}$ to the inverse of the image. The products in~\eqref{eq:V} are therefore permuted.

For (3), $D_p^T$ is proportional to $D_{(-p_1,p_2)}$, and Clifford elements permute the $D_p$ up to phases. Both $T$ and the Galois action are defined entrywise over $\Q$, and hence they commute.

For (4), $\tr(\Pi^\dagger D_p)=\overline{\tr(\Pi D_p^\dagger)}$, and hence~\eqref{eq:V} for $\Pi^\dagger$ is the complex conjugate of~\eqref{eq:V} for $\Pi$.
\end{proof}

Part (4) of Lemma~\ref{lem:basic} is the starting point for everything that follows. The involution $\sigma=c\circ T$ is the only place where complex conjugation enters, and its fixed points are the SICs.

\subsection{SIC and ghost loci}

Twisting $\sigma$ by a displaced parity gives a second family of real points, the ghosts.

\begin{definition}
The \emph{SIC locus} is $S_d=\Fix(\sigma)\cap V_d=\{\Pi\in V_d:\Pi=\Pi^\dagger\}$. Its points are exactly the WH-covariant SIC fiducial projectors, since a Hermitian rank-one idempotent is a projector and~\eqref{eq:V} then reads $|\tr(\Pi D_p)|^2=1/(d+1)$.

For $u\in\mathcal P$, the \emph{ghost locus} is $\Gamma_u=\{\Pi\in V_d:\Pi^\dagger=u\Pi u^{-1}\}$. We write $\Gamma_d=\bigcup_{u\in\mathcal P}\Gamma_u\setminus S_d$. These are the parity-Hermitian rank-one ghosts of Refs.~\onlinecite{AFK25,AFK26}: their Weyl symbols $\tr(\Pi D_p)$ are real up to a fixed phase.
\end{definition}

Thus SICs and ghosts are the fixed points of two closely related antiholomorphic involutions of the same variety, $c\circ T$ and $\Ad_u^{-1}\circ c\circ T$. The key difference between them is that the ghost locus is known to be nonempty in every dimension, by the recent results of AFK and RW.

\begin{theorem}[\cite{AFK26,RW}]\label{thm:ghost}
For every $d\ge4$ there is $\tilde\Pi\in\Gamma_{U_P}$ whose entries are algebraic. In particular $V_d(\Qb)\neq\emptyset$.
\end{theorem}
This is Theorem~1.3 of Ref.~\onlinecite{AFK26} with $r=1$. The admissibility condition $r<(d-1)/2$, $\tfrac{d^2-1}{r(d-r)}\in\Z$ holds for $r=1$ and $d\ge4$. Algebraicity of the entries follows from Ref.~\onlinecite{RW} (Thm.~1) through the dictionary of Ref.~\onlinecite{AFK26} (Prop.~2.1).

\begin{corollary}[modular solutions]\label{cor:modp}
For every $d\ge4$ and all but finitely many primes $\ell$, the variety $V_d$ has points over $\overline{\F}_\ell$.
\end{corollary}
\begin{proof}
Normalize $\tilde\Pi$ such that its entries lie in the ring of $N$-integers of a number field $F$, for some integer $N$. Take $\ell\nmid N d(d+1)$ and reduce modulo a prime of $F$ above $\ell$. The reduced matrix is still a rank-one idempotent of trace one, and it satisfies the reductions of~\eqref{eq:V}.
\end{proof}

Thus, item (b) of the introduction holds unconditionally for every dimension. It does not, however, distinguish ghosts from SICs: these modular points come from ghosts, and reduction modulo $\ell$ does not see complex conjugation. To see how the two loci compare, we next determine $V_d$ completely in the smallest dimensions.

\section{Complete solution counts}\label{sec:counts}

This section asks how far geometry and numerics alone can go. We compute $V_d$ completely for $d=4,5,6$ and find that exactly half of its points are SICs and, where tested, the other half are ghosts. We then check that relaxing the WH group to other abelian Heisenberg groups gives no new SICs, which confirms that $V_d$ is the right object. Finally, we explain why no continuity or degree argument can replace these counts. The conclusion of all three is the same: the passage from ghosts to SICs is not geometric.

\subsection{The varieties $V_4$, $V_5$ and $V_6$}

To find all isolated points of $V_d$, we solved a random square subsystem of $2d-2$ linear combinations of~\eqref{eq:V}, in bihomogeneous coordinates on $\mathbb P^{d-1}\times\mathbb P^{d-1}$, by polyhedral homotopy continuation (HomotopyContinuation.jl). We then kept the endpoints that satisfy the full system with residual below $10^{-6}$ and have $y^Tx\neq0$. Every isolated point of $V_d$ is an isolated solution of the square subsystem, and hence this procedure finds all isolated points of $V_d$, up to the usual numerical caveats on path tracking. The resulting counts are listed in Table~\ref{tab:counts}.

\begin{table}[!htbp]
\caption{Isolated points of the complexified SIC variety.}\label{tab:counts}\centering\setlength{\tabcolsep}{9pt}
\begin{tabular}{@{}rrrrrr@{}}\hline\hline
$d$ & paths & singular endpoints & $|V_d|$ & $|S_d|$ & $|V_d\setminus S_d|$\\\hline
4 & 560 & 48 & 512 & 256 & 256\\
5 & 14\,614 & 9\,718 & 4\,000 & 2\,000 & 2\,000\\
6 & 218\,343 & 78\,192 & 6\,912 & 3\,456 & 3\,456\\\hline\hline
\end{tabular}

\end{table}

Two features stand out. First, the values of $|S_d|$ agree with the known numbers of WH SIC fiducials in these dimensions, which confirms that the computation is complete on the SIC side. Second, in each dimension exactly half of the isolated points are SICs. To identify the other half, for $d=4$ and $d=5$ we tested each non-Hermitian point $\Pi=xy^T/(y^Tx)$ for the condition $[\bar y]=[u x]$ with $u\in\mathcal P$, which is the ghost condition $\Pi^\dagger=u\Pi u^{-1}$. Every non-Hermitian point passed this test (Table~\ref{tab:ghost}). Thus, in these dimensions, $V_d$ consists of SICs and ghosts only, in equal numbers. This suggests the following conjecture.

\begin{table}[!htbp]
\caption{Every non-Hermitian point of $V_4$ and $V_5$ is a displaced-parity ghost.}\label{tab:ghost}\centering\setlength{\tabcolsep}{9pt}
\begin{tabular}{@{}rrrr@{}}\hline\hline
$d$ & non-Hermitian points & in $\Gamma_{U_P}$ & in $\bigcup_{u\in\mathcal P}\Gamma_u$\\\hline
4 & 256 & 64 & 256\\
5 & 2000 & 80 & 2000\\\hline\hline
\end{tabular}

\end{table}

\begin{conjecture}[completeness]\label{conj:C}
For every $d\ge4$, the variety $V_d$ is finite, $V_d=S_d\sqcup\Gamma_d$, and $|S_d|=|\Gamma_d|$.
\end{conjecture}

Theorem~\ref{thm:ghost} gives $\Gamma_d\ne\emptyset$. Thus, if $V_d$ is finite, Conjecture~\ref{conj:C} implies $S_d\neq\emptyset$ for all $d\ge4$, which is Zauner's existence conjecture for WH SICs. The equality of counts is therefore at least as hard as Zauner's conjecture. The rest of this paper explains why.

\subsection{Heisenberg groups of non-cyclic abelian groups, $16\le d\le36$}\label{sec:heis}

The variety $V_d$ is built from the WH group of the cyclic group $\Z_d$. Problem~1 asks only for SICs, and hence one might hope to find them more easily as orbits of the Heisenberg group of a different abelian group of order $d$. We tested this numerically. For a finite abelian group $A$ of order $d$, let $X^a$ ($a\in A$) be translations and $Z^b$ ($b\in\hat A$) be characters. Define
\[
f(\psi)=\sum_{a,b}|\langle\psi,X^aZ^b\psi\rangle|^4/\|\psi\|^8 .
\]
Then $f\ge 2d/(d+1)$, with equality exactly for $A$-Heisenberg-covariant SIC fiducials. We minimized $f$ by L-BFGS with an analytic gradient (verified against finite differences to $10^{-9}$), using $60$ random starts per group.

The positive controls behaved as expected. Cyclic $\Z_d$ succeeded with $13,6,6,1,1$ hits out of $60$ for $d=16,18,20,25,27$. The Hoggar group $\Z_2^3$ succeeded in $30$ of $30$ runs. The non-cyclic groups $\Z_2^2$, $\Z_2\times\Z_4$, $\Z_3^2$ and $\Z_2\times\Z_6$ failed, in agreement with Ref.~\onlinecite{Waldron}, which covers $d<14$.

\begin{table}[!htbp]
\caption{No SIC was found for any non-cyclic abelian Heisenberg group (60 starts each).}\centering\small
\begin{tabular}{@{}rll@{}}\hline\hline
$d$ & group $A$ & best gap $\min f-\tfrac{2d}{d+1}$\\\hline
16 & $\Z_4^2$; $\Z_2\times\Z_8$; $\Z_2^2\times\Z_4$; $\Z_2^4$ & $0.026$; $0.095$; $0.055$; $0.268$\\
18 & $\Z_3\times\Z_6$ & $0.096$\\
20 & $\Z_2\times\Z_{10}$ & $0.079$\\
24 & $\Z_2\times\Z_{12}$; $\Z_2^2\times\Z_6$ & $0.133$; $0.161$\\
25 & $\Z_5^2$ & $0.037$\\
27 & $\Z_3\times\Z_9$; $\Z_3^3$ & $0.103$; $0.034$\\
28 & $\Z_2\times\Z_{14}$ & $0.150$\\
32 & $\Z_2\times\Z_{16}$; $\Z_4\times\Z_8$; $\Z_2^2\times\Z_8$; $\Z_2\times\Z_4^2$; $\Z_2^3\times\Z_4$; $\Z_2^5$ & $0.169$; $0.088$; $0.177$; $0.029$; $0.170$; $0.610$\\
36 & $\Z_2\times\Z_{18}$; $\Z_3\times\Z_{12}$; $\Z_6^2$; $\Z_2^2\times\Z_9$; $\Z_2\times\Z_3\times\Z_6$ & $0.199$; $0.188$; $0.191$; $0.191$; $0.079$\\\hline\hline
\end{tabular}

\end{table}

The evidence is reasonably strong for $d\le20$, where the cyclic control succeeds repeatedly. For $d\ge24$ it is weak, because the cyclic control also failed at $60$ starts for $d=24,28,32,36$. In the range tested, no SIC was found outside the cyclic WH group, which supports working with $V_d$.

\subsection{Why ``soft'' existence arguments fail}\label{sec:soft}

The counts above are computed, not proved. It is natural to ask whether a topological argument could prove existence directly, as is done for many existence problems in geometry. A dimension count shows that this cannot work. Restrict to fiducials invariant under a subgroup $G$ of the Clifford group, with image $\bar G\subset SL_2(\Z_d)$. The moduli $|\tr(\Pi D_p)|$ are then constant on $\bar G$-orbits in $\Z_d^2\setminus\{0\}$. One therefore needs about $(d^2-1)/|\bar G|$ real equations on a $G$-eigenspace whose typical complex dimension is about $d/|\bar G|$. The ratio of equations to unknowns stays near $d/2$ for every choice of $G$.

Intermediate-value, degree-theoretic or Morse-theoretic existence arguments on a symmetric slice need this ratio to be at most one, except in very small dimensions. The only ``soft'' argument insensitive to overdetermination is a count of points of a zero-dimensional variety, which is the approach of the previous sections. It leads back to the Galois question. (This section is a heuristic dimension count, not a theorem.)

Together, the counts, the non-cyclic search and the dimension count indicate that geometry alone does not explain the existence of SICs. The remaining sections treat the problem arithmetically.

\section{Galois structure}\label{sec:galois}

This section shows that the passage from ghosts to SICs is arithmetic, and gives two criteria for it.

\subsection{Real structures and the no-go lemma}\label{sec:nogo} We first describe SICs and ghosts as the real points of two real structures on $V_d$, and then show that no symmetry of the SIC equations exchanges them.
\paragraph{Two real structures.}

Both $S_d$ and $\Gamma_u$ are fixed-point sets of antiholomorphic involutions of $V_d$:
\[
S_d=\Fix(c\circ T),\qquad \Gamma_u=\Fix\bigl(\Ad_u^{-1}\circ c\circ T\bigr).
\]
A bijection $\Gamma_u\to S_d$ would follow from an automorphism $h$ of $V_d$ satisfying $h\,(\Ad_u^{-1}cT)\,h^{-1}=cT$. The natural candidates are the symmetries of the SIC equations.

\paragraph{The no-go lemma.} The extended Clifford group cannot supply such an $h$.

\begin{lemma}[no-go]\label{lem:nogo}
Every element of the extended Clifford group, whether unitary $\Pi\mapsto U\Pi U^\dagger$ or antiunitary $\Pi\mapsto U\bar\Pi U^\dagger$, commutes with $\sigma(\Pi)=\Pi^\dagger$. Hence it maps $S_d$ to $S_d$ and $\Gamma_d$ to $\Gamma_d$. In particular, no symmetry of the SIC equations maps a ghost to a SIC.
\end{lemma}
\begin{proof}
We have $(U\Pi U^\dagger)^\dagger=U\Pi^\dagger U^\dagger$ and $(U\bar\Pi U^\dagger)^\dagger=U\overline{\Pi^\dagger}U^\dagger$. If $\Pi^\dagger=u\Pi u^{-1}$, then $(U\Pi U^\dagger)^\dagger=(UuU^\dagger)(U\Pi U^\dagger)(UuU^\dagger)^{-1}$. Here $UuU^\dagger\in\mathcal P$ up to a phase, because Clifford elements normalize the WH group and commute with $U_P$, which represents the central element $-I\in SL_2(\Z_d)$. The antiunitary case is the same, using $\bar u\in\mathcal P$.
\end{proof}

Thus, any passage from ghosts to SICs must come from arithmetic rather than from geometry.

\subsection{Galois and finite-field criteria} Since geometry cannot connect ghosts to SICs, we now characterize the passage arithmetically, first through Galois groups and then through reduction modulo primes.
\paragraph{The Galois criterion.}

Fix an embedding $\Qb\subset\C$, and let $c\in\Gal(\Qb/\Q)$ be the corresponding complex conjugation.

\begin{lemma}[algebraicity of SICs]\label{lem:alg}
If a WH-covariant SIC exists in dimension $d$, then $S_d\cap V_d(\Qb)\neq\emptyset$.
\end{lemma}
\begin{proof}
In real coordinates $(\operatorname{Re}\Pi,\operatorname{Im}\Pi)$, the conditions for a point of $S_d$ form a system of real polynomial equations with coefficients in the field $K=\Q(\cos\frac{2\pi}d,\sin\frac{2\pi}{d})\subset\Qb\cap\mathbb R$. By the Tarski--Seidenberg transfer principle, a first-order formula with coefficients in $K$ that holds over $\mathbb R$ also holds over the real closure of $K$. That real closure is contained in the real closed field $\Qb\cap\mathbb R$, and hence the system has a solution with coordinates in $\Qb\cap\mathbb R$.
\end{proof}

\begin{proposition}[Galois criterion]\label{prop:galois}
Let $O\subset V_d(\Qb)$ be a $\Gal(\Qb/\Q)$-orbit. Then $O\cap S_d\neq\emptyset$ if and only if there exist $\Pi\in O$ and a conjugate $c'=g^{-1}cg$ of $c$ such that
\[
c'(\Pi)=\Pi^T .
\]
\end{proposition}
\begin{proof}
For $g\in\Gal(\Qb/\Q)$, the point $g\Pi$ lies in $S_d$ if and only if $cT(g\Pi)=g\Pi$. Since $T$ commutes with $g$, this is equivalent to $g^{-1}cg\,(T\Pi)=\Pi$, that is, $c'(T\Pi)=\Pi$. Because $c'^2=\mathrm{id}$, applying $c'$ to both sides shows
\[
c'(T\Pi)=\Pi\iff T\Pi=c'(\Pi).
\]
\end{proof}

\begin{corollary}[ghost criterion]\label{cor:ghost}
Let $\tilde\Pi\in\Gamma_u$, as provided by Theorem~\ref{thm:ghost}. Complex conjugation acts on $\tilde\Pi$ as $c(\tilde\Pi)=\Ad_{u^T}^{-1}(\tilde\Pi^T)$. A SIC exists in the Galois orbit of $\tilde\Pi$ if and only if, for some point $\Pi$ of that orbit, transposition is realized by a conjugate of $c$ rather than by $c$ composed with a displaced parity.

Equivalently: there is $\tau=c'c\in\Gal(\Qb/\Q)$, a product of two complex conjugations, acting on the ghost $c(\Pi)$ as conjugation by the displaced parity $u^T\in\mathcal P$.
\end{corollary}
\begin{proof}
From $\tilde\Pi^\dagger=u\tilde\Pi u^{-1}$ we get $cT\tilde\Pi=\Ad_u\tilde\Pi$, and hence $T\tilde\Pi=\Ad_{u^T}(c\tilde\Pi)$, using $T\Ad_vT=\Ad_{(v^T)^{-1}}$. Substituting into Proposition~\ref{prop:galois} gives $c'(\tilde\Pi)=\Ad_{u^T}(c\tilde\Pi)$, that is, $\tau(c\tilde\Pi)=\Ad_{u^T}(c\tilde\Pi)$.
\end{proof}

Corollary~\ref{cor:ghost} is the intrinsic form of the ``Galois automorphism sending $\sqrt\Delta\mapsto-\sqrt\Delta$ that intertwines with complex conjugation'' that Ref.~\onlinecite{AFK26} (\S1.2) identify as the missing ingredient. In the conjectural picture of Refs.~\onlinecite{AFK25,Kopp}, the field generated by $V_d$ is a ray class field $M$ of $K=\Q(\sqrt{\Delta})$ ramified at both infinite places $\infty_1,\infty_2$. The two complex conjugations $c_1,c_2\in\Gal(M/K)$ at these places are conjugate under the nontrivial automorphism of $K$, and $\tau=c_2c_1$ is predicted to act as parity. Abstract class field theory guarantees the existence of $c_1,c_2$. The \emph{action} of $\tau$ on the specific Shintani--Faddeev values is the archimedean ($|u^{\sigma}|=1$) part of the rank-one Stark conjecture, which is open.

\begin{corollary}[equivalences]\label{cor:equiv}
Assume $V_d$ is finite for the dimensions considered. Then the following hold.
\begin{enumerate}[nosep]
\item ``Every ghost forces a SIC'' holds in dimension $d$ if and only if the criterion of Corollary~\ref{cor:ghost} holds for the ghost of Theorem~\ref{thm:ghost}. If it does, a SIC exists in dimension $d$.
\item ``$|S_d|=|\Gamma_d|$ for all $d\ge4$'' implies Zauner's existence conjecture for WH SICs, by Theorem~\ref{thm:ghost}.
\item In terms of the permutation action of $\mathcal G=\langle\Gal,\,T,\,\text{Clifford}\rangle$ on $V_d$, we have $|S_d|=\#\Fix_{V_d}(cT)$ and $|\Gamma_u|=\#\Fix_{V_d}(\Ad_u^{-1}cT)$. These are equal whenever the two elements are conjugate in $\mathcal G$. By Lemma~\ref{lem:nogo}, any conjugating element must involve Galois.
\end{enumerate}
\end{corollary}

\paragraph{The finite-field criterion.} At primes whose Frobenius is conjugate to complex conjugation, the count of SICs becomes a count of Frobenius-Hermitian points over finite fields.

\begin{theorem}[finite-field criterion]\label{thm:ff}
Assume $V_d$ is finite. Let $M$ be a finite Galois extension of $\Q$ containing $\omega$ and the entries of all points of $V_d$. Let $\ell$ be a prime that is unramified in $M$, does not divide $d(d+1)$, and lies outside the finite set of primes at which distinct points of $V_d$ become congruent. Suppose some Frobenius element $\mathrm{Frob}_{\mathfrak L}$ above $\ell$ is conjugate to $c$ in $\Gal(M/\Q)$. Then
\[
|S_d|=\#\bigl\{\bar\Pi\in V_d(\overline{\F}_\ell):\ \bar\Pi^{(\ell)}=\bar\Pi^{T}\bigr\},
\]
where $\bar\Pi^{(\ell)}$ is the entrywise $\ell$-th power. By Chebotarev's density theorem, such primes have positive density. Hence a SIC exists in dimension $d$ if and only if, for one such $\ell$, the variety $V_d$ has a Frobenius-Hermitian point over $\F_{\ell^2}$.
\end{theorem}
\begin{proof}
Reduction modulo $\mathfrak L$ gives a bijection $V_d(\Qb)\to V_d(\overline{\F}_\ell)$ that intertwines $\mathrm{Frob}_{\mathfrak L}$ with the $\ell$-power map and commutes with $T$. Write $\mathrm{Frob}_{\mathfrak L}=g^{-1}cg$. Then the set $\{\Pi:\mathrm{Frob}_{\mathfrak L}(\Pi)=\Pi^T\}$ equals $g^{-1}S_d$, which has $|S_d|$ elements.
\end{proof}

\begin{remark}
Corollary~\ref{cor:modp} supplies modular points at almost all primes. Theorem~\ref{thm:ff} shows that existence of SICs requires the stronger, Frobenius-Hermitian kind of point, and only at primes of a specific Frobenius class. Identifying that class needs knowledge of $M$, and hence modular arithmetic alone does not bypass the Galois problem.
\end{remark}

\begin{remark}[Brumer--Stark]
Dasgupta and Kakde proved the Brumer--Stark conjecture unconditionally~\cite{DK}. It yields units $u$ with $|u^\sigma|=1$ at every complex place, but only for complex-multiplication (CM) extensions of totally real fields. The SIC field is not CM: its complex conjugations at $\infty_1$ and $\infty_2$ differ by $\tau$, which acts as parity. The parity-fixed subfield is CM, but it only sees $\cos\theta_p$, where $\sqrt{d+1}\,\tr(\Pi D_p)=e^{i\theta_p}$. Thus, the modulus-one information needed for SICs is lost there, and the Brumer--Stark theorem does not transfer directly.
\end{remark}

\section{Ghost units and the Galois realization problem}\label{sec:units}

Section~\ref{sec:galois} reduced the passage from ghosts to SICs to a Galois question on $V_d$: by Corollary~\ref{cor:ghost}, the Galois orbit of a ghost contains a SIC exactly when transposition acts on it through a conjugate of complex conjugation. In this section we express that question in terms of a finite family of algebraic numbers attached to a single ghost, its Weyl coefficients. They turn out to be real and to come in inverse pairs, and the Galois question becomes whether one field automorphism inverts all of them at once. This is the form in which the problem is passed to the finite quantum dilogarithms in Sec.~\ref{sec:GR}.

\subsection{Weyl coefficients of ghosts}

Use the convention $D_{-p}=D_p^{-1}=D_p^\dagger$. Let $\tilde\Pi\in\Gamma_{U_P}$ be a ghost, as provided by Theorem~\ref{thm:ghost}, and let $\nu_p=\tr(\tilde\Pi D_p)$ be its Weyl coefficients. Since the $D_p$ form a basis of $M_d(\C)$, the $\nu_p$ determine $\tilde\Pi$. They have the following elementary properties.

\begin{lemma}\label{lem:nu}
\begin{enumerate}[nosep]
\item $\nu_p\in\mathbb R$ for all $p$.
\item $\nu_{-p}=1/((d+1)\nu_p)$ for $p\neq0$.
\item $U_P\tilde\Pi U_P$ has Weyl coefficients $\nu_{-p}$.
\end{enumerate}
\end{lemma}
\begin{proof}
For (1), $\overline{\nu_p}=\tr(\tilde\Pi^\dagger D_p^\dagger)=\tr(U_P\tilde\Pi U_PD_{-p})=\tr(\tilde\Pi D_p)$. Part (2) is~\eqref{eq:V}. For (3), $U_PD_pU_P=D_{-p}$.
\end{proof}

Parts (2) and (3) together say that conjugation by the parity operator acts on the Weyl coefficients as inversion, up to the constant factor $d+1$. For a SIC the coefficients would instead satisfy $|\nu_p|^2=1/(d+1)$. The difference between the two cases is therefore whether inversion of the $\nu_p$ is realized by complex conjugation (a SIC) or only algebraically (a ghost).

\subsection{Galois realization of transposition}

Corollary~\ref{cor:ghost} combines two different requirements:
\begin{enumerate}[label=(\roman*),nosep]
\item transposition acts on the orbit through \emph{some} Galois element;
\item that element is conjugate to $c$.
\end{enumerate}
Requirement (ii) is the archimedean part of the problem, and it is the subject of the rest of the paper. We first isolate requirement (i), which is purely algebraic.

\begin{conjecture}[GR]\label{conj:GR}
Let $O\subset V_d(\Qb)$ be a Galois orbit containing a ghost. Then there is $h_O\in\Gal(\Qb/\Q)$ with $\Pi^T=h_O(\Pi)$ for some (equivalently, every) $\Pi\in O$.
\end{conjecture}
The two forms are equivalent because $T$ commutes with Galois: if $T\Pi=h\Pi$, then $T(g\Pi)=(ghg^{-1})(g\Pi)$. By Proposition~\ref{prop:galois}, the orbit $O$ contains a SIC if and only if $h_O$ can be chosen conjugate to $c$.

For the ghost $\tilde\Pi$, Lemma~\ref{lem:nu} turns Conjecture~\ref{conj:GR} into a statement about its Weyl coefficients alone.

\begin{proposition}\label{prop:A}
Let $h\in\Gal(\Qb/\Q)$ fix $\tau$. Then $U_P\tilde\Pi U_P=h(\tilde\Pi)$ if and only if $h(\nu_p)=1/((d+1)\nu_p)$ for all $p\neq0$. In that case $T(\tilde\Pi)=(c\,h)(\tilde\Pi)$, and hence Conjecture~\ref{conj:GR} holds on the orbit of $\tilde\Pi$, with $h_O=ch$.
\end{proposition}
\begin{proof}
Since $h$ fixes every $D_p$, the matrix $h(\tilde\Pi)$ has Weyl coefficients $h(\nu_p)$. Compare these with Lemma~\ref{lem:nu}(2),(3). Since $\{D_p\}_{p\in\Z_d^2}$ is a basis of $M_d(\C)$, orthogonal for the Hilbert--Schmidt inner product, a matrix is uniquely determined by its Weyl coefficients $\tr(\,\cdot\,D_p)$. Finally, $T\tilde\Pi=c(\tilde\Pi^\dagger)=c(U_P\tilde\Pi U_P)=c\,h(\tilde\Pi)$.
\end{proof}

To state the result in its final form, write $\nu_p=\varepsilon_p/\sqrt{d+1}$. We call the $\varepsilon_p$ the \emph{overlap units} of the ghost. By Lemma~\ref{lem:nu} they are real and satisfy $\varepsilon_{-p}=\varepsilon_p^{-1}$. Proposition~\ref{prop:A} then says that Conjecture~\ref{conj:GR} asks for a field automorphism that inverts the whole family $(\varepsilon_p)$ simultaneously, up to sign. Zauner's conjecture asks in addition that this automorphism, composed with $c$, be a complex conjugation; equivalently, that $|\varepsilon_p|=1$ for all $p$ under some embedding of $\Qb$ into $\C$. Thus the existence of a SIC in dimension $d$ is a question about one explicit finite family of algebraic numbers. In Sec.~\ref{sec:GR} we identify these numbers with special values of finite quantum dilogarithms, where Stark arithmetic becomes available.

\section{Finite quantum dilogarithms and Stark arithmetic}\label{sec:GR}

This section is the core of the paper. Subsection~\ref{sec:dict} transfers the problem to the RW special values. Subsections~\ref{sec:struct} and~\ref{sec:reality} give structure theorems and the reality and signature conditions. Subsections~\ref{sec:PL}--\ref{sec:CFT} develop the place law, which subsumes those conditions. Subsection~\ref{sec:NG} recasts the place law as a unitarity statement for Galois conjugates of the near-group category attached to the Stark solution.

\subsection{The dictionary}\label{sec:dict}

The ghost overlaps are built from the special values $F^\pm_\gamma(u)$ of Ref.~\onlinecite{RW}. Under the dictionary of Ref.~\onlinecite{AFK26} (Prop.~2.1, eqs.~(8)--(10)), the map $p\mapsto-p$ corresponds to $u\mapsto-u$. We therefore study the RW variety directly.

Let $(G,\langle\cdot\rangle)$ be a metric group of order $N$, and let $Z=Z(G,\langle\cdot\rangle)$ be its finite set of finite quantum dilogarithms (Ref.~\onlinecite{RW}, Def.~4, Thm.~8). Set $K_0=\Q(\zeta_{4N})$, which contains $\sqrt N$ and the values of $\langle\cdot\rangle$. Every $E\in Z$ satisfies $E(0)^2=\sqrt N E(0)+1$ and $E(u)E(-u)=\langle u\rangle^{-1}$ for $u\ne0$ (Ref.~\onlinecite{RW}, Thm.~5). Let $Z_\pm=\{E:E(0)=(\sqrt N\pm\sqrt{N+4})/2\}$ and $L=K_0(\sqrt{N+4})$, and assume $\sqrt{N+4}\notin K_0$. The reflection $\rho E=E\circ(-1)$ maps $Z$ to itself (Ref.~\onlinecite{RW}, Prop.~6) and commutes with Galois. For the solution $E_*$ introduced below, we write R2 (reflection realized by Galois) for the statement that some $h\in\Gal(\Qb/L)$ satisfies $h(E_*)=\rho E_*$. Under the dictionary, R2 is the form taken by GR (Conjecture~\ref{conj:GR}).

Following (Ref.~\onlinecite{RW}, \S6), call $E$ \emph{real} if $\overline{E(x)}=\langle x\rangle E(x)$ (the ghost type), and \emph{unitary} if $\overline{E(x)}=\langle -x\rangle E(-x)$, which gives $|E|=1$ (the SIC type). A solution that is both real and unitary is even, that is, $\rho E=E$. Let $\sigma\in\Gal(\Qb/K_0)$ satisfy $\sigma(\sqrt{N+4})=-\sqrt{N+4}$.

\paragraph{From unitary solutions to SICs.} The passage from a unitary-type Galois conjugate of $E_*$ to a SIC does not require transferring the TCI or the order-$(2m+1)$ symmetry to the conjugate datum. Galois stability of $V_d$ does the work, through the following results.

\begin{lemma}[Hermiticity criterion]\label{lem:herm}
A point $\Pi\in V_d$ is Hermitian, that is, $\Pi\in S_d$, if and only if $(d+1)|\tr(\Pi D_p)|^2=1$ for all $p\ne0$.
\end{lemma}
\begin{proof}
Use the convention $D_p^\dagger=D_{-p}$, and put $a_p=\tr(\Pi D_p)$. Then $\tr(\Pi^\dagger D_p)=\overline{\tr(\Pi D_{-p})}=\overline{a_{-p}}$. Since $\{D_p\}$ is a basis of $M_d(\C)$, $\Pi$ is Hermitian if and only if $\overline{a_{-p}}=a_p$ for all $p$; for $p=0$ both sides equal $1$. By~\eqref{eq:V}, $(d+1)a_pa_{-p}=1$ for $p\neq0$.

If $(d+1)|a_p|^2=1$ for all $p\ne0$, then $\overline{a_{-p}}=1/((d+1)a_{-p})=a_p$, and hence $\Pi$ is Hermitian. Conversely, if $\Pi$ is Hermitian then $\overline{a_p}=a_{-p}$, and $(d+1)|a_p|^2=(d+1)a_pa_{-p}=1$.
\end{proof}

\begin{corollary}[unit-circle criterion]\label{cor:unit}
Let $\tilde\Pi\in V_d(\Qb)$ be a ghost with normalized overlaps $\varepsilon_p=\sqrt{d+1}\,\tr(\tilde\Pi D_p)$. If some $\sigma\in\Gal(\Qb/\Q)$ satisfies $|\sigma(\varepsilon_p)|=1$ for all $p\ne0$, then $\sigma(\tilde\Pi)$ is a WH-covariant SIC fiducial projector in dimension $d$.
\end{corollary}
\begin{proof}
By Lemma~\ref{lem:basic}(2), $\sigma(\tilde\Pi)\in V_d$. The automorphism $\sigma$ maps each $D_p$ to a root of unity times $D_{\pi(p)}$ for a permutation $\pi$ of $\Z_d^2\setminus\{0\}$. Hence the moduli $|\tr(\sigma(\tilde\Pi)D_q)|$ are the numbers $|\sigma(\tr(\tilde\Pi D_p))|=|\sigma(\varepsilon_p)|/\sqrt{d+1}$, using $\sigma(\sqrt{d+1})=\pm\sqrt{d+1}$. All of these equal $1/\sqrt{d+1}$, and hence Lemma~\ref{lem:herm} applies.
\end{proof}

Corollary~\ref{cor:unit} is an unconditional form of the observation in Ref.~\onlinecite{RW} (Rem.~7). To connect it with the finite quantum dilogarithm statements we use the AFK dictionary in the following form.

\begin{lemma}[dictionary]\label{lem:H}
Let $d\ge4$ and let $\tilde\Pi$ be the AFK ghost of dimension $d$ (rank $r=1$), with $\gamma=A_t=L^{2m+1}$, its attracting fixed point $\tau$, and $E_*=F^+_\gamma$ on $G_d=\Z^2/\Z^2(\gamma-I)$. For every $p\not\equiv0\pmod d$ there are a root of unity $\zeta_p$ and an element $u_p\neq0$ of $G_d$ such that
\[
\varepsilon_{p}=\zeta_p\,E_*(u_p)^{-1},
\]
where $\varepsilon_p$ is the normalized overlap $\sqrt{d+1}\,\tr(\tilde\Pi D_{p'})$ for the index $p'$ obtained from $p$ by the twist matrix $G$ of Ref.~\onlinecite{AFK25}, Eq.~(1.50). The roots of unity $\zeta_p$ and the map $p\mapsto u_p$ are given by formulas with rational coefficients, independent of the embedding.
\end{lemma}
\begin{proof}
By Ref.~\onlinecite{AFK25} (Def.~1.11, (1.12), (1.46), (1.47)) with $r=1$, the ghost fiducial is
\[
\tilde\Pi=\frac1d I+\frac{1}{d\sqrt{d+1}}\sum_{p\notin d\Z^2}\tilde\nu_{Gp}D_p,
\]
with normalized ghost overlaps
\[
\tilde\nu_p=\phi_p\,\mathbf v_{A_t}^{d^{-1}p}(\rho_t).
\]
Here $\phi_p=(-1)^{s_d(p)}e^{-\pi i\Psi(A_t)/12}\xi_d^{-f_{jm}Q(p)/f}$ is a root of unity (Ref.~\onlinecite{AFK25}, Def.~1.30 and Eq.~(1.44)), $\mathbf v$ is the Shintani--Faddeev modular cocycle, and $\rho_t$ is the fixed point of $A_t$. The cocycle is evaluated only at the non-integral arguments $d^{-1}p$, $p\not\equiv0$.

By Ref.~\onlinecite{AFK26} (Prop.~2.1), for $u\notin(\gamma^\top-I)\Z^2$, that is $u\neq0$ in $G_d$, one has
\[
F^\pm_\gamma(u)=e^{\pi i\Psi(\gamma)/12}\,\mathbf v_\gamma^{\frac1N(I-\gamma^{-1})Su}(\tau)^{-1},
\]
and the integral arguments correspond exactly to $u=0$. The computation in Ref.~\onlinecite{AFK26} (\S5), namely $\tfrac1N(I-\gamma^{-1})S(L-I)^\top=-d^{-1}L^{-m-1}S$, shows that every argument $d^{-1}p$ is of the form $\tfrac1N(I-\gamma^{-1})Su$ modulo $\Z^2$, for a unique class $u=u_p$, which is nonzero. Periodicity of $\mathbf v$ in its argument at real-multiplication points (Ref.~\onlinecite{KoppSF}, Prop.~4.35) completes the identification, and $F^+=F^-$ away from $u=0$.
\end{proof}

\begin{remark}
Lemma~\ref{lem:H} is assembled from the statements of Refs.~\onlinecite{AFK25,AFK26} as quoted above. The matching of conventions (orientation of $\tau$, the matrix $S$, the twist $G$, and row versus column vectors) is the bookkeeping carried out in Ref.~\onlinecite{AFK26} (\S2, \S5). A reader should check it against the original. The essential content is simple: each overlap is a single cocycle value at a non-integral argument, and hence $E_*(0)$ never occurs.

The use of Lemma~\ref{lem:H} in Theorem~\ref{thm:HSRUSIC} is insensitive to these conventions. The proof needs only two structural facts:
\begin{enumerate}[label=(\roman*),nosep]
\item each normalized overlap is a root of unity times $E(u_p)^{\pm1}$ for a single solution value with $u_p\neq0$;
\item the solution $E$ lies in the Galois orbit of $E_*$.
\end{enumerate}
A convention mismatch of the usual kind does not affect either fact. Examples are $u_p\mapsto\pm u_p$, $G\mapsto G^{-1}$, $S\mapsto S^\top$, $\mathbf v\mapsto\mathbf v^{-1}$, or evaluation at the other fixed point of $A_t$, which replaces $E_*$ by a Galois conjugate. HS and RU are assumed for all Galois conjugates, and hence the conclusion is unchanged.
\end{remark}

As a consequence, all arithmetic of the ghost overlaps is carried by the finite quantum dilogarithm values $E_*(u)$, $u\ne0$.

\subsection{Structure theorems}\label{sec:struct}

\begin{lemma}\label{lem:CM}
$L$ is a CM field. If $E\in Z_+$ is real and all its values lie in $L$, then $\sigma E\in Z_-$ is real. If moreover $E$ is not even, neither is $\sigma E$.
\end{lemma}
\begin{proof}
$K_0$ is CM and $\sqrt{N+4}$ is real, and hence complex conjugation is central in $\Gal(L/\Q)$. Hence $c(\sigma E)=\sigma(cE)=\sigma(\langle\cdot\rangle E)=\langle\cdot\rangle\sigma E$, using that $\sigma$ fixes $\langle\cdot\rangle$. Evenness is preserved since $\sigma\rho=\rho\sigma$.
\end{proof}

\begin{theorem}\label{thm:C}
Suppose $Z_+=\{E_*,\rho E_*\}$ with $E_*$ real and not even. Then at least one of the following holds:
\begin{enumerate}[label=(\alph*),nosep]
\item there is $h\in\Gal(\Qb/L)$ with $h(E_*)=\rho E_*$;
\item $Z_-$ contains a real solution that is not even.
\end{enumerate}
In particular, if $Z_-$ contains no non-even real solution, then reflection is realized by Galois on $E_*$.
\end{theorem}
\begin{proof}
Every $g\in\Gal(\Qb/L)$ fixes $K_0$, and hence $g$ maps $Z$ to itself. It also fixes $\sqrt{N+4}$, hence fixes $E(0)$, and hence $g$ permutes $Z_+$. By hypothesis $Z_+$ consists of exactly the two elements $E_*$ and $\rho E_*$.

Suppose first that some $g\in\Gal(\Qb/L)$ satisfies $g(E_*)\ne E_*$. Since $g(E_*)\in Z_+=\{E_*,\rho E_*\}$, necessarily $g(E_*)=\rho E_*$, which is~(a).

Otherwise every $g\in\Gal(\Qb/L)$ fixes $E_*$, that is, fixes each value $E_*(u)$. By Galois theory every value of $E_*$ then lies in $L$, and Lemma~\ref{lem:CM} shows that $\sigma E_*$ is a non-even real solution in $Z_-$, which is~(b).
\end{proof}

\begin{theorem}\label{thm:Cprime}
Let $E_*\in Z_+$ be real. If $\sigma E_*$ is unitary, then $h:=(\sigma^{-1}c\sigma)\,c$ lies in $\Gal(\Qb/L)$ and satisfies $h(E_*)=\rho E_*$.
\end{theorem}
\begin{proof}
Put $c'=\sigma^{-1}c\sigma$. From $c(\sigma E_*)=\langle\cdot\rangle\rho\sigma E_*=\sigma(\langle\cdot\rangle\rho E_*)$ we get $c'(E_*)=\langle\cdot\rangle\rho E_*$, and by definition $c(E_*)=\langle\cdot\rangle E_*$. Both $c$ and $c'$ invert roots of unity and fix $\sqrt{N+4}$, and hence $h=c'c$ fixes $L$. Then $h(E_*)=c'(\langle\cdot\rangle E_*)=\langle\cdot\rangle^{-1}\langle\cdot\rangle\rho E_*=\rho E_*$.
\end{proof}

Theorem~\ref{thm:Cprime} is the finite-dilogarithm form of ``Zauner $\Rightarrow$ GR''. Theorem~\ref{thm:C} is genuinely weaker in its hypothesis. It does not ask that the solutions with $E(0)<0$ be unitary, only that they not be real.

Theorem~\ref{thm:C} assumes that $Z_+$ is a single reflection pair. The next result removes that hypothesis when the relevant orbit has odd size. Let $c$ denote complex conjugation and put $\tilde c(E)=\langle\cdot\rangle^{-1}\,c(E)$. By Ref.~\onlinecite{RW} (Thm.~5(iv), Rem.~4), $\tilde c$ is an involution of $Z$. Real solutions are its fixed points, and unitary solutions are the fixed points of $\rho\tilde c$.

\begin{theorem}[orbit version]\label{thm:Corbit}
Let $E_*\in Z_+$ be real, and let $O=\Gal(\Qb/L)\cdot E_*$. Suppose $\rho O\neq O$, that is, reflection is not realized on $E_*$ by any element of $\Gal(\Qb/L)$. Then $\sigma(O)\subset Z_-$ contains no unitary solution and no even solution. Moreover, $\tilde c$ restricts to an involution of $\sigma(O)$ whose fixed points are exactly the real solutions in $\sigma(O)$. Consequently
\[
\#\{\text{real solutions in }\sigma(O)\}\equiv |O|\pmod 2 .
\]
In particular, if $|O|$ is odd, then $Z_-$ contains a non-even real solution.
\end{theorem}
\begin{proof}
Let $\Gamma=\Gal(\Qb/K_0)$ and $G_L=\Gal(\Qb/L)$, which has index $2$ in $\Gamma$. Complex conjugation normalizes both groups.

Take $g\in G_L$ and put $x=\sigma g$. Since $E_*$ is real and $cxc\in\Gamma$ fixes $\langle\cdot\rangle$,
\[
\tilde c(xE_*)=\langle\cdot\rangle^{-1}(cxc)(cE_*)=(cxc)E_*=x\,h E_*,\qquad h:=x^{-1}cxc .
\]
Both $x$ and $cxc$ flip $\sqrt{N+4}$, and hence $h\in G_L$. Hence $\tilde c$ maps $\sigma(O)$ to itself.

If $xE_*$ were unitary, then $\tilde c(xE_*)=\rho xE_*=x\rho E_*$, and hence $hE_*=\rho E_*$. That would put $\rho E_*$ in $O$, and hence $\rho O=O$, a contradiction.

If $xE_*$ were even, then $\rho E_*=E_*$, since $\rho$ commutes with Galois. Again $\rho O=O$, a contradiction.

The parity statement holds because an involution of a finite set fixes a number of points congruent to the size of the set modulo $2$.
\end{proof}
Theorem~\ref{thm:C} is the case $|O|=1$, since in that case $\rho O\ne O$ forces $E_*$ to be $L$-rational. When $|O|$ is even, the conclusion is only that $\sigma(O)$ consists of non-unitary solutions, each real or of mixed type. Excluding that possibility requires essentially the full unitarity of $Z_-$. Thus, the cardinality hypothesis in Theorem~\ref{thm:C} is doing real work, and an unconditional orbit-wise statement would need a new idea.

\paragraph{Complete solution sets in small cases.}
Theorem~5(iii) of Ref.~\onlinecite{RW} places $E$ in an eigenspace of $W=\langle\cdot\rangle^{-1}\circ\mathcal F$, with eigenvalue $\lambda$ satisfying $\lambda^3=\int\langle x\rangle^{-1}dx$. This reduces the pentagon system to a few quadratic equations in eigenspace coordinates, which we solved completely by homotopy continuation.

\begin{table}[!htbp]
\caption{All finite quantum dilogarithms in small cases.}\label{tab:fqd}\centering\small
\begin{tabular}{|@{}|l|l|r|l|l|@{}|}\hline\hline
$N$ & metric group & $|Z|$ & $Z_+$ & $Z_-$\\\hline
5 & $\gamma$ of cycle $(7)$ & 5 & 4 real, two $\rho$-pairs & 1 even (real and unitary)\\
6--9 & cyclic, $\langle x\rangle=\zeta_{2N}^{x(x+N)}$ & 4 & 1 real $\rho$-pair & 1 unitary non-even $\rho$-pair\\
16 & $\gamma=\alpha^3$, $\alpha=\left(\begin{smallmatrix}3&-1\\1&0\end{smallmatrix}\right)$ (SIC, $d=4$) & 8 & 4 real, two $\rho$-pairs & 4 unitary non-even\\\hline\hline
\end{tabular}

\end{table}

For $N=6,7,8,9$ the hypotheses of Theorem~\ref{thm:C} hold. A PSLQ test at $300$--$400$ digits shows that $E_*(1)\notin L$, while $E_*(1)+E_*(-1)\in L$ with relation height at most $4$. Thus, $\{E_*,\rho E_*\}$ is a single $\Gal(\Qb/L)$-orbit.

For the $d=4$ SIC group ($N=16$), let $\theta$ be a generic rational combination of the coordinates. At $120$ digits, $\theta(E)\notin L$ for every $E\in Z_+$, and $\theta(E)+\theta(\rho E)\in L$. The sums over the non-reflection pairs are not in $L$, while the sum over all four solutions is. Hence the orbit of $E_*$ contains $\rho E_*$.

In all these cases, then, reflection is realized by Galois.

\paragraph{A larger scan.} We applied the same solver to all $94$ conjugacy classes of hyperbolic $\gamma\in SL_2(\Z)$ with $5\le N\le30$. In every class, each solution is either real or unitary. Every solution in $Z_+$ is real, and every solution in $Z_-$ is unitary. No non-even real solution with $E(0)<0$ occurs; the only real solution with $E(0)<0$ is even ($N=5$). The size $|Z_+|$ ranges over $2,4,\dots,12$. The set $Z_-$ is empty for some classes with $N=12,14,21,28$. In every such class $\sqrt{N+4}\in K_0$, as it must be: otherwise $\sigma E_*\in Z_-$. The converse fails; for example, $N=12$ with cycle $(4,4)$ has $Z_-\ne\emptyset$. When $\sqrt{N+4}\in K_0$, any automorphism sending $d_-\mapsto d_+$ must move the cyclotomic coefficients. The pseudo-unitary companion predicted by Thornton's theorem (Ref.~\onlinecite{Tho}, Thm.~IV.3.6) then lies in the solution set of a Galois-conjugate metric datum, not in $Z_-$ itself.

In Izumi's construction the dimension of $\rho$ is $-\sqrt N/E(0)$ (Ref.~\onlinecite{RW}, \S5.4). Hence $E(0)<0$ corresponds to positive (Frobenius--Perron) dimension. Unitary-type solutions with $E(0)<0$ are exactly Izumi's input data for unitary near-group categories. The exceptional case of Theorem~\ref{thm:C} is therefore a pseudo-unitary solution of real type.

\subsection{Reality and signature conditions}\label{sec:reality}

\begin{conjecture}[real-or-unitary dichotomy]\label{conj:RU}
For the metric groups arising from hyperbolic $\gamma\in SL_2(\Z)$, every finite quantum dilogarithm is of real type or of unitary type. Equivalently, the involution $\tilde c$ of $Z$ satisfies $\tilde c(E)\in\{E,\rho E\}$ for every $E\in Z$.
\end{conjecture}
This holds in all $94$ classes with $N\le30$; no mixed-type solution was found. Combined with Theorem~\ref{thm:Corbit}, Conjecture~\ref{conj:RU} gives the following. If R2 fails on an orbit $O$ of any size, then $\sigma(O)$ consists entirely of non-even real solutions with $E(0)<0$.

Conjecture~\ref{conj:RU} together with the sign statement (no non-even real solution with $E(0)<0$) is equivalent to unitarity of all non-even solutions in $Z_-$. That is the full dichotomy, and it is of Zauner strength. The point of separating the two statements is that each alone is strictly weaker.

\paragraph{A reduced system, and total reality.} Put $x_u=s_uE(u)$, where $s_u^2=\langle u\rangle$ and $s_{-u}=s_u$. In these coordinates:
\begin{itemize}[nosep]
\item $\tilde c$ is ordinary complex conjugation, and $\rho$ is $x_u\mapsto x_{-u}$;
\item the pairing relation becomes $x_ux_{-u}=1$ for $u\neq 0$;
\item the eigenvalue condition $\hat E=\lambda\langle\cdot\rangle E$ says that $x$ lies in a subspace $X_\lambda$, the image of the eigenspace of the order-three unitary $W=\langle\cdot\rangle^{-1}\mathcal F$ (essentially Izumi's $R_{c'}$ (Ref.~\onlinecite{Izu}, Rem.~9.2)).
\end{itemize}
Both $\tilde c$ and $\rho$ preserve $X_\lambda$. Real type means $x\in\mathbb R^G$, and unitary type means $|x_u|=1$ for all $u\neq0$. Either way, every $t_u=x_u+x_u^{-1}$ is real.

\begin{conjecture}[total reality]\label{conj:TR}
Let $(G,\langle\cdot\rangle)$ be a metric group and $\lambda^3=\int\langle x\rangle^{-1}dx$. Every $x\in X_\lambda$ with $x_0^2=\sqrt N x_0+1$ and $x_ux_{-u}=1$ for all $u\ne0$ is real or unimodular. Equivalently, the even part $t$ of $x$ is real.
\end{conjecture}
\emph{Caution on interpretation.} Despite its linear-algebraic appearance, Conjecture~\ref{conj:TR} is \emph{not} a property of the Weil eigenspace $X_\lambda$. The control experiments below show that total reality fails for random subspaces with the same symmetries, and also fails for the true eigenspace once any of the pairing equations is dropped. TR is a property of the full, overdetermined solution variety, which is nonempty only because of the RW identities. Proposition~\ref{prop:TRarith} below shows that it is in fact an arithmetic statement.

The reduced system omits the pentagon equation. In every computed case it has the same solutions as the full RW system, apart from two classes, $N=9$ cycle $(2,2,5)$ and $N=16$ cycle $(3,3,3)$, where extra solutions appear; these are also totally real.

We tested Conjecture~\ref{conj:TR} on all $268$ nondegenerate cyclic metrics $\langle x\rangle=\zeta_{2N}^{a x(x+N)}$ with $5\le N\le21$, and on the $54$ $SL_2(\Z)$ classes with $N\le20$. Every solution is real or unimodular, and none is of mixed type. Conjecture~\ref{conj:TR} implies Conjecture~\ref{conj:RU} coordinatewise.

\begin{proposition}[arithmetic form of TR]\label{prop:TRarith}
Let $\mathcal S$ be the union of the solution sets of Conjecture~\ref{conj:TR} over all Galois conjugates of the metric datum $(\langle\cdot\rangle,\sqrt N,\lambda)$. Then $\mathcal S$ is $\Gal(\Qb/\Q)$-stable. Conjecture~\ref{conj:TR} holds on all of $\mathcal S$ if and only if, for every solution, the field $\Q(t_u:u\in G)$ is totally real. Equivalently, every Galois conjugate of every $x_u$ is real or of absolute value $1$.
\end{proposition}
\begin{proof}
Galois automorphisms act coordinatewise and map the data $s_u$, $\langle\cdot\rangle$ and $\lambda$ to conjugate data, and hence $\mathcal S$ is stable and $\sigma(t_u(E))=t_u(\sigma E)$. Hence every conjugate of $t_u$ is again a $t$-coordinate of a solution, and it is real exactly when TR holds there. Finally, $x+x^{-1}\in\mathbb R$ if and only if $x\in\mathbb R$ or $|x|=1$.
\end{proof}
For the Stark solution $E_*$, the rank-one Stark conjecture predicts exactly such a pattern: each conjugate is real at the distinguished real place and of absolute value $1$ elsewhere. Thus, Conjecture~\ref{conj:TR}, across all conjugate metric data, is a \emph{weakened archimedean Stark statement}. It asserts that every conjugate is real or unimodular, without saying which. The sign--type correlation is precisely the missing information about which conjugates are which. TR is therefore not independent of Stark arithmetic, but it is strictly weaker than the full archimedean Stark statement.

\emph{Salem-type behaviour.} The condition in Conjecture~\ref{conj:TR} has the shape of the defining property of Salem numbers: all conjugates real or unimodular, with totally real trace $\tau+\tau^{-1}$. We computed the minimal polynomials of $t_u$ for the Stark solutions with $N=6,7,9$. In every case they are totally real, which confirms TR at the level of minimal polynomials, independently of solver completeness. Exactly half of the roots lie outside $[-2,2]$. Thus, the units $x_u$ are Salem-type but not Salem: Salem numbers would have exactly one root outside. The half--half split is what the rank-one Stark conjecture predicts, since the places of the ray class field lie equally over the two infinite places of the real quadratic base field. Section~\ref{sec:PL} makes this precise: the base field is $F=\Q(\sqrt{N(N+4)})$, and the split is by the place of $F$ (Definition~\ref{def:PL}).

\begin{conjecture}[half split, HS]\label{conj:HS}
For the Stark solution $E_*$ and every $u\neq0$, all Galois conjugates of $t_u=x_u+x_u^{-1}$ are real. Moreover, exactly half of them satisfy $|t|>2$ and the other half satisfy $|t|<2$.
\end{conjecture}
Conjecture~\ref{conj:HS} strengthens Conjecture~\ref{conj:TR} for $E_*$. Up to the boundary case $|t|=2$, the two halves are the conjugates at which $x_u$ is real and those at which it is unimodular. HS records the archimedean \emph{signature} predicted by Stark, but not which embedding is the distinguished one. Section~\ref{sec:PL} identifies the two halves: they are the embeddings above the two infinite places of the real quadratic field of $\gamma$. Its relation to the other statements is as follows.

\begin{proposition}\label{prop:HSRU}
Suppose $\sqrt{N+4}\notin K_0$. If Conjecture~\ref{conj:HS} holds for $E_*$ and Conjecture~\ref{conj:RU} holds for all Galois-conjugate metric data, then some Galois conjugate of $E_*$ is of unitary type.
\end{proposition}
\begin{proof}
By HS, some $\sigma\in\Gal(\Qb/\Q)$ gives $|\sigma(t_u)|<2$ for some $u$. Then $\sigma(x_u)$ is unimodular and not real. The conjugate $\sigma E_*$ solves the system for the conjugate metric datum, and it is not of real type because $\sigma(x_u)\notin\mathbb R$. By Conjecture~\ref{conj:RU} it is therefore of unitary type.
\end{proof}

\begin{theorem}[HS and RU give SICs]\label{thm:HSRUSIC}
Let $d\ge4$ with $d+1$ neither a square nor twice a square, that is, with $\sqrt{N+4}\notin K_0$ (Proposition~\ref{prop:samemetric}). If Conjecture~\ref{conj:HS} holds for $E_*$ and Conjecture~\ref{conj:RU} holds for all Galois-conjugate metric data of $G_d$, then a WH-covariant SIC exists in dimension $d$.
\end{theorem}
\begin{proof}
By Proposition~\ref{prop:HSRU} there is $\sigma$ with $\sigma E_*$ of unitary type. For $u\ne0$, unitary type gives
\[
|\sigma E_*(u)|^2=\langle u\rangle'\,\sigma E_*(u)\,\sigma E_*(-u)=1,
\]
where $\langle\cdot\rangle'$ is the conjugate metric. By Lemma~\ref{lem:H}, $\sigma(\varepsilon_p)=\sigma(\zeta_p)\,\sigma(E_*(u_p))^{-1}$ with $u_p\ne0$. Hence $|\sigma(\varepsilon_p)|=1$ for every $p\not\equiv0$, and Corollary~\ref{cor:unit} gives the SIC.
\end{proof}
Conversely, Stark's archimedean conjecture implies HS. HS is therefore not ``easier than Zauner'' in any absolute sense: it is Zauner-strength modulo the dichotomy RU. What HS isolates is the counting part of the archimedean information, and RU isolates the coordinatewise purity part.

\emph{Control experiments.} Total reality is not a generic feature of such systems. We ran two controls.
\begin{enumerate}[nosep]
\item Replace $X_\lambda$ by a random $\rho$- and conjugation-invariant subspace with dimensions chosen such that the pairing system is square. Then non-real solutions are common, for example only $16$ of $64$ solutions are real or unimodular for $N=13$.
\item Keep the true Weil eigenspace but impose only a random square subset of the pairing equations. Again non-real solutions appear in most cases, along with mixed solutions in which some coordinates are real and others unimodular.
\end{enumerate}
Thus, the phenomenon requires both the Weil eigenspace and the \emph{full} overdetermined set of pairing equations. Any proof must use the whole system, and no square subsystem can be totally real in general. In particular, a representation-theoretic proof based on the order-three Weil operator alone cannot work. A total-reality mechanism of Shapiro type, such as a positive-definite trace form on the coordinate ring, would have to be built from the complete overdetermined system.

Izumi already observed this phenomenon in small cases (Ref.~\onlinecite{Izu}, Rem.~9.2). He noted that his $C^*$ equations split into a Galois-stable part and two extra equations, $Jb=b$ and $d>0$, which ``come from the unitarity''. He also noted that the Galois conjugates with $d<0$ satisfy all equations except those two.

\begin{proposition}[categorical form of R2]\label{prop:cat}
Let $\mathcal C_E$ be the Izumi near-group category attached to $E$ (Ref.~\onlinecite{RW}, Thm.~12), and let $h\in\Gal(\Qb/K_0)$. Then $h(E)=\rho E$ if and only if the Galois-conjugate category $h(\mathcal C_E)$ is equivalent to $\mathcal C_E$ by an equivalence that acts on the invertible objects by $g\mapsto -g$.
\end{proposition}
\begin{proof}
By Ref.~\onlinecite{RW} (\S5.5), $E$ is recovered from the category up to relabelling $G$ by an automorphism, and the relabelling is read off from the action on the invertible objects. The category $h(\mathcal C_E)$ is built from the data $(\langle\cdot\rangle,h(E))$, since $h$ fixes $K_0$. Relabelling by $-1$ preserves $\langle\cdot\rangle$, and hence $\mathcal C_{\rho E}\simeq\mathcal C_E$ via $g\mapsto-g$.
\end{proof}
Thus, R2 says that $\mathcal C_E$ is Galois-stable with a prescribed action on its invertibles. By Ref.~\onlinecite{Izu} (Cor.~13.5), $\mathrm{Aut}(\mathcal C_E)$ consists of the automorphisms of $G$ that preserve $(\langle\cdot\rangle,E)$. Hence for non-even $E$ the relabelling $-1$ is not an autoequivalence of $\mathcal C_E$, and the equivalence in Proposition~\ref{prop:cat} is necessarily Galois-twisted. Note that reflection does not produce a new category: $\mathcal C_{\rho E}\simeq\mathcal C_E$. The arithmetic content lies entirely in whether a Galois automorphism realizes this autoequivalence. In the same way, real type corresponds to $\tilde c$ fixing $E$, which is the complex conjugate datum, and unitary type to $\tilde c$ acting as $g\mapsto -g$. This is the form in which Izumi's $C^*$-condition appears.

\begin{remark}[relation to the Haagerup--Izumi programme]\label{rem:HI}
Gannon, Schopieray and Yadav~\cite{GSY} construct cyclic Haagerup--Izumi (HI) categories from Barnes double-sine values. They work on the branch $\dim\rho=d_-$, and their pseudo-unitary $d_+$ categories are Galois conjugates. Such a conjugate is unitary exactly when its HI matrix is Hermitian (Ref.~\onlinecite{EG}, Thm.~2(c)). Exact Hermiticity checks verify this for every odd $3\le N\le31$ (Ref.~\onlinecite{GSY}, Rem.~3.8), and the introduction of Ref.~\onlinecite{GSY} reports verification for all odd $N\le43$. The uniform statement remains open (Ref.~\onlinecite{GSY}, Rem.~3.8).

The near-group situation here is closely parallel. Unitary-type solutions with $E(0)<0$ are Izumi's data for unitary near-group categories. The dichotomy ``$Z_-$ is unitary'' is the near-group analogue of the open HI unitarity question. Moreover, for every odd $N\ge3$, the equivariantization of an untwisted cyclic HI category for $\Z/N$ is a near-group category of type $(G\times\widehat G,\,N^2)$ with $G=\Z/N$ and quadratic form $q(x,\chi)=\chi(x)$ (Ref.~\onlinecite{GSY}, Cor.~6.5). This is the Izumi type $(G,|G|)$ considered here. It does not, however, reach the SIC groups directly. The SIC metric group $G_d$ has order $d^2(d-3)$, which is a square $n^2$ only when $d-3=k^2$, and then $n=dk$ is even, since $d=k^2+3$ is even whenever $k$ is odd. Thus, the odd-$N$ family of Ref.~\onlinecite{GSY} never contains a SIC group. Transferring HI unitarity results to $Z_-$ for $G_d$ would require extending those methods to other metric groups, including groups of even order. Section~\ref{sec:NG} makes the near-group side of this analogy precise: for a pseudo-unitary conjugate, unitarity reduces to Izumi's Hermiticity condition, which is exactly unitary type (Proposition~\ref{prop:izumi}).
\end{remark}

\begin{remark}[on the strength of this evidence]
In every computed case, $Z_-$ consists of unitary solutions. By Theorem~\ref{thm:Cprime} this already forces the conclusion. Thus, the computations are consistent with Conjecture~\ref{conj:GR}, but they do not show that GR is easier than Zauner's conjecture. What does separate the two is the weaker hypothesis of Theorem~\ref{thm:C}: when $|Z_+|=2$, R2 follows as soon as $Z_-$ contains no non-even real solution. An earlier hope that $|Z_+|=2$ always holds is false: $|Z_+|\in\{2,4,6,8,10,12\}$ all occur (Table~\ref{tab:data}).
\end{remark}

\paragraph{The sign statement and the sign--type correlation.} Two further statements recur below. Let $\sigma\in\Gal(\Qb/K_0)$ with $\sigma(\sqrt{N+4})=-\sqrt{N+4}$, as in Lemma~\ref{lem:CM}.
\begin{itemize}[nosep]
\item The \emph{sign statement}: no non-even real solution has $E(0)<0$. This is the hypothesis under which Theorem~\ref{thm:C} yields R2.
\item The \emph{sign--type correlation}: every solution with $E(0)<0$ is of unitary type, or at least the specific conjugate $\sigma E_*$ is. This implies $\mathrm{Z}(d)$ directly, through Lemma~\ref{lem:H} and Corollary~\ref{cor:unit}, since $\sigma E_*$ always exists. Given TR, it also implies HS.
\end{itemize}

\paragraph{How these conditions are related.} Section~\ref{sec:PL} introduces the place law PL and shows that it implies each of the conditions of this subsection. The implications are:
\[
\begin{array}{r@{\ \Longrightarrow\ }l@{\qquad}l}
\mathrm{PL} & \mathrm{HS}\ \Longrightarrow\ \mathrm{TR} & (\text{Prop.~\ref{prop:PLforms}})\\
\mathrm{PL} & \text{sign--type correlation} & (\text{Thm.~\ref{thm:PLtype}, Lemma~\ref{lem:dimsign}})\\
\mathrm{PL} & \text{RU on the Galois orbit of }E_* & (\text{Thm.~\ref{thm:PLtype}})\\
\mathrm{PL} & \mathrm{GR} & (\text{Thm.~\ref{thm:Cprime}})
\end{array}
\]
In addition, the parity of PL is detected $2$-adically (Corollary~\ref{cor:2adic}). HS, TR, RU and the sign--type correlation are therefore best read as shadows of PL: HS records the count, TR the reality, RU the purity, and the sign--type correlation the restriction to automorphisms fixing $K_0$.

\paragraph{A counting form of HS.} Put $B_u=\{\sigma:|\sigma(t_u)|>2\}$, and let $A$ be the set of $\sigma$ at which $\sigma E_*$ has negative categorical dimension. By Lemma~\ref{lem:dimsign}, $A=\{\sigma:\sigma(\sqrt D)>0\}$. On $\Gal(\Qb/K_0)$ this is the set where $\sigma E_*(0)>0$, but over the full Galois group the sign of $E(0)$ is \emph{not} the right invariant (Sec.~\ref{sec:PL}).
\begin{itemize}[nosep]
\item $|A|$ is exactly half of the Galois group, for trivial reasons. Thus, given TR, the whole content of HS is that $|B_u|$ is also half: $|A|=|B_u|$.
\item The place law is the equality $A=B_u$ for every $u$.
\end{itemize}
HS is thus the counting shadow of PL, and the converse fails in general. Showing that two counts agree is the kind of statement global Galois or regulator arguments can often reach. However, HS alone does not give SICs without RU, whereas PL does (Theorem~\ref{thm:PLSIC}). In the data, the equality $A=B_u$ holds in every case (Appendix~\ref{app:PL}).

\subsection{The place law and its single-unit and 2-adic forms}\label{sec:PL}\label{sec:2adic} This subsection states the place law and proves that it gives SICs. It then reduces the archimedean input to a single unit, and finally shows which part of that input is visible $2$-adically.
\paragraph{Which conjugates are unitary.}
The statements HS, RU and the sign--type correlation all concern the same question: at which embeddings is a conjugate of $E_*$ real, and at which is it unitary? This subsection answers it in a Galois-invariant form. The answer is governed by the real quadratic field of $\gamma$, not by the sign of $E(0)$ alone. It yields a single statement that implies Zauner's conjecture without RU (Theorem~\ref{thm:PLSIC}), and a reduction of the archimedean input to one unit (Theorem~\ref{thm:single}).

Throughout, $E_*\in Z_+$ is the real-type Stark solution of Sec.~\ref{sec:GR}. We fix $\Qb\subset\C$ with $\sqrt N>0$ and $\sqrt{N+4}>0$, and we regard every $\sigma\in\Gal(\Qb/\Q)$ as an embedding $\Qb\to\C$. Put
\[
D=N(N+4)=(\operatorname{tr}\gamma)^2-4,\qquad F=\Q(\sqrt D),\qquad \sqrt D=\sqrt N\sqrt{N+4}>0 .
\]
Thus, $F$ is the real quadratic field generated by the fixed point of $\gamma$. Since $(N+2)^2-4$ is never a square for $N\ge1$, $F$ is a genuine quadratic field. Its two infinite places $\infty_\pm$ are $\sqrt D\mapsto\pm\sqrt D$. As before, $x_u=s_uE_*(u)$ and $t_u=x_u+x_u^{-1}$. Recall $x_{-u}=x_u^{-1}$, and $x_u=\pm1$ when $2u=0$.

\paragraph{Types of conjugates.} A conjugate $\sigma E_*$ solves the RW system for the conjugate datum, in which $\langle\cdot\rangle$, $\sqrt N$ and $E(0)$ are replaced by $\sigma\langle\cdot\rangle$, $\sigma(\sqrt N)=\pm\sqrt N$ and $\sigma(E_*(0))$. Since $|\sigma(s_u)|=1$ and $\sigma(s_u)^2=\sigma\langle u\rangle$, a direct computation from the definitions gives:
\begin{itemize}[nosep]
\item $\sigma E_*$ is of real type if and only if $\sigma(x_u)\in\mathbb R$ for all $u$;
\item $\sigma E_*$ is of unitary type if and only if $|\sigma(x_u)|=1$ for all $u\ne0$.
\end{itemize}
For example, if $|\sigma(x_u)|=1$, then
\[
\overline{\sigma E_*(u)}=\frac{\sigma(s_u)}{\sigma(x_u)}=\sigma\langle u\rangle\,\frac{\sigma(x_{-u})}{\sigma(s_{-u})}=\sigma\langle u\rangle\,\sigma E_*(-u).
\]

\begin{lemma}[sign of the categorical dimension]\label{lem:dimsign}
For $\sigma\in\Gal(\Qb/\Q)$, the categorical dimension $d_\rho(\sigma E_*)=-\sigma(\sqrt N)/\sigma(E_*(0))$ of the conjugate datum satisfies
\[
\operatorname{sign} d_\rho(\sigma E_*)=-\operatorname{sign}\sigma(\sqrt D).
\]
In particular, for $\sigma\in\Gal(\Qb/K_0)$ one has $d_\rho(\sigma E_*)>0$ if and only if $\sigma(E_*(0))<0$.
\end{lemma}
\begin{proof}
$\sigma(E_*(0))=(\sigma(\sqrt N)+\sigma(\sqrt{N+4}))/2$ has the sign of $\sigma(\sqrt{N+4})$, because $\sqrt{N+4}>\sqrt N$. Hence $\operatorname{sign} d_\rho=-\operatorname{sign}\sigma(\sqrt N)\operatorname{sign}\sigma(\sqrt{N+4})=-\operatorname{sign}\sigma(\sqrt D)$. If $\sigma$ fixes $K_0\ni\sqrt N$, the first factor is $+1$.
\end{proof}

Lemma~\ref{lem:dimsign} shows that the condition ``$E(0)<0$'' of the earlier sections is the correct one only for automorphisms fixing $K_0$. Lemma~\ref{lem:ngdim} below gives the exact value $d_\rho(\sigma E_*)=(N-\sigma(\sqrt D))/2$, and shows that positive categorical dimension is the same as pseudo-unitarity of the conjugate near-group category. Over the full Galois group, the invariant quantity is the sign of the categorical dimension, or equivalently the place of $F$ below $\sigma$. This matters as soon as $\sigma$ moves $\sqrt N$, and the data confirm it. For $N=6$ and $u=2$, $\sqrt{N+4}$ and $\sqrt N$ both lie in $\Q(t_u)$, but each splits the conjugates of $t_u$ $4/4$ against the pattern $|t|\gtrless2$, whereas $\sqrt D$ splits them $8/0$ (Appendix~\ref{app:PL}).

\begin{definition}[place law]\label{def:PL}
Let $u\in G_d$ with $2u\ne0$ and $t_u\ne\pm2$. We say that \emph{$E_*$ satisfies the place law at $u$}, written $\mathrm{PL}(u)$, if for every $\sigma\in\Gal(\Qb/\Q)$ the number $\sigma(t_u)$ is real and
\[
\big(\sigma(t_u)^2-4\big)\,\sigma(\sqrt D)>0 .
\]
That is, $|\sigma(t_u)|>2$ above $\infty_+$ and $|\sigma(t_u)|<2$ above $\infty_-$.
\end{definition}
\begin{conjecture}[place law]\label{conj:PL}
For every $d\ge4$, the Stark solution $E_*$ of the SIC group $G_d$ satisfies $\mathrm{PL}(u)$ for every $u\in G_d$ with $2u\ne0$.
\end{conjecture}
By Theorem~\ref{thm:PLSIC} below, Conjecture~\ref{conj:PL} implies Zauner's existence conjecture for WH-covariant SICs in every dimension $d\ge4$. It is implied by the archimedean part of the rank-one Stark conjecture.

\begin{proposition}[equivalent forms of PL]\label{prop:PLforms}
Let $2u\neq0$ and $t_u\neq\pm2$, and put $y_u=D^{1/4}(x_u-x_u^{-1})$ for a fixed real fourth root $D^{1/4}$. The following are equivalent.
\begin{enumerate}[label=(\roman*),nosep]
\item $\mathrm{PL}(u)$.
\item $\Q(t_u)$ is totally real, $\sqrt D\in\Q(t_u)$, and $(t_u^2-4)\sqrt D$ is totally positive in $\Q(t_u)$.
\item The field $\Q(t_u,y_u)$ is totally real.
\end{enumerate}
If they hold, then exactly $m_u=[\Q(t_u):F]$ conjugates of $t_u$ satisfy $|t|>2$, and exactly $m_u$ satisfy $|t|<2$. In particular $\mathrm{PL}(u)$ implies HS at $u$.
\end{proposition}
\begin{proof}
(i)$\Rightarrow$(ii). If $\sqrt D\notin\Q(t_u)$, some $\sigma,\sigma'$ agree on $t_u$ but $\sigma'(\sqrt D)=-\sigma(\sqrt D)$, which contradicts~(i). The rest of (ii) is a restatement of (i). (ii)$\Rightarrow$(i) is immediate.

(i)$\Leftrightarrow$(iii). We have $y_u^2=(t_u^2-4)\sqrt D$. If (i) holds, every $\sigma(y_u)^2$ is positive and every $\sigma(t_u)$ is real, and hence $\Q(t_u,y_u)$ is totally real. Conversely, if $\Q(t_u,y_u)$ is totally real, then $\sigma(y_u)^2>0$, since $y_u\ne0$. Also $\sigma(\sqrt D)=\sigma(y_u)^2/\sigma(t_u^2-4)$ is real and nonzero, and hence $(\sigma(t_u)^2-4)\sigma(\sqrt D)=\sigma(y_u)^2>0$.

For the count, $F\subset\Q(t_u)$, and exactly half of the $[\Q(t_u):\Q]=2m_u$ embeddings of $\Q(t_u)$ lie above each of $\infty_\pm$.
\end{proof}

Form (iii) shows that the place law is a total-reality statement of the same kind as TR (Proposition~\ref{prop:TRarith}). TR asks that the even part $x_u+x_u^{-1}$ generate a totally real field. PL asks the same, in addition, for the odd part $x_u-x_u^{-1}$ twisted by $D^{1/4}$.

\begin{theorem}[the place law determines every type]\label{thm:PLtype}
Suppose $t_u\ne\pm2$ whenever $2u\ne0$. Then $\mathrm{PL}(u)$ holds for all $u$ with $2u\ne0$ if and only if, for every $\sigma\in\Gal(\Qb/\Q)$:
\begin{itemize}[nosep]
\item $\sigma E_*$ is of real type when $\sigma(\sqrt D)>0$;
\item $\sigma E_*$ is of unitary type when $\sigma(\sqrt D)<0$.
\end{itemize}
Equivalently, by Lemma~\ref{lem:dimsign}, a conjugate of $E_*$ is unitary if and only if its categorical dimension is positive.
\end{theorem}
\begin{proof}
Suppose PL holds for all $u$, and fix $\sigma$.
\begin{itemize}[nosep]
\item If $\sigma(\sqrt D)>0$, then each $\sigma(t_u)$ is real with $|\sigma(t_u)|>2$. Thus, $\sigma(x_u)$, a root of $X^2-\sigma(t_u)X+1$, is real. Hence $\sigma E_*$ is of real type.
\item If $\sigma(\sqrt D)<0$, then $\sigma(t_u)\in(-2,2)$, and hence $|\sigma(x_u)|=1$ for every $u\ne0$. Hence $\sigma E_*$ is of unitary type.
\end{itemize}
The two indices $u$ with $2u=0$ and $u=0$ impose no condition. The type is decided by $\sigma|_F$ alone, and hence it is the same for all $u$ simultaneously.

Conversely, real type gives $\sigma(x_u)\in\mathbb R$ and hence $|\sigma(t_u)|\ge2$. Unitary type gives $|\sigma(x_u)|=1$ and hence $\sigma(t_u)\in[-2,2]$. The value $\pm2$ is excluded by hypothesis, and hence the inequalities are strict and match the sign of $\sigma(\sqrt D)$.
\end{proof}

\begin{theorem}[the place law gives SICs]\label{thm:PLSIC}
Let $d\ge4$ and let $E_*$ be the Stark solution for $G_d$. If $\mathrm{PL}(u)$ holds for every $u\in G_d$ with $2u\ne0$, then a WH-covariant SIC exists in dimension $d$. No assumption of RU, and no assumption on $d+1$, is needed.
\end{theorem}
\begin{proof}
$\sqrt D$ is irrational, and hence some $\sigma$ satisfies $\sigma(\sqrt D)<0$. By Theorem~\ref{thm:PLtype}, $\sigma E_*$ is of unitary type, and hence $|\sigma(E_*(u))|=1$ for $u\ne0$. The proof of Theorem~\ref{thm:HSRUSIC} (Lemma~\ref{lem:H} and Corollary~\ref{cor:unit}) then gives the SIC.
\end{proof}

\paragraph{Relation to the earlier statements.} The place law is the Galois-invariant completion of the conjectures of this section.
\begin{itemize}[nosep]
\item \emph{TR.} PL implies TR for every conjugate of $E_*$, by form~(ii).
\item \emph{HS.} PL implies HS, and it identifies the two halves: they are the embeddings above $\infty_+$ and above $\infty_-$ (Proposition~\ref{prop:PLforms}).
\item \emph{RU.} By Theorem~\ref{thm:PLtype}, PL for all $u$ implies RU on the Galois orbit of $E_*$, which is all that Theorem~\ref{thm:HSRUSIC} uses. The danger that different $u$ choose different types, which RU was introduced to exclude, cannot occur, because under PL the type depends only on $\sigma|_F$.
\item \emph{Sign--type correlation.} Restricted to $\sigma\in\Gal(\Qb/K_0)$, Theorem~\ref{thm:PLtype} is exactly the sign--type correlation of Sec.~\ref{sec:summary}, by Lemma~\ref{lem:dimsign}. PL is that correlation extended to automorphisms moving $\sqrt N$.
\item \emph{R2 and GR.} Suppose $\sqrt{N+4}\notin K_0$. Take $\sigma\in\Gal(\Qb/K_0)$ with $\sigma(\sqrt{N+4})=-\sqrt{N+4}$. Then $\sigma(\sqrt D)<0$, and hence PL makes $\sigma E_*$ unitary, and Theorem~\ref{thm:Cprime} gives R2 with the explicit element $h=(\sigma^{-1}c\sigma)c$. Thus PL implies Conjecture~\ref{conj:GR} as well.
\item \emph{Stark.} PL says that each $x_u$ is real at every place above $\infty_+$ and has absolute value $1$ at every place above $\infty_-$. This is exactly the shape of the archimedean part of the rank-one Stark conjecture over $F$. Thus, PL is not weaker than Stark's archimedean statement; it is that statement, with the distinguished place made explicit.
\end{itemize}
Accordingly, Theorem~\ref{thm:PLSIC} replaces the pair HS $+$ RU of Theorem~\ref{thm:HSRUSIC} by a single statement. Moreover, each $\mathrm{PL}(u)$ concerns one algebraic number at a time.

\paragraph{Evidence.} For $39$ of the $94$ classes of Table~\ref{tab:fullscan} ($5\le N\le18$), including the $d=4$ SIC group ($N=16$, cycle $(3,3,3)$), we identified the minimal polynomials of the traces $t_u$ of the real solutions in $Z_+$. These are $100$ distinct polynomials, of degree up to $32$ (Table~\ref{tab:PL}). Each was tested exactly in PARI/GP for three properties:
\begin{itemize}[nosep]
\item total reality;
\item whether $\sqrt D$ lies in $\Q(t_u)$;
\item the sign of $(\sigma(t_u)^2-4)\sigma(\sqrt D)$ at every embedding $\sigma$.
\end{itemize}
In all $100$ cases the place law holds, with no exception. It also holds in the SIC dimension $d=5$ ($N=50$), at all $16$ embeddings (Sec.~\ref{sec:PL}, end). In contrast, $\Q(\sqrt{N+4})$ and $\Q(\sqrt N)$ are often not contained in $\Q(t_u)$, and when they are, they do not predict the split.

\paragraph{Reduction of the archimedean input to a single unit.}
Corollary~\ref{cor:unit} needs \emph{all} overlap units to be unimodular under one embedding. Theorem~\ref{thm:PLSIC} obtains this from PL at every $u$. The next result shows that algebraic information transports unimodularity from one unit to all of them.

\begin{theorem}[single-unit reduction]\label{thm:single}
Let $u_0\in G_d$, and suppose:
\begin{enumerate}[label=(G\arabic*),nosep]
\item \emph{(generation)} $x_v\in\Q(x_{u_0})$ for every $v\in G_d$;
\item \emph{(inversion)} every field embedding $\kappa:\Q(x_{u_0})\to\Qb$ with $\kappa(x_{u_0})=x_{u_0}^{-1}$ satisfies $\kappa(x_v)=x_v^{-1}$ for all $v$.
\end{enumerate}
If some Galois conjugate of $x_{u_0}$ lies on the unit circle, then the corresponding conjugate $\sigma E_*$ is of unitary type, and a WH-covariant SIC exists in dimension $d$.
\end{theorem}
\begin{proof}
Let $\sigma\in\Gal(\Qb/\Q)$ with $|\sigma(x_{u_0})|=1$, and put $c_\sigma=\sigma^{-1}c\sigma\in\Gal(\Qb/\Q)$. Then
\[
c_\sigma(x_{u_0})=\sigma^{-1}\big(\overline{\sigma(x_{u_0})}\big)=\sigma^{-1}\big(\sigma(x_{u_0})^{-1}\big)=x_{u_0}^{-1}.
\]
Thus, $c_\sigma$ restricts to an embedding $\kappa$ of $\Q(x_{u_0})$ as in~(G2), and $c_\sigma(x_v)=x_v^{-1}$ for every $v$, by~(G1) and~(G2). Applying $\sigma$ gives $\overline{\sigma(x_v)}=\sigma(c_\sigma x_v)=\sigma(x_v)^{-1}$, that is, $|\sigma(x_v)|=1$ for all $v\ne0$. Hence $\sigma E_*$ is of unitary type, and the SIC follows as in Theorem~\ref{thm:PLSIC}.
\end{proof}

\begin{corollary}[R2 version]\label{cor:singleR2}
Assume R2, that is, $h(E_*)=\rho E_*$ for some $h\in\Gal(\Qb/L)$, and assume~(G1) for some $u_0$. If some conjugate of $x_{u_0}$ is unimodular, then a SIC exists in dimension $d$.
\end{corollary}
\begin{proof}
Since $h$ fixes $s_v\in K_0$, we have $h(x_v)=s_vE_*(-v)=x_{-v}=x_v^{-1}$ for all $v$. An embedding of $\Q(x_{u_0})$ is determined by the image of $x_{u_0}$, and hence every $\kappa$ as in~(G2) equals $h|_{\Q(x_{u_0})}$, and~(G2) holds.
\end{proof}

Theorem~\ref{thm:Cprime} gives the converse direction, $\mathrm{Z}\Rightarrow\mathrm{R2}$. Thus, modulo the algebraic conditions (G1)--(G2), the archimedean content of Zauner's conjecture is concentrated in one algebraic number: \emph{does $x_{u_0}$ have a conjugate on the unit circle?} The next corollary turns this into the sign of a rational integer.

\begin{corollary}[integer criterion]\label{cor:integer}
Let $P$ be the minimal polynomial of $t_{u_0}$, of degree $n$, and suppose $[\Q(x_{u_0}):\Q(t_{u_0})]=2$. Then
\[
P(2)P(-2)=\prod_{i}(t_i^2-4)=(-1)^{n}\,N_{\Q(x_{u_0})/\Q}\big(x_{u_0}^2-1\big),
\]
where $t_i$ runs over the roots of $P$. If $P(2)P(-2)<0$, then $P$ has a real root in $(-2,2)$, and hence $x_{u_0}$ has a non-real unimodular conjugate. Under (G1)--(G2), a SIC then exists in dimension $d$. Under $\mathrm{PL}(u_0)$, the number of roots of $P$ in $(-2,2)$ is $m_{u_0}=n/2$, and hence $P(2)P(-2)<0$ holds exactly when $m_{u_0}$ is odd.
\end{corollary}
\begin{proof}
$P(2)P(-2)=\prod(2-t_i)(-2-t_i)=\prod(t_i^2-4)$. Non-real roots occur in conjugate pairs and contribute positive factors, real roots outside $[-2,2]$ contribute positive factors, and real roots inside contribute negative ones. Thus, the sign is $(-1)^{\#\{i:\,t_i\in(-2,2)\}}$. For the norm identity, note that $N_{\Q(x)/\Q(t)}(x-1)=(x-1)(x^{-1}-1)=2-t$ and $N_{\Q(x)/\Q(t)}(x+1)=2+t$. Hence $P(2)=N_{\Q(x)/\Q}(x-1)$ and $P(-2)=(-1)^nN_{\Q(x)/\Q}(x+1)$.
\end{proof}

\paragraph{Exact checks.} The conditions (G1) and (G2) are exact statements about the minimal polynomial of $x_{u_0}$. We verified them in PARI/GP for $18$ classes, including the $d=4$ SIC group (Table~\ref{tab:gen}). In every class some $u_0$ satisfies (G1), and (G2) holds for every such $u_0$.
\begin{itemize}[nosep]
\item For $d=4$, all six admissible $u_0$ generate, with $[\Q(x_{u_0}):\Q]=8$ and $m_{u_0}=2$. Thus, the integer criterion does not apply there, but a Sturm count of the roots of $P$ in $(-2,2)$ does.
\item In four classes ($N=9$ with cycles $(2,2,5)$ and $(11)$, and $N=12$ with cycles $(2,3,2,3)$ and $(4,4)$), a generating $u_0$ has $m_{u_0}$ odd and $P(2)P(-2)<0$. There, Theorem~\ref{thm:single} certifies the existence of a unitary-type conjugate from the sign of a single integer.
\end{itemize}

\paragraph{A 2-adic shadow of the place law.} The single-unit criterion asks whether one totally real algebraic integer has a conjugate in $(-2,2)$. The parity of the number of such conjugates can be read off from two integers.

\begin{corollary}[2-adic signature criterion]\label{cor:2adic}
Let $P\in\Z[X]$ be monic with $P(0)$ odd, and let $t_1,\dots,t_n$ be its roots. Then $P(2)P(-2)\equiv1\pmod4$, and consequently
\[
\#\{i:\ t_i\in(-2,2)\}\ \equiv\ \frac{|P(2)|\,|P(-2)|-1}{2}\pmod 2 .
\]
In particular, let $P$ be the minimal polynomial of $t_{u_0}$, with $[\Q(x_{u_0}):\Q(t_{u_0})]=2$ as in Corollary~\ref{cor:integer}, and hence $|P(2)|=|N_{\Q(x_{u_0})/\Q}(x_{u_0}-1)|$ and $|P(-2)|=|N_{\Q(x_{u_0})/\Q}(x_{u_0}+1)|$. Suppose
\begin{enumerate}[label=(\roman*),nosep]
\item (G1) and (G2) hold for $u_0$ (for instance, R2 and (G1));
\item $N_{\Q(x_{u_0})/\Q}(x_{u_0}-1)$ is odd;
\item $|N_{\Q(x_{u_0})/\Q}(x_{u_0}-1)|\cdot|N_{\Q(x_{u_0})/\Q}(x_{u_0}+1)|\equiv3\pmod4$.
\end{enumerate}
Then a WH-covariant SIC exists in dimension $d$.
\end{corollary}
\begin{proof}
Write $P(X)=Q(X^2)+X\,R(X^2)$ with $Q,R\in\Z[X]$. Then
\[
P(2)P(-2)=Q(4)^2-4R(4)^2\equiv Q(4)^2\pmod 4 .
\]
Also $Q(4)\equiv Q(0)=P(0)\pmod4$ is odd, and hence $Q(4)^2\equiv1\pmod 8$, and $P(2)P(-2)\equiv1\pmod4$. An odd integer $M\equiv1\pmod 4$ is negative exactly when $|M|\equiv3\pmod4$. The parity formula then follows from the sign rule in the proof of Corollary~\ref{cor:integer}.

For the second part, $P(0)=\pm N_{\Q(t)/\Q}(t)$ and $t=(x^2+1)/x$ with $x$ a unit, and hence $P(0)$ is odd if and only if $x^2+1\equiv(x-1)^2$ is prime to $2$, that is, if and only if $N(x-1)$ is odd. By (iii), $P(2)P(-2)<0$, and Corollary~\ref{cor:integer} gives the SIC.
\end{proof}

\paragraph{What the criterion detects.} Corollary~\ref{cor:2adic} is the first statement in this paper in which the surviving archimedean obstruction leaves a non-archimedean footprint. Hypotheses (ii) and (iii) involve only the prime-ideal factorizations of $(x_{u_0}-1)$ and $(x_{u_0}+1)$ and a condition at $2$. No embedding into $\mathbb R$ or $\C$ is used. Yet they determine one bit of the archimedean signature, namely the parity of the number of unimodular conjugates of $x_{u_0}$. This is an elementary instance of Hilbert reciprocity: the sign of a rational number is determined by its local data at the finite primes.

The criterion detects only this parity. Under PL the number of unimodular conjugates is $m_{u_0}=[\Q(t_{u_0}):F]$, and hence the criterion can succeed only when $m_{u_0}$ is odd. It says nothing about how the conjugates are distributed among the places of $F$, which is the content of PL. The two statements live at different levels:
\begin{gather*}
\text{finite primes}\ \longrightarrow\ |N(x_u\pm1)|\ \longrightarrow\ \text{parity of the unitary count},\\
\mathrm{PL}\ \longrightarrow\ \text{full distribution among the infinite places}.
\end{gather*}
When $P(0)$ is even, the sign of $P(2)P(-2)$ is still determined by $2$-adic data: by $|P(2)P(-2)|$ together with $P(2)P(-2)$ modulo $2^{v_2(P(2)P(-2))+2}$, that is, by congruences of $t_{u_0}$ modulo a power of $2$. In that case, however, it is no longer a function of the two norms alone.

\paragraph{Data.} Across all $100$ polynomials, $P(2)P(-2)<0$ holds exactly for the $13$ with $m_u$ odd, as Corollary~\ref{cor:integer} predicts under PL.
\begin{itemize}[nosep]
\item $P(0)$ is odd for every polynomial with $N$ odd, for $35$ of the $38$ with $N\equiv2\pmod4$, and for none with $N\equiv0\pmod4$.
\item The congruence $P(2)P(-2)\equiv1\pmod4$ holds for all $57$ polynomials with $P(0)$ odd, as it must.
\item For the SIC groups, $N=d^2(d-3)\equiv2\pmod4$ exactly when $d\equiv1\pmod4$ ($d=5,9,13,\dots$). This is where hypothesis~(ii) is expected to hold, and it does for $d=5$ (below). For $d=4$, $N=16$ and $P(0)$ is even.
\end{itemize}

The parity that the criterion reads off is in fact an invariant of a field, independent of the chosen unit.

\begin{proposition}[the parity is a field invariant]\label{prop:parityinv}
Let $M=\Q(x_v:v\in G_d)$, and suppose (G1)--(G2) hold for $u_0$, with $x_{u_0}\ne x_{u_0}^{-1}$. Then $\kappa$ is an automorphism of $M$ of order $2$, and $\Q(t_{u_0})=M^\kappa$ has degree $[M:\Q]/2$. If moreover $\mathrm{PL}(u_0)$ holds, the number of conjugates of $t_{u_0}$ in $(-2,2)$ is $[M:\Q]/4$. Consequently, under PL, Corollaries~\ref{cor:integer} and~\ref{cor:2adic} can succeed for a generating unit only if $[M:\Q]\equiv4\pmod 8$. Whether they succeed does not depend on which generating unit is used.
\end{proposition}
\begin{proof}
By (G1) we have $M=\Q(x_{u_0})$, and by (G2) $\kappa(x_v)=x_v^{-1}\in M$ for all $v$. Thus, $\kappa(M)=M$ and $\kappa^2=\mathrm{id}$. The element $t_{u_0}$ is $\kappa$-invariant, and $x_{u_0}$ is a root of $X^2-t_{u_0}X+1$, and hence $[M:\Q(t_{u_0})]\le2$. Since $\kappa\ne\mathrm{id}$, this forces $\Q(t_{u_0})=M^\kappa$. Under PL, exactly half of the $[M:\Q]/2$ embeddings of $\Q(t_{u_0})$ lie above $\infty_-$, and these are the conjugates in $(-2,2)$.
\end{proof}

\paragraph{The case $d=5$.} The RW system for $N=50$ is too large for our complete solver, and hence we used the ghost side instead. We took one of the $80$ $U_P$-Hermitian ghosts in the complete solution set of $V_5$ (Table~\ref{tab:ghost}) and refined it to $400$ digits. Its normalized overlaps $\varepsilon_p$ are real by Lemma~\ref{lem:nu}. Since $x_{u_p}=s_{u_p}\zeta_p\varepsilon_p^{-1}$ (Lemma~\ref{lem:H}) and both $x_{u_p}$ and $\varepsilon_p$ are real, $s_{u_p}\zeta_p=\pm1$. Hence $t_{u_p}=\pm(\varepsilon_p+\varepsilon_p^{-1})$, and the minimal polynomial is determined up to $X\mapsto-X$.

We validated this method on $d=4$: the ghost reproduces exactly the polynomial $X^4-6X^2+4$ obtained from the RW solution for cycle $(3,3,3)$, together with (G1) and (G2) for every generating $u_0$.

For $d=5$ we have $D=2700$ and $F=\Q(\sqrt3)$. The results are as follows.
\begin{itemize}[nosep]
\item All $24$ overlaps give one minimal polynomial,
\begin{multline*}
P(X)=X^{16}-32X^{14}+423X^{12}-2984X^{10}+12105X^8\\
-28304X^6+36023X^4-21432X^2+3721 .
\end{multline*}
\item $P$ is totally real, with $8$ roots outside and $8$ inside $[-2,2]$.
\item $\sqrt3\in\Q(t)$, and the place law holds at all $16$ embeddings.
\item $P(0)=61^2$ is odd, as predicted for $d\equiv1\pmod4$.
\item $P(2)=P(-2)=5^2$.
\item (G1) and (G2) hold exactly for $u_0=(0,1)$, with $[\Q(x_{u_0}):\Q]=32$ (Table~\ref{tab:gen}).
\end{itemize}
Thus $m=8$ and $[M:\Q]=32$, and by Proposition~\ref{prop:parityinv} the parity criteria cannot apply in dimension $5$, whichever unit is used.

The $d=5$ computation therefore confirms PL in the first SIC dimension beyond $d=4$. It also shows a limitation: the 2-adic criterion is controlled by the $2$-part of $[M:F]$. The method succeeds only when $[M:F]\equiv2\pmod 4$, and this fails for $d=4$ ($[M:F]=4$) and $d=5$ ($[M:F]=16$).

The norms $|N(x_u\pm1)|$ are not prime powers. Examples are $5^4\cdot13$ ($N=13$), $17\cdot101^4$ ($N=17$) and $7^2\cdot839^2$ ($N=14$). They are nevertheless highly structured. Apart from primes dividing $N$, almost every prime factor satisfies $p\equiv\pm1\pmod N$: the primes $53$, $19$, $83$, $839$, $61$, $31$, $1471$, $101$, $8059$, $107$ and $17$ occur for $N=9,10,14,14,15,16,16,17,17,18,18$. Such primes split completely in $\Q(\zeta_N)^+$. The exceptions are $7$ for $N=16$ (with $7\equiv-1\pmod 8$, occurring as $7^4$) and $3$ for $N=5$. This is the pattern expected if the norms are governed by splitting in the relevant ray class field. A formula for $|N(x_u\pm1)|$, if one exists, is therefore more likely an Euler-product or Gross--Stark-type description over split primes than a closed expression. Together with (G1)--(G2), such a description would prove $\mathrm{Z}(d)$ for every $d$ satisfying hypotheses (ii)--(iii) of Corollary~\ref{cor:2adic} for some generating $u_0$. The full table of $P(\pm2)$ is included in the supplementary material.

\subsection{A class-field form of the place law}\label{sec:CFT}
The place law is an archimedean statement, but Proposition~\ref{prop:parityinv} shows that norms of $x_u\pm1$ see only one bit of it. We now show that, under the algebraic part of the Stark picture, the \emph{whole} place law is equivalent to a congruence at a single prime. We then verify this congruence for the SIC groups $d=4$ and $d=5$.

Let $M=\Q(x_v:v\in G_d)$ and assume $t_u\ne\pm2$ whenever $2u\neq0$. By Proposition~\ref{prop:PLforms}, $\mathrm{PL}$ forces $F\subset M$; we assume this throughout.

\begin{proposition}[class-field form of PL]\label{prop:CFT}
Suppose that $M/F$ is abelian. Then the following hold.
\begin{enumerate}[label=(\alph*),nosep]
\item Every place of $M$ above $\infty_+$ is real. All complex conjugations at places of $M$ above $\infty_-$ restrict to a single element $c_-\in\Gal(M/F)$ with $c_-^2=1$.
\item $\mathrm{PL}(u)$ holds for every $u$ with $2u\ne0$ if and only if $c_-(x_v)=x_v^{-1}$ for every $v$, that is, if and only if $c_-=\kappa$ on $M$.
\item Let $\mathfrak f$ be the finite part of the conductor of $M/F$, and let $\beta\in\mathcal O_F$ satisfy $\beta\equiv1\pmod^\times\mathfrak f$, $\beta>0$ at $\infty_+$ and $\beta<0$ at $\infty_-$. Then $c_-$ is the Artin symbol of the ideal $(\beta)$. Consequently $\mathrm{PL}$ holds if and only if, for one (equivalently every) prime $\mathfrak p\nmid\mathfrak f$ of $F$ in the ray class of $(\beta)$ at which the reduction of the minimal polynomial of $x_{u_0}$ is separable, and every prime $\mathfrak P\mid\mathfrak p$ of $M$,
\[
x_v^{\,N\mathfrak p+1}\equiv1\pmod{\mathfrak P}\qquad\text{for all }v .
\]
Under (G1), it suffices to check $v=u_0$.
\end{enumerate}
\end{proposition}
\begin{proof}
(a) The identity embedding of $M$ is real, because $E_*$ is of real type, and hence every $x_v$ is real. It lies above $\infty_+$, since $\sqrt D>0$ there. Because $M/F$ is Galois, every place above $\infty_+$ is a conjugate of the identity place, and its decomposition group is conjugate to the trivial one. Hence every such place is real. The complex conjugation at a place above $\infty_-$ fixes $F$, and hence it lies in $\Gal(M/F)$. These elements are conjugate in $\Gal(M/F)$, which is abelian, and hence they coincide.

(b) Let $\sigma:M\to\C$ lie above $\infty_-$. Then $\overline{\sigma(y)}=\sigma(c_-y)$ for $y\in M$. If $c_-(x_v)=x_v^{-1}$, then $|\sigma(x_v)|=1$, and hence $\sigma(t_v)\in[-2,2]$, and in fact $\sigma(t_v)\in(-2,2)$ because $t_v\neq\pm2$. Above $\infty_+$, $\sigma(x_v)$ is real by~(a), and hence $|\sigma(t_v)|>2$. This is $\mathrm{PL}$. Conversely, $\mathrm{PL}$ gives $|\sigma(x_v)|=1$ above $\infty_-$. Therefore $\sigma(c_-x_v)=\overline{\sigma(x_v)}=\sigma(x_v)^{-1}$, and hence $c_-(x_v)=x_v^{-1}$.

(c) The idele whose only nontrivial component is $-1$ at $\infty_-$ maps to $c_-$ under the Artin map. The principal idele $\beta$ has trivial Artin image. Its components at the primes dividing $\mathfrak f$ are $\equiv1$, and its component at $\infty_+$ is positive. Hence the Artin symbol of the ideal $(\beta)$ equals $c_-^{-1}=c_-$, and every prime $\mathfrak p$ in the ray class of $(\beta)$ has $\mathrm{Frob}_{\mathfrak p}=c_-$. Since $\mathrm{Frob}_{\mathfrak P}(y)\equiv y^{N\mathfrak p}\pmod{\mathfrak P}$, the identity $c_-=\kappa$ is equivalent to $x_v^{N\mathfrak p}\equiv x_v^{-1}\pmod{\mathfrak P}$, given separability. Combine this with~(b).
\end{proof}

Proposition~\ref{prop:CFT} moves the place law from the infinite places to the finite ones. Under the algebraic hypothesis that $M/F$ is abelian, the full archimedean distribution asserted by PL, and not merely its parity, is equivalent to a single congruence: \emph{at primes of $F$ in the ray class of $(\beta)$, the overlap units reduce into the norm-one torus.} This is the finite-field analogue of the unit circle. It refines the finite-field criterion of Theorem~\ref{thm:ff} by naming the relevant primes through the sign of a generator at the two infinite places of $F$.

Proposition~\ref{prop:CFT} is a translation, not a new conjecture. It restates, for the overlap units, the classical archimedean Stark relation $\varepsilon^{1+c_-}=1$, where $c_-$ is complex conjugation at the ramified infinite place. The involution $\kappa:x_v\mapsto x_v^{-1}$ plays the role of $\varepsilon\mapsto\varepsilon^{-1}$, and hence the place law becomes the identity $c_-=\kappa$. The hypothesis that $M/F$ is abelian is the algebraic (non-archimedean) part of Stark's conjecture in this setting. It is not proved here.

\paragraph{Test for $d=4$ and $d=5$.} We used the generating unit $x_{u_0}$ of Sec.~\ref{sec:PL} (from the $U_P$-ghost, Table~\ref{tab:gen}), computed its relative minimal polynomial over $F$, and applied PARI/GP (\texttt{nfgaloisconj}, \texttt{rnfconductor}, \texttt{bnrisprincipal}). For every degree-one prime $\mathfrak p$ of $F$ with $N\mathfrak p\le5000$ not dividing the conductor, we compared two independent quantities:
\begin{itemize}[nosep]
\item the action of $\mathrm{Frob}_{\mathfrak p}$ on $x_{u_0}$, read off from the factorization of the relative minimal polynomial modulo $\mathfrak p$;
\item the ray class of $\mathfrak p$.
\end{itemize}
The results are in Table~\ref{tab:CFT}.
\begin{itemize}[nosep]
\item In both dimensions, $M/F$ is abelian, $M$ is not Galois over $\Q$, and the conductor is ramified at $\infty_-$ only.
\item The subgroup cut out by $M$ is trivial, and hence $M$ is the \emph{full} ray class field of $F$ modulo $\mathfrak f\infty_-$, with $\mathfrak f=(8)=(2)^3$ for $d=4$ and $\mathfrak f=\mathfrak p_2^4(5)$ for $d=5$ (Table~\ref{tab:CFT}).
\item At every one of the $1305$ primes tested, $\mathrm{Frob}_{\mathfrak p}$ acts as $\kappa$ exactly when $\mathfrak p$ lies in the class of $(\beta)$, acts trivially exactly in the trivial class, and acts in neither way otherwise.
\item The opposite sign pattern ($\beta'\equiv1$, $\beta'<0$ at $\infty_+$, $\beta'>0$ at $\infty_-$) gives the trivial class, as it must, since $\infty_+$ is unramified.
\end{itemize}
Thus, the congruence distinguishes the two infinite places of $F$. Together with Proposition~\ref{prop:CFT}, this confirms that $c_-=\kappa$, which is the place law, in both SIC dimensions. It confirms it through finite primes alone.

\begin{table}[!htbp]
\caption{Class-field test of the place law in the SIC dimensions $d=4,5,9,13$. In every case the unit field is abelian over $F$ and ramified at $\infty_-$ only. Here $M$ is the field of the overlap units and $M_2$ the field of their squares (the sign-free unit field). Panel (b) tests $c_-=\kappa$ (Proposition~\ref{prop:CFT}) at the degree-one primes $\mathfrak p\nmid\mathfrak f$ of $F$; entries are matches/primes tested, and there are no mismatches. The class of $c_-$ is the ray class of $(\beta)$.}\label{tab:CFT}\centering\small\setlength{\tabcolsep}{6pt}
(a) Structure of the unit field\par\smallskip
\begin{tabular}{@{}rlllll@{}}\hline\hline
$d$ & $F$ & field & Galois group over $F$ & conductor & identification\\\hline
4 & $\Q(\sqrt5)$ & $M$ & $(\Z/2)^2$ & $(8)\,\infty_-$ & exact\\
5 & $\Q(\sqrt3)$ & $M$ & $\Z/8\times\Z/2$ & $\mathfrak p_2^4(5)\,\infty_-$ & exact\\
9 & $\Q(\sqrt{15})$ & $M_2$ & $\Z/12\times\Z/3$ & $(9)\,\infty_-$ & exact\\
13 & $\Q(\sqrt{35})$ & $M_2$ & $\Z/24\times\Z/4$ & $(13)\,\infty_-$ & Frobenius orders$^{\dagger}$\\\hline\hline
\end{tabular}

\bigskip
(b) Frobenius test of $c_-=\kappa$\par\smallskip
\begin{tabular}{@{}rcccc@{}}\hline\hline
$d$ & primes & class of $c_-$: & trivial class: & other classes:\\
 & tested & $\mathrm{Frob}_{\mathfrak p}=\kappa$ & $\mathrm{Frob}_{\mathfrak p}=1$ & $\mathrm{Frob}_{\mathfrak p}\notin\{1,\kappa\}$\\\hline
4 & $653$ ($N\mathfrak p\le5000$) & 164/164 & 156/156 & 333/333\\
5 & $652$ ($N\mathfrak p\le5000$) & 36/36 & 34/34 & 582/582\\
9 & --- & \multicolumn{3}{l}{not needed: $c_-=\kappa$ follows from Proposition~\ref{prop:uniqueinv}}\\
13 & $2230$ ($N\mathfrak p\le20000$) & 15/15 & \multicolumn{2}{c}{order of $\mathrm{Frob}_{\mathfrak p}$ = order of class: 2230/2230}\\\hline\hline
\end{tabular}

\smallskip
\raggedright\footnotesize $^{\dagger}$For $d=13$ the trace field $M_2^\kappa$ (conductor $(13)$, group $\Z/12\times\Z/4$) was computed exactly; $M_2$ was identified by comparing the order of $\mathrm{Frob}_{\mathfrak p}$ with the order of the class of $\mathfrak p$ in $\mathrm{Cl}((13)\infty_-)$. Ghost overlaps were refined to $400$ ($d=5$), $2000$ ($d=9$) and $20500$ ($d=13$) digits.
\end{table}

\paragraph{Conductor.} For $d=5$, the finite conductor $\mathfrak p_2^4(5)$ of $\Q(\varepsilon_p)$ has a $2$-part that comes only from the sign ambiguity of the units, which reflects the phase convention for the displacement operators. The sign-free combinations $\varepsilon_p^2$, $\varepsilon_p\varepsilon_q$ and $\varepsilon_p/\varepsilon_q$ generate the full ray class field of $\Q(\sqrt3)$ modulo $(5)\infty_-$, with Galois group $\Z/8$. For $d=4$, the units themselves give $(8)\infty_-$, and their squares give $(4)\infty_-$. In both cases the conductor is $d'\infty_-$ for the appropriate version of the unit field, where $d'=d$ for $d$ odd and $d'=2d$ for $d$ even is the integer that appears in the AFK construction~\cite{AFK25}. Thus, for $d=4,5$, the overlap-unit field of the RW/Izumi construction and the conductor of the AFK construction describe the same ray class field, ramified at $\infty_-$ alone.

\paragraph{The next SIC dimensions, $d=9$ and $d=13$.} The same pattern persists (Table~\ref{tab:CFT}). For $d=9$, the sign-free unit field $M_2$ is exactly the ray class field of $\Q(\sqrt{15})$ modulo $(9)\infty_-$, with $\Gal(M_2/F)\cong\Z/12\times\Z/3=\mathrm{Cl}((9)\infty_-)$. For $d=13$, the trace field $M_2^\kappa$ is exactly the ray class field of $\Q(\sqrt{35})$ modulo $(13)$, with Galois group $\Z/12\times\Z/4$ and no infinite ramification. The Frobenius test identifies $M_2$ itself with the ray class field modulo $(13)\infty_-$, of degree $96$ over $F$. At all $15$ tested primes in the class of $c_-$, the Frobenius acts as $\kappa$. In both dimensions PL was also checked directly: for every conjugate of $t=x_{u_0}^2+x_{u_0}^{-2}$ ($36$ for $d=9$, $96$ for $d=13$), the sign of $t^2-4$ agrees with the sign of the corresponding conjugate of $\sqrt D$, and exactly half of the conjugates lie in $(-2,2)$. Refining the norm pattern observed above, the prime factors of $P_t(\pm2)$ outside the conductor all satisfy $p\equiv\pm1\pmod{4d}$. For example, they are $37,71,181,431,1151,295129$ for $d=9$. Thus, in every SIC dimension computed, the sign-free unit field is the full ray class field of $F$ of conductor $d'\infty_-$, and its $\kappa$-fixed subfield is the ray class field of conductor $d'$. For $d=17$ the units were not identified at $20500$ digits, and we estimate that about $45000$ digits are needed.

The place law depends only on $|\sigma(x_v)|$, and hence Proposition~\ref{prop:CFT} and its proof apply verbatim to $M_2$ in place of $M$. In particular, $\mathrm{PL}$ is equivalent to $c_-=\kappa$ on $M_2$ whenever $M_2/F$ is abelian.

\begin{proposition}[unique involution]\label{prop:uniqueinv}
Let $M'$ be $M$ or $M_2$, and suppose the following hold:
\begin{enumerate}[label=(\roman*),nosep]
\item $M'/F$ is Galois;
\item inversion $\kappa:x_v\mapsto x_v^{-1}$ is an automorphism of $M'$ fixing $F$, and $\kappa\neq\mathrm{id}$;
\item $\infty_-$ ramifies in $M'$;
\item $\Gal(M'/F)$ has exactly one element of order~$2$.
\end{enumerate}
Then every complex conjugation of $M'$ above $\infty_-$ equals $\kappa$. Hence $\mathrm{PL}$ holds, and therefore $\mathrm{Z}(d)$ holds.
\end{proposition}
\begin{proof}
A complex conjugation at a place of $M'$ above $\infty_-$ fixes $F$. By (iii) it is nontrivial, and hence it is an element of order $2$ in $\Gal(M'/F)$. By (ii), $\kappa$ is also an element of order $2$ in $\Gal(M'/F)$. By (iv) they coincide. The proof of Proposition~\ref{prop:CFT}(a) uses only that $M'/F$ is Galois, and hence every place above $\infty_+$ is real. The argument of Proposition~\ref{prop:CFT}(b) then gives $\mathrm{PL}$, and Theorem~\ref{thm:PLSIC} gives $\mathrm{Z}(d)$.
\end{proof}

Proposition~\ref{prop:uniqueinv} removes the archimedean input entirely in such dimensions. What remains, (i) and (ii), is purely algebraic: it is the algebraic part of the Stark picture together with R2. Hypothesis (ii) also shows that $\Gal(M'/F)$ always has even order whenever inversion is a field automorphism, and hence families with $\Gal(M'/F)$ of odd order cannot occur.

\emph{When does (iv) hold?} Suppose, as in all computed cases, that $M_2$ is the full ray class field of $F$ modulo $d'\infty_-$. Then (iii) holds exactly when $|\mathrm{Cl}(d'\infty_-)|=2\,|\mathrm{Cl}(d')|$. Condition (iv) then says that the $2$-Sylow subgroup of $\mathrm{Cl}(d'\infty_-)$ is cyclic. Among $4\le d\le2000$ with $\infty_-$ ramified, this happens for exactly $24$ dimensions:
\[
\begin{aligned}
&5,\,7,\,9,\,11,\,15,\,17,\,19,\,27,\,47,\,53,\,67,\,71,\,99,\\
&191,\,199,\,431,\,449,\,467,\,787,\,881,\,971,\,1061,\,1619,\,1681.
\end{aligned}
\]
The counts by range are $15/247$ for $4\le d\le250$, then $3$, $0$, $3$, $1$, $0$, $2$, $0$ in successive blocks of $250$ up to $2000$. For $d=4$ and $d=13$ the $2$-Sylow is not cyclic, $(\Z/2)^2$ and $\Z/8\times\Z/4$ respectively, but $c_-=\kappa$ holds there numerically (Table~\ref{tab:CFT}). Genus theory, the contribution of $(\mathcal O_F/d')^\times$, and the sign at $\infty_-$ force $2$-torsion whenever $\mathrm{disc}\,F$ or $d'$ has many prime factors. By the Erd\H{o}s--Kac theorem the number of prime factors of the square-free part of $(d-3)(d+1)$ grows like $\log\log d$, so the set of $d$ with cyclic $2$-Sylow is expected to have density zero (unlike the odd part of class groups, which Cohen--Lenstra heuristics model). Infinitude is still plausible under Bateman--Horn or Schinzel-type conjectures on prime-pattern families (for instance prime $d$ with $d+1=2m^2$ and $m^2-2$ prime), but it is unproved, and even an infinite such set would still require the algebraic hypotheses (i)--(ii) of Proposition~\ref{prop:uniqueinv} in those dimensions.

\begin{proposition}[the unit field is not cyclotomic over $F$]\label{prop:noncyc}
Let $K$ be a finite extension of $F$ contained in $F\cdot\Q^{\mathrm{ab}}$, the compositum of $F$ with all cyclotomic fields. Then the complex conjugations of $K$ above $\infty_+$ and above $\infty_-$ coincide. In particular, $K$ is ramified at $\infty_+$ if and only if it is ramified at $\infty_-$. Consequently, if $M$ is ramified at $\infty_-$ but not at $\infty_+$, as for $d=4,5$, then $M\not\subset F\cdot\Q^{\mathrm{ab}}$.
\end{proposition}
\begin{proof}
Let $c$ be complex conjugation for the identity embedding, which lies above $\infty_+$, and let $g\in\Gal(\Qb/\Q)$ with $g(\sqrt D)=-\sqrt D$. A complex conjugation above $\infty_-$ is $gcg^{-1}$. On $F\cdot\Q^{\mathrm{ab}}$, both $c$ and $gcg^{-1}$ fix $F$, because $F$ is real at both places. Restricting further to $\Q^{\mathrm{ab}}$, where the Galois group is abelian, gives $gcg^{-1}|_{\Q^{\mathrm{ab}}}=c|_{\Q^{\mathrm{ab}}}$. An element of $\Gal(F\Q^{\mathrm{ab}}/F)$ is determined by its restriction to $\Q^{\mathrm{ab}}$. Therefore $c$ and $gcg^{-1}$ agree on $F\Q^{\mathrm{ab}}$, and hence on $K$.
\end{proof}
Proposition~\ref{prop:noncyc} explains why the remaining step cannot be elementary. The field $M$, and with it the identity $c_-=\kappa$, cannot be reached through roots of unity, Gauss sums, or Kronecker--Weber. Any proof must construct abelian extensions of the real quadratic field $F$ that are ramified at exactly one real place, which is the content of Hilbert's twelfth problem for $F$.

\paragraph{What this changes.} For a given $d$, the place law, and hence $\mathrm{Z}(d)$ by Theorem~\ref{thm:PLSIC}, follows from two inputs:
\begin{enumerate}[label=(\roman*),nosep]
\item the \emph{algebraic} statement that the overlap-unit field $M$ is abelian over $F$, which is expected to be a ray class field of conductor $\mathfrak f\infty_-$;
\item the \emph{congruence} $x_{u_0}^{N\mathfrak p+1}\equiv1\pmod{\mathfrak P}$ at a single prime $\mathfrak p$ in the class of $c_-$.
\end{enumerate}
When $\Gal(M/F)$ has a unique element of order $2$, input (ii) is automatic once $\kappa$ is an automorphism of $M$ fixing $F$ and $\infty_-$ ramifies (Proposition~\ref{prop:uniqueinv}). Neither input refers to an embedding into $\C$. The remaining difficulty is not removed: proving (ii) uniformly in $d$ requires knowing the reductions of the RW special values modulo such primes. It is, however, moved to the setting of Gross--Stark-type and $p$-adic methods, where comparable statements have been proved.

\subsection{Total reality in place of the unique involution}\label{sec:TRinv}
Proposition~\ref{prop:uniqueinv} reaches $c_-=\kappa$ without any archimedean input, but it pays for this with hypothesis~(iv), a condition on $\mathrm{Cl}(d'\infty_-)$ that holds for only $24$ dimensions below $2000$ and is expected to have density zero. We now show that~(iv) can be traded for total reality (TR), the weakest-looking archimedean statement of Sec.~\ref{sec:reality}. Under the algebraic hypotheses this does not weaken the archimedean input, because TR then turns out to be equivalent to PL (Corollary~\ref{cor:collapse}); what changes is that no condition on the class group remains, and that the ramification hypothesis becomes unconditional on a set of dimensions of positive density.

\begin{proposition}[total reality replaces the unique involution]\label{prop:TRinv}
Let $M'$ be $M$ or $M_2$, and suppose:
\begin{enumerate}[label=(\roman*),nosep]
\item $M'/F$ is Galois;
\item inversion $\kappa$ is an automorphism of $M'$ fixing $F$, and $\kappa\ne\mathrm{id}$;
\item $\infty_-$ ramifies in $M'$;
\item\label{it:TRhyp} the fixed field $M'^{\kappa}$ is totally real.
\end{enumerate}
Then every complex conjugation of $M'$ at a place above $\infty_-$ equals $\kappa$. Hence $\mathrm{PL}$ holds, and therefore $\mathrm{Z}(d)$ holds.
\end{proposition}
\begin{proof}
Let $w$ be a place of $M'$ above $\infty_-$, let $\sigma:M'\to\C$ induce it, and let $c_w$ be the associated complex conjugation. Since $F$ is real at $\infty_-$, $c_w$ fixes $F$, and hence $c_w\in\Gal(M'/F)$. Because $M'/F$ is Galois, all places above $\infty_-$ are conjugate, and hence by~(iii) all of them are complex; in particular $c_w\ne\mathrm{id}$.

By~\ref{it:TRhyp}, $\sigma(M'^{\kappa})\subset\mathbb R$, and hence $c_w$ acts trivially on $M'^{\kappa}$. Therefore $c_w\in\Gal(M'/M'^{\kappa})$, and $\Gal(M'/M'^{\kappa})=\langle\kappa\rangle$ by Artin's lemma, a group of order $2$ by~(ii). Since $c_w\ne\mathrm{id}$, we get $c_w=\kappa$. Consequently $\overline{\sigma(x_v)}=\sigma(\kappa x_v)=\sigma(x_v)^{-1}$, that is, $|\sigma(x_v)|=1$ for every $v$. The proof of Proposition~\ref{prop:CFT}(a) uses only that $M'/F$ is Galois, and hence every place above $\infty_+$ is real; the argument of Proposition~\ref{prop:CFT}(b) then gives $\mathrm{PL}$, and Theorem~\ref{thm:PLSIC} gives $\mathrm{Z}(d)$.
\end{proof}

Hypothesis~\ref{it:TRhyp} is exactly Conjecture~\ref{conj:TR}. Indeed, under~(G1) one has $[M:\Q(t_{u_0})]=2$ and $t_v\in M^{\kappa}$ for every $v$, so that
\[
M^{\kappa}=\Q(t_v:v\in G_d)=\Q(t_{u_0}),
\]
and total reality of that field is TR in the form of Proposition~\ref{prop:TRarith}. For $M_2$ the same applies with $t_v^2-2$ in place of $t_v$.

\begin{corollary}[the archimedean hierarchy collapses]\label{cor:collapse}
Assume (i)--(iii) of Proposition~\ref{prop:TRinv} together with~(G1). Then
\[
\mathrm{TR}\ \Longleftrightarrow\ \mathrm{HS}\ \Longleftrightarrow\ \mathrm{PL}.
\]
\end{corollary}
\begin{proof}
$\mathrm{PL}\Rightarrow\mathrm{HS}\Rightarrow\mathrm{TR}$ is Proposition~\ref{prop:PLforms}. The converse is Proposition~\ref{prop:TRinv}.
\end{proof}

\begin{remark}
Propositions~\ref{prop:uniqueinv} and~\ref{prop:TRinv} are not comparable hypothesis by hypothesis: the former removes the archimedean input altogether, whereas the latter retains TR. What is gained is reach, not a weaker archimedean hypothesis: under (i)--(iii), TR is equivalent to PL by Corollary~\ref{cor:collapse}. TR was verified for all $268$ nondegenerate cyclic metrics with $5\le N\le21$ and all $54$ $SL_2(\Z)$ classes with $N\le20$, whereas the cyclic $2$-Sylow condition of Proposition~\ref{prop:uniqueinv} holds for only $24$ dimensions $d\le2000$. Under Corollary~\ref{cor:collapse} the separation between TR, HS and PL recorded in Sec.~\ref{sec:reality} disappears: the three differ only in the absence of the algebraic hypotheses. In particular, proving TR in these dimensions is exactly as hard as proving PL.
\end{remark}

\paragraph{The remaining hypothesis is about units, not class groups.} Assume, as in all computed cases, that $M_2$ is the ray class field of $F$ modulo $d'\infty_-$. Hypothesis~(iii) is then the statement that this conductor really involves $\infty_-$, that is, $|\mathrm{Cl}(d'\infty_-)|=2\,|\mathrm{Cl}(d')|$. It admits a criterion free of class numbers.

\begin{proposition}[unit criterion for ramification at $\infty_-$]\label{prop:ramcrit}
Put $\mathfrak m_0=d'\mathcal O_F$ and $\mathfrak m=\mathfrak m_0\infty_-$, and let
\[
U_1=\{\eta\in\mathcal O_F^{\times}:\eta\equiv1\ (\mathrm{mod}\ d')\}.
\]
The natural surjection $\mathrm{Cl}(\mathfrak m)\to\mathrm{Cl}(\mathfrak m_0)$ has kernel of order $2$ if every $\eta\in U_1$ is positive at $\infty_-$, and is an isomorphism otherwise. Hence $\infty_-$ ramifies in the ray class field of conductor $\mathfrak m$ if and only if
\[
\mathrm{(C}_d)\qquad\text{no unit of $F$ congruent to $1$ modulo $d'$ is negative at $\infty_-$.}
\]
\end{proposition}
\begin{proof}
The kernel is $P_{\mathfrak m_0,1}/P_{\mathfrak m,1}$, where $P_{\mathfrak m_0,1}$ consists of the principal ideals with a generator $\equiv1\pmod{d'}$, and $P_{\mathfrak m,1}$ requires in addition a generator positive at $\infty_-$. Two generators of one ideal that are both $\equiv1\pmod{d'}$ differ by an element of $U_1$, and hence $(\alpha)\mapsto\mathrm{sgn}_{\infty_-}(\alpha)$ induces an isomorphism $P_{\mathfrak m_0,1}/P_{\mathfrak m,1}\cong\{\pm1\}/\mathrm{sgn}_{\infty_-}(U_1)$. The sign map is onto $\{\pm1\}$, because $\alpha=1-nd'$ is $\equiv1\pmod{d'}$ and negative at $\infty_-$ for $n$ large. By class field theory the ray class field of conductor $\mathfrak m$ strictly contains that of conductor $\mathfrak m_0$ exactly when the kernel is nontrivial, and the additional ramification is then at $\infty_-$.
\end{proof}

Throughout this part put $\Delta=(d-3)(d+1)=(d-1)^2-4$; then $F=\Q(\sqrt\Delta)$, since $D=d^2(d-2)^2\Delta$. The point is that the unit $\varepsilon_\alpha=\frac12(d-1+\sqrt\Delta)$, the eigenvalue of $\alpha$ that exceeds $1$ at $\infty_+$, reduces modulo $d$ to a root of unity of \emph{odd} order.

\begin{lemma}[$\varepsilon_\alpha$ is a cube root of unity modulo $d$]\label{lem:zeta3}
$\varepsilon_\alpha\in\mathcal O_F$, $N_{F/\Q}(\varepsilon_\alpha)=1$, and $\varepsilon_\alpha>0$ at both infinite places. Moreover $\varepsilon_\alpha^2=(d-1)\varepsilon_\alpha-1$, and hence
\[
\varepsilon_\alpha^{3}=d(d-2)\,\varepsilon_\alpha+(1-d).
\]
Consequently:
\begin{enumerate}[label=(\alph*),nosep]
\item for $d$ odd, $\varepsilon_\alpha^{3}\equiv1\pmod{d}$; in fact $\Delta\equiv-3\pmod d$, so the image of $\varepsilon_\alpha$ in $\mathcal O_F/d$ is $\frac12(-1+\sqrt{-3})$, a primitive cube root of unity;
\item for $d$ even, $\varepsilon_\alpha^{3}\equiv1-d\pmod{2d}$ and $(1-d)^2\equiv1\pmod{2d}$, so $\varepsilon_\alpha$ has order dividing $6$ in $(\mathcal O_F/2d)^{\times}$.
\end{enumerate}
\end{lemma}
\begin{proof}
Since $2\varepsilon_\alpha-(d-1)=\sqrt\Delta$ and $\Delta=(d-1)^2-4\equiv(d-1)^2\pmod4$, $\varepsilon_\alpha$ is an algebraic integer, with $N(\varepsilon_\alpha)=\frac{(d-1)^2-\Delta}{4}=1$; and $\Delta<(d-1)^2$ gives $\frac12(d-1-\sqrt\Delta)>0$. The quadratic relation is the characteristic equation of $\alpha$, and cubing it gives $\varepsilon_\alpha^3=((d-1)^2-1)\varepsilon_\alpha-(d-1)=d(d-2)\varepsilon_\alpha+(1-d)$. For $d$ odd, $d\mid d(d-2)$ and $1-d\equiv1\pmod d$, which is~(a); and $\Delta=(d-3)(d+1)\equiv(-3)(1)=-3\pmod d$, so modulo $d$ the element $\sqrt\Delta$ is a square root of $-3$ and $\varepsilon_\alpha=\frac12(-1+\sqrt{-3})$, whose cube is $1$. For $d$ even, $d-2$ is even, so $2d\mid d(d-2)$, and $2d\mid d^2$, so $(1-d)^2=1-2d+d^2\equiv1\pmod{2d}$, which is~(b).
\end{proof}

\begin{theorem}[ramification from the unit group]\label{thm:ramgen}
If $\mathcal O_F^{\times}=\langle-1,\varepsilon_\alpha\rangle$, then (C$_d$) holds for every $d\ge4$; equivalently, $\infty_-$ ramifies in the ray class field of $F$ of conductor $d'\infty_-$.
\end{theorem}
\begin{proof}
Every unit is $\pm\varepsilon_\alpha^{\,n}$. By Lemma~\ref{lem:zeta3}, $\varepsilon_\alpha>0$ at $\infty_-$, and hence so is $\varepsilon_\alpha^{\,n}$; such units never violate (C$_d$). A unit negative at $\infty_-$ is therefore of the form $-\varepsilon_\alpha^{\,n}$, and it is $\equiv1\pmod{d'}$ precisely when $\varepsilon_\alpha^{\,n}\equiv-1\pmod{d'}$. Cubing that congruence gives $(\varepsilon_\alpha^{3})^{n}\equiv-1\pmod{d'}$.

If $d$ is odd, then $\varepsilon_\alpha^{3}\equiv1\pmod{d}$, so $1\equiv-1\pmod d$, which is impossible for $d\ge4$. If $d$ is even, then $(\varepsilon_\alpha^{3})^{n}\in\{1,1-d\}$ modulo $2d$ by Lemma~\ref{lem:zeta3}(b), so either $-1\equiv1\pmod{2d}$, or $-1\equiv1-d\pmod{2d}$, that is, $d\equiv2\pmod{2d}$; both are impossible for $d\ge4$.
\end{proof}

\begin{theorem}[an unconditional family of positive density]\label{thm:RD}
Let $d=2m+1$ with $m\ge2$ even and $m^2-1$ squarefree. Then $F=\Q(\sqrt{m^2-1})$, $\mathcal O_F=\Z[\sqrt{m^2-1}]$, and
\[
\mathcal O_F^{\times}=\langle-1,\ m+\sqrt{m^2-1}\rangle=\langle-1,\varepsilon_\alpha\rangle .
\]
Hence (C$_d$) holds and $\infty_-$ ramifies in the ray class field of conductor $d\infty_-$. The set $\mathcal D$ of such $d$ has natural density
\[
\tfrac14\prod_{p>2}\bigl(1-2p^{-2}\bigr)\approx0.16132 ,
\]
and in particular it is infinite.
\end{theorem}
\begin{proof}
Since $d$ is odd, $\Delta=(d-3)(d+1)=4(m^2-1)$, so the squarefree part of $\Delta$ is $m^2-1$ and $\varepsilon_\alpha=\frac12(2m+2\sqrt{m^2-1})=m+\sqrt{m^2-1}$. As $m$ is even, $m^2-1\equiv3\pmod4$, and hence $\mathcal O_F=\Z[\sqrt{m^2-1}]$ and $x^2-(m^2-1)y^2=-1$ is insoluble, since modulo $4$ it reads $x^2+y^2\equiv3$. Therefore $\mathcal O_F^{\times}=\langle-1,\eta\rangle$ with $\eta$ the fundamental solution of $x^2-(m^2-1)y^2=1$. The continued fraction $\sqrt{m^2-1}=[\,m-1;\overline{1,\,2m-2}\,]$ has period $2$, with first period convergent $m/1$, and hence $\eta=m+\sqrt{m^2-1}=\varepsilon_\alpha$. Theorem~\ref{thm:ramgen} applies. This is the Richaud--Degert case $D_0=n^2-1$.

For the density, $d\mapsto m=(d-1)/2$ is a bijection from the odd integers onto $\Z$, and the conditions are that $m$ be even, of density $\frac14$ in $d$, and that $(m-1)(m+1)$ be squarefree. For $m$ even the factors $m\pm1$ are odd and coprime, so $p^2\mid m^2-1$ for an odd prime $p$ if and only if $m\equiv\pm1\pmod{p^2}$, which excludes two residues modulo $p^2$; and $p^2\le m^2-1$ forces $p\le m$, so the sieve is finite. Inclusion--exclusion gives the stated density, by Mirsky's theorem on squarefree values of products of distinct linear forms.
\end{proof}

\begin{corollary}[a reformulation on the family $\mathcal D$]\label{cor:TRfinal}
Let $d\in\mathcal D$ and suppose:
\begin{enumerate}[label=(S\arabic*),nosep]
\item $M_2$ is the ray class field of $F$ of conductor $d\infty_-$, the algebraic part of the Stark picture;
\item $\mathrm{R2}$ holds, and (G1) holds for some $u_0$;
\item $\mathrm{TR}$ holds for $E_*$, that is, $\Q(t_u:u\in G_d)$ is totally real.
\end{enumerate}
Then $\mathrm{Z}(d)$ holds. Since $\mathcal D$ is infinite, (S1)--(S3) for all $d\in\mathcal D$ would imply Problem~1.
\end{corollary}
\begin{proof}
(S1) makes $M_2/F$ abelian, hypothesis~(i), and by Theorem~\ref{thm:RD} its conductor involves $\infty_-$, hypothesis~(iii). (S2) gives $\kappa\in\Gal(M_2/F)$, hypothesis~(ii), and identifies $M_2^{\kappa}$ with the trace field; (S3) is hypothesis~\ref{it:TRhyp}. Apply Proposition~\ref{prop:TRinv}.
\end{proof}

Corollary~\ref{cor:TRfinal} is conditional and does not reduce the archimedean difficulty. By Corollary~\ref{cor:collapse}, under (S1) and (S2) hypothesis (S3) is equivalent to the place law itself, so the corollary is a reformulation of Theorem~\ref{thm:PLSIC} on the family $\mathcal D$, not a proof of Problem~1 from weaker input. What it does achieve is the following. No condition on $\mathrm{Cl}(d'\infty_-)$ survives, no congruence has to be checked, and the one hypothesis of Proposition~\ref{prop:uniqueinv} that is not algebraic in nature, ramification at $\infty_-$, is proved unconditionally on a set of positive density rather than assumed. Compared with Proposition~\ref{prop:uniqueinv}, which needs no archimedean input but applies only to $24$ dimensions below $2000$, the present route applies on a set of positive density but needs the full archimedean input, in the form TR.

\paragraph{Verification.} Proposition~\ref{prop:ramcrit}, Lemma~\ref{lem:zeta3} and Theorem~\ref{thm:ramgen} were checked by exact integer arithmetic in $\mathcal O_F=\Z[w]$, with $w=\frac12(1+\sqrt{D_0})$ or $\sqrt{D_0}$ for $D_0$ the squarefree part of $\Delta$, for every $4\le d\le20000$ (Table~\ref{tab:ram}). The fundamental unit of $F$ was computed from the continued fraction of $\sqrt{D_0}$, corrected by the index-$3$ test $a(a^2\mp3)=2x_1$ when $D_0\equiv1\pmod4$, and spot-checked against $\varepsilon(\sqrt5)=\frac12(1+\sqrt5)$, $\varepsilon(\sqrt{15})=4+\sqrt{15}$, $\varepsilon(\sqrt{35})=6+\sqrt{35}$ and $\varepsilon(\sqrt{21})=\frac12(5+\sqrt{21})$. Condition (C$_d$) held in every one of the $19\,997$ dimensions, with no exception, which also extends to $d\le20000$ the census of Proposition~\ref{prop:uniqueinv}(iii) reported above for $d\le2000$. Independently, for every $4\le d\le2000$ we confirmed $|\mathrm{Cl}(d'\infty_-)|=2\,|\mathrm{Cl}(d')|$ directly with PARI/GP (\texttt{bnrclassno}), with no exception.

\begin{table}[!htbp]
\caption{Verification of the unit criterion (C$_d$) and of Lemma~\ref{lem:zeta3}, by exact arithmetic in $\mathcal O_F$, for every $4\le d\le20000$. Here $\mathcal D$ is the Richaud--Degert family of Theorem~\ref{thm:RD}, whose members need no verification.}\label{tab:ram}\centering\small
\begin{tabular}{@{}lrr@{}}\hline\hline
Statement checked, $4\le d\le20000$ & count & fraction\\\hline
dimensions examined & $19\,997$ & \\
(C$_d$) holds, i.e.\ $\infty_-$ ramifies & $19\,997$ & all\\
$\mathcal O_F^{\times}=\langle-1,\varepsilon_\alpha\rangle$ (Thm.~\ref{thm:ramgen} applies) & $19\,697$ & $98.50\%$\\
$d\in\mathcal D$ (Thm.~\ref{thm:RD}, unconditional) & $3\,229$ & $16.15\%$\\
$\mathrm{ord}(\varepsilon_\alpha)=3$ in $(\mathcal O_F/d)^{\times}$, $d$ odd & $9\,998$ & all odd $d$\\
$\mathrm{ord}(\varepsilon_\alpha)=6$ in $(\mathcal O_F/2d)^{\times}$, $d$ even & $9\,999$ & all even $d$\\\hline\hline
\end{tabular}
\end{table}

\begin{figure}[!htbp]
\centering
\includegraphics[width=\textwidth]{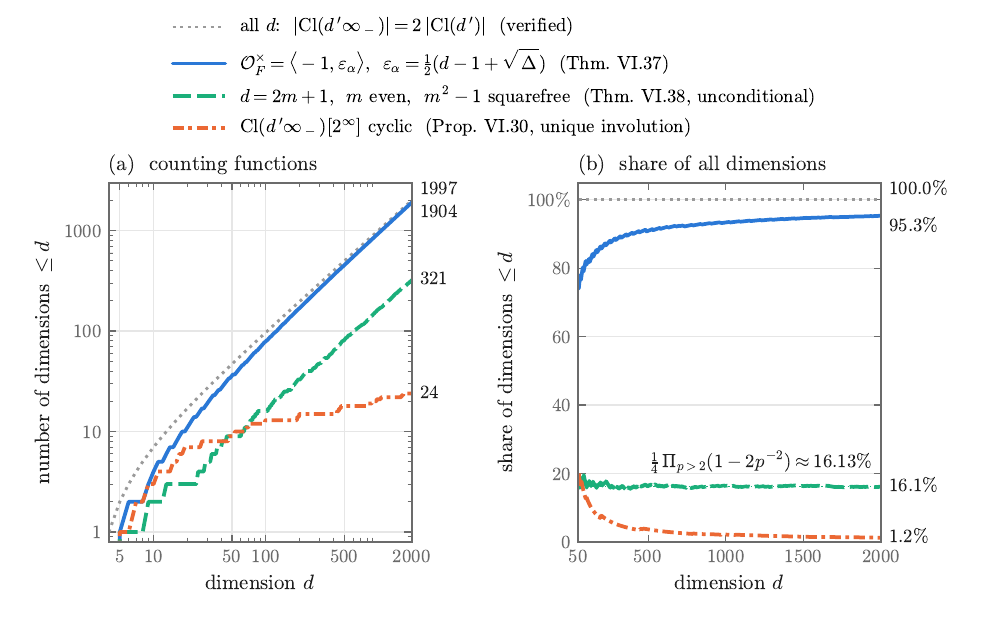}
\caption{Dimensions $4\le d\le2000$ covered by the hypotheses of this subsection. Dotted gray: all $d$; condition (C$_d$), and hence ramification at $\infty_-$, holds for each of them (Table~\ref{tab:ram}). Solid blue: $\mathcal O_F^\times=\langle-1,\varepsilon_\alpha\rangle$, where Theorem~\ref{thm:ramgen} proves (C$_d$). Dashed green: the family $\mathcal D$ of Theorem~\ref{thm:RD}, where (C$_d$) is proved unconditionally. Dash-dotted orange: the $24$ dimensions with cyclic $2$-Sylow subgroup of $\mathrm{Cl}(d'\infty_-)$, where Proposition~\ref{prop:uniqueinv} applies (census of Sec.~\ref{sec:CFT}). (a) Counting functions on logarithmic axes: a set of positive density appears as a line of slope $1$, parallel to the gray reference, and a set of density zero bends away from it. (b) The same sets as a share of all dimensions $\le d$; the dotted green line is the predicted density $\frac14\prod_{p>2}(1-2p^{-2})\approx0.1613$ of $\mathcal D$. The figure concerns hypothesis~(iii) only; the archimedean input (TR, equivalently PL) is not affected.}\label{fig:ramcov}
\end{figure}

The smallest members of $\mathcal D$ are 5, 9, 13, 25, 29, 33, 37, 41, 45, 61, 65, 69, 73, 77, 81, 85, 105, 109, 113, 117, 121, \dots, and they include $d=5,9,13$, the odd dimensions in which the overlap-unit field has been identified exactly (Table~\ref{tab:CFT}). The density of $\mathcal D$ observed up to $20000$ is $0.16147$, against the predicted $0.16132$. Figure~\ref{fig:ramcov} compares the three sets of dimensions discussed in this subsection with the set of all $d$.

\subsection{An explicit reciprocity law for the overlap units}\label{sec:RL}
Proposition~\ref{prop:CFT} identifies the place law with the action of a single Galois element, $c_-$, on the overlap units. Class field theory describes every element of $\Gal(M_2/F)$ as an Artin symbol, and hence the place law would follow from a rule computing the Artin symbol of an ideal \emph{directly} from a generator. For special values at $s=0$ of partial zeta functions such a rule is known: Arakawa~\cite{Arakawa} (Thms.~A and~B), building on Coates--Sinnott and Deligne--Ribet, shows that the idele $s$ acts on $\exp(2\pi i\,\zeta(0,\cdot))$ through the regular representation $F^\times\to GL_2(\Q)$ of $F$, $z\mapsto s^{-1}z$, for ideles $s$ that are \emph{positive at both infinite places}. For the values at $s=0$ of the derivatives, which are the Stark--Shintani units relevant here, the analytic half of the corresponding statement is known: the invariant of the class $\mathcal R\mathcal A$ is the inverse of the invariant of $\mathcal A$, where $\mathcal R$ is the class of $c_-$ (Ref.~\onlinecite{KoppSF}, Cor.~8.3; Ref.~\onlinecite{YamamotoMA}, Thm.~6.1). The missing Galois half is a reciprocity law. In this subsection we state such a law in explicit form, show that it implies $c_-=\kappa$, and verify it exactly in the SIC dimensions $d=5$ and $d=9$.

\paragraph{The matrix of a generator.} Let $d\ge5$ be odd, and let $\tilde\Pi$ be a ghost in dimension $d$ with normalized overlaps $\varepsilon_p$, $p\in\Z_d^2\setminus\{0\}$, which are real with $\varepsilon_{-p}=\varepsilon_p^{-1}$. Put $e_p=\varepsilon_p^2$, and let $M_\varepsilon=\Q(e_p:p\ne0)$ be the sign-free field of the ghost overlaps. The group $SL_2(\Z_d)$ acts on the labels by $p\mapsto gp$, with $p$ regarded as a column vector. The ghosts constructed in Refs.~\onlinecite{AFK25,AFK26} have Zauner symmetry: there is an element $\Lambda\in SL_2(\Z_d)$ of order~$3$ with $\operatorname{tr}\Lambda\equiv d-1\pmod d$ and
\[
e_{\Lambda p}=e_p\qquad\text{for all }p\ne0 .
\]
Let
\[
\epsilon_\alpha=\tfrac12\big(d-1+\sqrt{(d-3)(d+1)}\big)
\]
be the eigenvalue of $\alpha$ (trace $d-1$) that is larger than $1$ at $\infty_+$, and let $\mathcal O_\alpha=\Z[\epsilon_\alpha]$. Since $\sqrt D=d(d-2)\sqrt{(d-3)(d+1)}$, $\mathcal O_\alpha$ is an order of $F$. The unit $\epsilon_\alpha$ is totally positive with $\epsilon_\alpha+\epsilon_\alpha^{-1}=d-1$. The matrix $\Lambda^{-1}$ satisfies the same relation, $\Lambda^{-2}-(d-1)\Lambda^{-1}+I=0$, by the Cayley--Hamilton theorem. Hence there is a unique ring homomorphism
\[
\phi:\mathcal O_\alpha\longrightarrow M_2(\Z_d),\qquad \phi(\epsilon_\alpha)=\Lambda^{-1}.
\]
Concretely, $\phi(a+b\,\delta)=aI+b\,(2\Lambda^{-1}-(d-1)I)$ for $\delta=2\epsilon_\alpha-(d-1)=\sqrt{(d-3)(d+1)}$. If $\varpi\in\mathcal O_\alpha$ is prime to $d$, then $\det\phi(\varpi)\equiv N_{F/\Q}(\varpi)\not\equiv0$, and $\phi(\varpi)\in GL_2(\Z_d)$. In $d=5$ and $d=9$, $\mathcal O_\alpha=\mathcal O_F$, and $\phi(\sqrt{D_0})=\Lambda^{-1}-\tfrac{d-1}2 I$ with $D_0=3$, resp.\ $15$.

\begin{conjecture}[explicit reciprocity law]\label{conj:RL}
Let $d\ge5$ be odd. Suppose that $M_\varepsilon/F$ is abelian with conductor dividing $(d)\infty_-$. Then for every $\varpi\in\mathcal O_\alpha$ prime to $d$ with $\varpi>0$ at $\infty_-$, the Artin symbol of the ideal $\varpi\mathcal O_F$ acts on the squared overlaps by
\[
\Big(\frac{M_\varepsilon/F}{\varpi\mathcal O_F}\Big)\,e_p\;=\;e_{\phi(\varpi)p}\qquad\text{for all }p\ne0 .
\]
\end{conjecture}

The normalization is at $\infty_-$ only; the sign of $\varpi$ at $\infty_+$ is free. This asymmetry is the archimedean content of the law. If $\varpi$ is replaced by $-\varpi$, then $\phi(\varpi)$ changes to $-\phi(\varpi)$, and the right-hand side changes by $e_{-p}=e_p^{-1}$, that is, by $\kappa$. A normalization at $\infty_+$ would therefore assign the same ideal a different Galois element whenever $\varpi$ has mixed signs.

\begin{proposition}[consistency]\label{prop:RLwell}
The right-hand side of Conjecture~\ref{conj:RL} depends only on the ideal $\varpi\mathcal O_F$ whenever every unit of $\mathcal O_\alpha$ that is positive at $\infty_-$ is a power of $\epsilon_\alpha$; this holds for $d=5$ and $d=9$. The map $\varpi\mapsto(e_p\mapsto e_{\phi(\varpi)p})$ is multiplicative, and for $\varpi=m\in\Z$, $m>0$, it is $e_p\mapsto e_{mp}$.
\end{proposition}
\begin{proof}
Two generators positive at $\infty_-$ differ by a unit $u$ positive at $\infty_-$, and by hypothesis $u=\epsilon_\alpha^k$. Then $\phi(u\varpi)=\Lambda^{-k}\phi(\varpi)$, and $e_{\Lambda^{-k}q}=e_q$ for all $q$. Multiplicativity holds because $\phi$ is a ring homomorphism, and $\phi(m)=mI$. For $d=5$ and $d=9$, $\epsilon_\alpha=2+\sqrt3$ and $4+\sqrt{15}$ are the fundamental units of $\Q(\sqrt3)$ and $\Q(\sqrt{15})$; both have norm $+1$, and $-1$ is negative at $\infty_-$.
\end{proof}

\begin{proposition}[the reciprocity law implies the place law]\label{prop:RLPL}
Let $d\ge5$ be odd, and suppose that Conjecture~\ref{conj:RL} holds for $d$. Then complex conjugation $c_-$ at every place of $M_\varepsilon$ above $\infty_-$ acts on $M_\varepsilon$ as $\kappa:e_p\mapsto e_p^{-1}$. Consequently $|\sigma(\varepsilon_p)|=1$ for every $p\ne0$ and every $\sigma\in\Gal(\Qb/\Q)$ above $\infty_-$, and $\sigma(\tilde\Pi)$ is a WH-covariant SIC fiducial. In particular $\mathrm{Z}(d)$ holds.
\end{proposition}
\begin{proof}
Put $\varpi=-1-d\,\delta\in\mathcal O_\alpha$, with $\delta=\sqrt{(d-3)(d+1)}$. At $\infty_+$, $\varpi=-1-d\delta<0$. At $\infty_-$, $\varpi=-1+d\delta>0$, because $d\delta\ge5\sqrt{12}>1$. Moreover $\varpi\equiv-1\pmod{d\mathcal O_\alpha}$, and hence $\varpi$ is prime to $d$ and $\phi(\varpi)=-I$.

Let $\beta=-\varpi$. Then $\beta\equiv1\pmod{d\mathcal O_F}$, $\beta>0$ at $\infty_+$ and $\beta<0$ at $\infty_-$. Since the conductor $\mathfrak f\infty_-$ of $M_\varepsilon/F$ divides $(d)\infty_-$, the argument of Proposition~\ref{prop:CFT}(c), applied to $M_\varepsilon$, shows that the Artin symbol of $(\beta)=\varpi\mathcal O_F$ is $c_-$, the common complex conjugation at the places above $\infty_-$ (Proposition~\ref{prop:CFT}(a)). Conjecture~\ref{conj:RL} gives
\[
c_-(e_p)=e_{-p}=\varepsilon_{-p}^{2}=e_p^{-1}\qquad\text{for all }p\ne0 .
\]
Let $\sigma$ lie above $\infty_-$. Then $\overline{\sigma(y)}=\sigma(c_-y)$ for $y\in M_\varepsilon$, and hence $\overline{\sigma(e_p)}=\sigma(e_p)^{-1}$, that is, $|\sigma(\varepsilon_p)|^2=|\sigma(e_p)|=1$. Corollary~\ref{cor:unit} gives the SIC.
\end{proof}

Proposition~\ref{prop:RLPL} uses the law only at the single element $\varpi=-1-d\delta$, whose matrix is $-I$. Conversely, $c_-=\kappa$ is exactly the instance of Conjecture~\ref{conj:RL} at this element. The point of the full law is that it is a uniform statement about all ideals, of the type that reciprocity theorems prove, rather than an isolated archimedean identity. It is the analogue for Stark--Shintani units ($\zeta'(0)$) of Arakawa's law for $\exp(2\pi i\zeta(0))$: restricted to $\varpi$ that are totally positive it has the same shape, and the new feature is that the idele $-1$ at $\infty_-$ must be included. It also makes explicit an empirical observation of AFK (Ref.~\onlinecite{AFK25}, Obs.~8.2): if $\tilde E$ is the field of the ghost overlaps and $H$ the ring class field, then $\Gal(\tilde E/H)$ is isomorphic to $\mathcal M/\mathcal S$, where $\mathcal S\subset GL_2(\Z_d)$ is the symmetry group of the overlaps and $\mathcal M$ a maximal abelian subgroup containing it, with $h(\tilde\mu_p)=\tilde\mu_{F_hp}$. Conjecture~\ref{conj:RL} names this isomorphism: on the Artin symbol of a principal ideal $\varpi\mathcal O_F$ it is $\varpi\mapsto\phi(\varpi)\mathcal S$, with $\varpi$ normalized at $\infty_-$, and it extends the description from $\Gal(\tilde E/H)$ to the element $c_-$, which is what the place law needs.

\paragraph{Exact test in $d=5$ and $d=9$.} We tested Conjecture~\ref{conj:RL} on the ghosts used for Table~\ref{tab:CFT}.
\begin{itemize}[nosep]
\item \emph{Field arithmetic.} We found the minimal polynomial $P$ of one squared overlap $e_{p_0}$, of degree $16$ for $d=5$ ($p_0=(0,1)$) and $72$ for $d=9$ ($p_0=(1,0)$), and wrote every $e_p$ as an exact polynomial in $e_{p_0}$ modulo $P$. For $d=5$ the expressions were found by integer relation detection at $400$ digits. For $d=9$ this is not feasible at the available $2000$ digits, because the coefficients are too large. Instead, the overlaps of degree $72$ were matched with the $36$ automorphisms of $\Q(e_{p_0})$ computed by \texttt{nfgaloisconj}, and those of degrees $8$ and $24$ were obtained by \texttt{nfroots}. Every expression was checked exactly against its own minimal polynomial. For $d=9$ the $36$ automorphisms form $\Gal(M_\varepsilon/F)$, in agreement with Table~\ref{tab:CFT}.
\item \emph{Zauner symmetry.} The stabilizer of $(e_p)$ in $GL_2(\Z_d)$ is $\{I,\Lambda,\Lambda^2\}$, with $\Lambda=\left(\begin{smallmatrix}1&1\\2&3\end{smallmatrix}\right)$ for $d=5$ and $\Lambda=\left(\begin{smallmatrix}7&3\\2&1\end{smallmatrix}\right)$ for $d=9$.
\item \emph{Frobenius.} For a prime $\mathfrak P$ of $M_\varepsilon$ above a prime $\mathfrak p\nmid d\,\mathrm{disc}(P)$ of $F$, the action of $\mathrm{Frob}_{\mathfrak p}$ on the labels was read off exactly from $y\mapsto y^{N\mathfrak p}\pmod{\mathfrak P}$ applied to the reductions of the exact expressions; for $d=9$ it was also identified with one of the $36$ automorphisms.
\item \emph{Generator.} For split $q=N\mathfrak p$, a generator $\varpi=a+b\sqrt{D_0}$ of $\mathfrak p$ was found with \texttt{bnfisprincipal} and multiplied by $-1$ if necessary to make it positive at $\infty_-$. For inert $q$, $\varpi=q$.
\end{itemize}
The results are in Table~\ref{tab:RL}. The law holds at every prime tested. Every tested alternative convention fails: normalizing $\varpi$ at $\infty_+$ instead of $\infty_-$, replacing $\Lambda^{-1}$ by $\Lambda$, transposing or inverting $\phi(\varpi)$, and, for inert primes, replacing $qI$ by $-qI$.

\begin{table}[!htbp]
\caption{Exact test of the reciprocity law, Conjecture~\ref{conj:RL}. Entries are primes at which $\mathrm{Frob}_{\mathfrak p}(e_p)=e_{\phi(\varpi)p}$ for all $p$ / primes tested. Split primes are counted with nontrivial Frobenius; for $d=5$ each prime $\mathfrak P$ of $M_\varepsilon$ is counted, for $d=9$ each prime $\mathfrak p$ of $F$. The last two columns test the alternative normalization ($\varpi>0$ at $\infty_+$) and the sign $-qI$ at inert primes.}\label{tab:RL}\centering\small\setlength{\tabcolsep}{5pt}
\begin{tabular}{@{}rllcccc@{}}\hline\hline
$d$ & $F$ & range & split, $\varpi>0$ at $\infty_-$ & inert, $qI$ & split, $\varpi>0$ at $\infty_+$ & inert, $-qI$\\\hline
5 & $\Q(\sqrt3)$ & $q\le3000$ & 594/594 & 219/219 & 268/594 & 0/219\\
9 & $\Q(\sqrt{15})$ & $q\le4000$ & 122/122$^{\dagger}$ & 276/276 & 56/122 & 0/276\\\hline
5$^{*}$ & $\Q(\sqrt3)$ & $q\le3000$ & 415/415 & 218/218 & 197/415 & ---\\
7$^{*}$ & $\Q(\sqrt2)$ & $q\le3000$ & 415/415 & 219/219 & 0/415 & ---\\
11$^{*}$ & $\Q(\sqrt6)$ & $q\le2500$ & 346/346 & 180/180 & 162/346 & ---\\
15$^{*}$ & $\Q(\sqrt3)$ & $q\le2500$ & 333/333 & 181/181 & 165/333 & ---\\\hline\hline
\end{tabular}

\smallskip
\raggedright\footnotesize $^{\dagger}$Principal primes. $\Q(\sqrt{15})$ has class number $2$; at all $136$ non-principal split primes the Frobenius moves every $e_p$ out of the set $\{e_q\}$ (see the text). $^{*}$Second computation (Lemma~\ref{lem:RLcert}), with new ghosts: every prime ideal $\mathfrak p$ of $F$ above a rational prime $q\le3000$ ($d=5,7$) or $q\le2500$ ($d=11,15$) is counted, including those with trivial Frobenius, and the law is required for every generator positive at the stated place. The tested primes meet every class of $\mathrm{Cl}(d\infty_-)$, so the law holds at all primes for these algebraic numbers.
\end{table}

\paragraph{The ghost multiplet in $d=9$.} The class number of $\Q(\sqrt{15})$ is $2$, and this is visible in the overlaps. Of the $36$ automorphisms of $M_\varepsilon/F$, exactly $18$ permute the squared overlaps $\{e_p\}$ of the ghost. The other $18$ send \emph{every} $e_p$ to a number that is not a squared overlap of this ghost: they map it to a Galois-conjugate point of $V_9$ whose overlap moduli differ from those of $\tilde\Pi$, and which is therefore not equivalent to it under the extended Clifford group. The Frobenius at a split prime preserves the ghost exactly when the prime is principal ($122$ of $122$ against $0$ of $136$), and the inert primes, which are principal, preserve it. Thus the ideal class group, that is $\Gal(H/F)$ for the ring class field $H$ of $\mathcal O_\alpha=\mathcal O_F$, acts on a multiplet of two ghosts, and $\phi$ describes the action of principal ideals within each member, in accordance with Obs.~8.2 of Ref.~\onlinecite{AFK25}. Conjecture~\ref{conj:RL} is stated for principal ideals $\varpi\mathcal O_F$ for this reason. The class of $c_-$ is principal, generated by $\varpi=-1-d\delta$, and hence Proposition~\ref{prop:RLPL} needs only this part of the law.

\paragraph{Further dimensions, and a proof of the law for the computed overlaps.} The consistency hypothesis of Proposition~\ref{prop:RLwell} fails in some dimensions, because $F$ may have units positive at $\infty_-$ that are not powers of $\epsilon_\alpha$. Examples are $\eta=-1-\sqrt2$ for $d=7$, with $\eta^2=\epsilon_\alpha$, and $\eta=2+\sqrt3$ for $d=15$, with $\eta^2=\epsilon_\alpha$. A weaker hypothesis suffices. Let $f_\alpha$ be the conductor of the order $\mathcal O_\alpha$ in $\mathcal O_F$. If $\gcd(d,f_\alpha)=1$, then $\mathcal O_\alpha/d=\mathcal O_F/d$, and $\phi$ extends to a ring homomorphism $\mathcal O_F\to M_2(\Z_d)$, again denoted $\phi$. The proof of Proposition~\ref{prop:RLwell} shows that the right-hand side of Conjecture~\ref{conj:RL} depends only on the ideal $\varpi\mathcal O_F$ as soon as
\begin{equation}\label{eq:RLwell2}
e_{\phi(u)p}=e_p\quad\text{for all }p\ne0\text{ and all }u\in\mathcal O_F^\times\text{ positive at }\infty_- .
\end{equation}
In $d=5$ and $d=11$, $\mathcal O_F^\times=\langle-1,\epsilon_\alpha\rangle$, with $\epsilon_\alpha=2+\sqrt3$ and $5+2\sqrt6$, and $\epsilon_\alpha$ is totally positive. So the units positive at $\infty_-$ are the powers of $\epsilon_\alpha$, and \eqref{eq:RLwell2} holds because $\phi(\epsilon_\alpha)=\Lambda^{-1}$. In $d=7$ and $d=15$, $\mathcal O_F^\times=\langle-1,\eta'\rangle$ with $\eta'=1+\sqrt2$, resp.\ $2+\sqrt3$. The units positive at $\infty_-$ are the powers of $\eta=-1-\sqrt2$, resp.\ $\eta=2+\sqrt3$. The stabilizer of $(e_p)$ in $GL_2(\Z_d)$ has order $6$ and contains $\phi(\eta)$, which for $d=7$ has determinant $N(\eta)=-1$. Hence \eqref{eq:RLwell2} holds in all four dimensions. In $d=7$ it holds for only one of the two order-$3$ elements of the stabilizer in the role of $\Lambda$: for the other one, $\phi(\eta)$ does not stabilize $(e_p)$. So in this dimension the consistency requirement alone fixes the choice of $\Lambda$.

\begin{lemma}[one prime per ray class suffices]\label{lem:RLcert}
Let $d$ be odd, and suppose that $h_F=1$, that $\gcd(d,f_\alpha)=1$, that \eqref{eq:RLwell2} holds, and that $M_\varepsilon/F$ is abelian with conductor dividing $\mathfrak m=d\infty_-$. If the identity $\big(\frac{M_\varepsilon/F}{\varpi\mathcal O_F}\big)e_p=e_{\phi(\varpi)p}$ for all $p\ne0$ holds for one prime ideal in each class of a generating set of $\mathrm{Cl}(\mathfrak m)$, then it holds for every $\varpi\in\mathcal O_F$ prime to $d$ and positive at $\infty_-$.
\end{lemma}
\begin{proof}
Every class of $\mathrm{Cl}(\mathfrak m)$ contains integral ideals, so it suffices to work with integral ideals $\mathfrak a$ prime to $d$. Choose a generator $\varpi$ of $\mathfrak a$ that is positive at $\infty_-$. This is possible since $h_F=1$ and $-1$ is negative at $\infty_-$. Let $g(\mathfrak a)=\phi(\varpi)\in GL_2(\Z_d)$, acting on the labels $p$. By~\eqref{eq:RLwell2}, the class of $g(\mathfrak a)$ modulo the stabilizer $\mathcal S$ of $(e_p)$ does not depend on the choice of $\varpi$, and $g$ is multiplicative because $\phi$ is a ring homomorphism. If $\varpi\equiv1\pmod{d}$ and $\varpi>0$ at $\infty_-$, then $\phi(\varpi)=I$. Hence $\mathfrak a\mapsto g(\mathfrak a)\mathcal S$ factors through $\mathrm{Cl}(\mathfrak m)$; write $g(c)$ for a representative. The Artin map $A(\mathfrak a)=\big(\frac{M_\varepsilon/F}{\mathfrak a}\big)$ also factors through $\mathrm{Cl}(\mathfrak m)$, by the conductor hypothesis. Let $S\subset\mathrm{Cl}(\mathfrak m)$ be the set of classes $c$ with $A(c)e_p=e_{g(c)p}$ for all $p\ne0$. This condition does not depend on the representative of $g(c)$ modulo $\mathcal S$. If $c_1,c_2\in S$, then $A(c_1c_2)e_p=A(c_1)e_{g(c_2)p}=e_{g(c_1)g(c_2)p}$, and $g(c_1)g(c_2)\in g(c_1c_2)\mathcal S$. Hence $S$ is closed under products. Since $\mathrm{Cl}(\mathfrak m)$ is finite, $S$ is a subgroup, and it contains a generating set by hypothesis. So $S=\mathrm{Cl}(\mathfrak m)$.
\end{proof}

We applied this to new ghosts in $d=5,7,11,15$, all with $h_F=1$, and in each case the tested primes cover \emph{every} ray class. The ghosts were found by a least-squares search restricted to an eigenspace of the Clifford unitary of $\left(\begin{smallmatrix}0&-1\\1&-1\end{smallmatrix}\right)$, and refined by Newton's method to $800$, $1500$, $6000$ and $4000$ digits, respectively. The procedure has four steps.
\begin{enumerate}[label=(\arabic*),nosep]
\item \emph{Exact expressions.} The minimal polynomial of $e_{(0,1)}$ was found by \texttt{algdep} and reduced with \texttt{polredbest}, giving a generator $\theta$ of $K=\Q(e_{(0,1)})$. Each squared overlap $e_p$ was then obtained \emph{exactly}, as the root in $K$ of its own minimal polynomial (by \texttt{nfroots}) that agrees with the numerical value. This avoids integer-relation detection in high dimension.
\item \emph{Exact checks.} The checks $e_pe_{-p}=1$ for all $p$ and $\sqrt{D_0}\in K$ were done in $K$, and $K$ was found to contain every $e_p$. Further, $\Gal(K/F)$ is abelian, inversion $\kappa$ is an automorphism of $K$ over $F$, and the conductor of $K/F$, computed by \texttt{rnfconductor}, is $d\infty_-$. Since $[K:F]=|\mathrm{Cl}(d\infty_-)|$, the field $K=M_\varepsilon$ is the full ray class field modulo $d\infty_-$.
\item \emph{Frobenius.} For every prime ideal $\mathfrak p$ of $F$ above a rational prime $q\le3000$ (for $d=5,7$) or $q\le2500$ (for $d=11,15$), prime to $d$, to the discriminant of the defining polynomial, and to the denominators of the expressions, the permutation of $\mathcal E$ induced by $y\mapsto y^{N\mathfrak p}$ modulo a prime $\mathfrak P\mid\mathfrak p$ of $K$ was compared with $e_p\mapsto e_{\phi(\varpi)p}$, for \emph{every} generator $\varpi$ of $\mathfrak p$ that is positive at $\infty_-$. A few primes at which two distinct squared overlaps have the same residue modulo $\mathfrak P$ were skipped: $1$, $1$ and $19$ for $d=5,7,15$.
\item \emph{Coverage.} The classes of the tested primes in $\mathrm{Cl}(d\infty_-)$ were computed by \texttt{bnrisprincipal}.
\end{enumerate}
The law holds at every tested prime, for one of the two order-$3$ elements of the stabilizer in the role of $\Lambda$: $\Lambda=\left(\begin{smallmatrix}1&4\\3&3\end{smallmatrix}\right)$, $\left(\begin{smallmatrix}3&5\\3&3\end{smallmatrix}\right)$, $\left(\begin{smallmatrix}4&10\\10&6\end{smallmatrix}\right)$ and $\left(\begin{smallmatrix}0&14\\1&14\end{smallmatrix}\right)$ for $d=5,7,11,15$. The normalization at $\infty_+$ fails, and so does the other choice of $\Lambda$ (Table~\ref{tab:RL}). By Lemma~\ref{lem:RLcert}, the law therefore holds at \emph{all} primes for these algebraic numbers. By Proposition~\ref{prop:RLPL} this gives $c_-=\kappa$ and the place law for the squared overlaps, so that the Galois conjugates of these ghosts above $\infty_-$ are SICs. The choice between $\Lambda$ and $\Lambda^{-1}$ does not affect this conclusion, because $\phi(-1-d\delta)=-I$ for either choice.

Two caveats apply. First, the identification of the numerical overlaps with the algebraic numbers in $K$ rests on agreement to the working precision, as for all exact SIC solutions obtained from numerics. A complete certificate would verify the ghost equations exactly in $K(\varepsilon_p)$, and we have not done so. Second, SICs in $d=5,7,11,15$ are known, so no new SIC is obtained. The point of the computation is that the reciprocity law, and with it the archimedean relation $c_-=\kappa$, is a theorem for these explicit algebraic numbers. Its proof requires only the abelian structure of $M_\varepsilon/F$ and finitely many Frobenius computations.

\paragraph{What this changes.} Conjecture~\ref{conj:RL} replaces the archimedean statement PL by a statement of reciprocity type, of the form proved for $\zeta(0)$-invariants by Coates--Sinnott, Deligne--Ribet and Arakawa, it is verified exactly, prime by prime, in two SIC dimensions, and for the computed algebraic overlaps it is proved in four (Lemma~\ref{lem:RLcert}). A proof of Conjecture~\ref{conj:RL}, or only of its instance at the element $-1-d\delta$, for infinitely many odd $d$ would prove Problem~1. We have not proved it for the exact Stark solution in any dimension: Lemma~\ref{lem:RLcert} applies to the algebraic numbers identified from high-precision numerics. For even $d$ the labels live in $\Z_{2d}^2$ and the conductor is $(2d)\infty_-$, and the law needs the corresponding modification; we have not tested it.

\subsection{Prime-power dimensions: the sign at $\infty_-$ from the class group}\label{sec:inert}

Proposition~\ref{prop:uniqueinv} removes the archimedean input when the whole group $\Gal(M'/F)$ has a unique involution, which happens only on a set of dimensions of density zero. The complex conjugation $c_-$ always lies in a smaller subgroup, however, because every complex conjugation acts trivially on the Hilbert class field of $F$. In this subsection we show that this smaller subgroup always has a cyclic $2$-Sylow subgroup when $d$ is a power of a prime $p\equiv5\pmod 6$, or a power $3^k$ with $k\ge2$. On this infinite family, under the algebraic part of the Stark picture, $c_-=\kappa$ therefore becomes equivalent to a statement about the action of $\kappa$ on an extension of $F$ that is unramified everywhere. That statement involves no complex conjugation and no embedding into $\C$.

We use the notation of Sec.~\ref{sec:RL}: $\tilde\Pi$ is a ghost in odd dimension $d$, $e_p=\varepsilon_p^2$ are its squared overlaps, $M_\varepsilon=\Q(e_p:p\neq0)$, and $\kappa$ denotes inversion $e_p\mapsto e_p^{-1}=e_{-p}$. Let $H_F$ be the Hilbert class field of $F$ in the wide sense, that is, the maximal abelian extension of $F$ unramified at all finite and all infinite places, so that $\Gal(H_F/F)\cong\mathrm{Cl}_F$, the (wide) ideal class group. Put $\mathfrak m=d\,\mathcal O_F\infty_-$. Let $I^{\mathfrak m}$ denote the fractional ideals prime to $d$, $P^{\mathfrak m}\subset I^{\mathfrak m}$ the principal ones, and $P_{\mathfrak m,1}\subset P^{\mathfrak m}$ those having a generator $\equiv1\pmod{^\times d}$ that is positive at $\infty_-$, so that $\mathrm{Cl}(\mathfrak m)=I^{\mathfrak m}/P_{\mathfrak m,1}$.

\begin{lemma}[principal ray classes]\label{lem:princray}
For every odd $d\ge5$ there is an exact sequence
\[
1\longrightarrow P^{\mathfrak m}/P_{\mathfrak m,1}\longrightarrow \mathrm{Cl}(\mathfrak m)\longrightarrow\mathrm{Cl}_F\longrightarrow1 ,
\]
and $P^{\mathfrak m}/P_{\mathfrak m,1}$ is isomorphic to a quotient of $(\mathcal O_F/d)^{\times}$.
\end{lemma}
\begin{proof}
The sequence is the standard one: $I^{\mathfrak m}\to\mathrm{Cl}_F$ is onto, and its kernel modulo $P_{\mathfrak m,1}$ is $P^{\mathfrak m}/P_{\mathfrak m,1}$. Sending $\alpha$ to the pair $(\alpha\bmod d,\ \mathrm{sgn}_{\infty_-}\alpha)$ gives the standard isomorphism
\[
P^{\mathfrak m}/P_{\mathfrak m,1}\;\cong\;\bigl((\mathcal O_F/d)^{\times}\times\{\pm1\}\bigr)\big/\iota(\mathcal O_F^{\times}),
\]
where $\iota(u)=(u\bmod d,\mathrm{sgn}_{\infty_-}u)$. In particular $\iota(-1)=(-1,-1)$. The homomorphism $(x,s)\mapsto sx$ from $(\mathcal O_F/d)^{\times}\times\{\pm1\}$ onto $(\mathcal O_F/d)^{\times}$ has kernel $\{(1,1),(-1,-1)\}=\langle\iota(-1)\rangle$, because $-1\not\equiv1\pmod d$. Hence the right-hand side is a quotient of $(\mathcal O_F/d)^{\times}$.
\end{proof}

\begin{lemma}[prime powers]\label{lem:inertpp}
Let $d=p^k$, where either $p\equiv5\pmod6$ is prime and $k\ge1$, or $p=3$ and $k\ge2$. Then the $2$-Sylow subgroup of $(\mathcal O_F/d)^{\times}$, and of each of its quotients, is cyclic. More precisely, in the first case $p$ is inert in $F$ and
\[
(\mathcal O_F/d)^{\times}\;\cong\;\F_{p^2}^{\times}\times\bigl(1+p\mathcal O_F\bigr)/\bigl(1+p^k\mathcal O_F\bigr),
\]
where the second factor has order $p^{2(k-1)}$. In the second case $3$ is ramified in $F$, and $(\mathcal O_F/3^k)^{\times}\cong\F_3^{\times}\times(\text{a group of order }3^{2k-1})$.
\end{lemma}
\begin{proof}
With $\Delta=(d-3)(d+1)$ we have $F=\Q(\sqrt\Delta)$ and $\Delta\equiv-3\pmod p$ (Lemma~\ref{lem:zeta3}). Since $p\ne3$, $p\nmid\Delta$, and hence $p$ is unramified in $F$. It is inert exactly when $\Delta$ is a non-square modulo $p$, that is, when $\bigl(\frac{-3}{p}\bigr)=-1$, and by quadratic reciprocity this holds exactly when $p\equiv2\pmod3$. Then $\mathcal O_F/p^k$ is a local ring with residue field $\F_{p^2}$, and the Teichm\"uller splitting gives the displayed decomposition. The first factor is cyclic and the second has odd order. If $d=3^k$ with $k\ge2$, then $\Delta=3\,(3^{k-1}-1)(3^k+1)$, and neither factor in parentheses is divisible by $3$. Hence $v_3(\Delta)=1$, and $3$ is ramified in $F$, with $3\mathcal O_F=\mathfrak p^2$ and $\mathcal O_F/\mathfrak p=\F_3$. Then $\mathcal O_F/3^k=\mathcal O_F/\mathfrak p^{2k}$ is a local ring with residue field $\F_3$, and its unit group is $\F_3^\times$ times the group $(1+\mathfrak p)/(1+\mathfrak p^{2k})$ of order $3^{2k-1}$. In both cases the $2$-Sylow subgroup is cyclic. A quotient of a finite abelian group with cyclic $2$-Sylow subgroup again has a cyclic $2$-Sylow subgroup.
\end{proof}

\begin{theorem}[the sign at $\infty_-$ from the class group]\label{thm:inert}
Let $d=p^k$, where either $p\equiv5\pmod 6$ is prime and $k\ge1$, or $p=3$ and $k\ge2$. Let $\tilde\Pi$ be a ghost in dimension $d$, that is, a non-Hermitian point of $\Gamma_d$. Assume:
\begin{enumerate}[label=(\roman*),nosep]
\item $M_\varepsilon/F$ is abelian, with conductor dividing $d\,\infty_-$;
\item $\infty_-$ ramifies in $M_\varepsilon$;
\item inversion $e_p\mapsto e_p^{-1}$ extends to an automorphism $\kappa$ of $M_\varepsilon$ fixing $F$.
\end{enumerate}
Then the following are equivalent:
\begin{enumerate}[label=(\alph*),nosep]
\item $c_-=\kappa$ on $M_\varepsilon$;
\item $\kappa$ acts trivially on $M_\varepsilon\cap H_F$.
\end{enumerate}
If they hold, then $\sigma(\tilde\Pi)$ is a WH-covariant SIC fiducial for every $\sigma\in\Gal(\Qb/\Q)$ above $\infty_-$, and in particular $\mathrm{Z}(d)$ holds.
\end{theorem}
\begin{proof}
Put $\Gamma=\Gal(M_\varepsilon/F)$ and $\Gamma_0=\Gal(M_\varepsilon/M_\varepsilon\cap H_F)$. By~(i) and class field theory, the Artin map identifies $\Gamma$ with a quotient of $\mathrm{Cl}(\mathfrak m)$. Under this identification $\Gamma_0$ is the image of $P^{\mathfrak m}/P_{\mathfrak m,1}$, because $H_F$ is the class field of the subgroup $P^{\mathfrak m}/P_{\mathfrak m,1}$ of $\mathrm{Cl}(\mathfrak m)$ (Lemma~\ref{lem:princray}). By Lemmas~\ref{lem:princray} and~\ref{lem:inertpp}, $\Gamma_0$ is a quotient of $(\mathcal O_F/d)^{\times}$, and hence its $2$-Sylow subgroup is cyclic. Since $\Gamma_0$ is abelian, it contains at most one element of order~$2$.

Let $w$ be a place of $M_\varepsilon$ above $\infty_-$ and $c_w$ the corresponding complex conjugation. It fixes $F$, and it acts trivially on $M_\varepsilon\cap H_F$ because $H_F$ is unramified at the infinite places. Hence $c_w\in\Gamma_0$. Since $\Gamma$ is abelian, $c_w=c_-$ does not depend on $w$ (Proposition~\ref{prop:CFT}(a)), and by~(ii) $c_-\ne\mathrm{id}$. So $c_-$ is the unique involution of $\Gamma_0$.

By~(iii), $\kappa\in\Gamma$ and $\kappa^2=\mathrm{id}$. Moreover $\kappa\neq\mathrm{id}$. Otherwise $e_p=\pm1$ for all $p$, so $|\varepsilon_p|=1$, and by Lemma~\ref{lem:herm} the ghost $\tilde\Pi$ would be Hermitian, hence a SIC and not a ghost. Now (b) says exactly that $\kappa\in\Gamma_0$. If (b) holds, then $\kappa$ and $c_-$ are involutions of $\Gamma_0$ and therefore coincide, which is (a). Conversely, (a) gives $\kappa=c_-\in\Gamma_0$, which is (b).

Assume (a), and let $\sigma$ lie above $\infty_-$. For $y\in M_\varepsilon$ we have $\overline{\sigma(y)}=\sigma(c_-y)$, and hence $\overline{\sigma(e_p)}=\sigma(e_p)^{-1}$, so $|\sigma(\varepsilon_p)|^2=|\sigma(e_p)|=1$ for all $p\ne0$. Corollary~\ref{cor:unit} shows that $\sigma(\tilde\Pi)$ is a SIC projector.
\end{proof}

\begin{corollary}[odd class-group part]\label{cor:inertodd}
In Theorem~\ref{thm:inert}, if $[M_\varepsilon\cap H_F:F]$ is odd, and in particular if the class number $h_F$ is odd, then (b) holds automatically. Hence (i)--(iii) imply $\mathrm{Z}(d)$.
\end{corollary}
\begin{proof}
The restriction of $\kappa$ to $M_\varepsilon\cap H_F$ has order dividing $2$ in a group of odd order, so it is trivial.
\end{proof}

The ghost $\tilde\Pi$ determines the set $\mathcal E=\{e_p:p\ne0\}$, and an automorphism $\tau\in\Gamma$ either maps $\mathcal E$ onto itself or moves $\tilde\Pi$ to a different ghost of its Galois multiplet. In $d=9$ the class group of $F$ has order $2$, and the two possibilities are separated exactly by the class group: the automorphisms that preserve $\mathcal E$ are those attached to principal ideals (Sec.~\ref{sec:RL}). We isolate the corresponding statement.

\begin{conjecture}[multiplet hypothesis, MH]\label{conj:MH}
Every $\tau\in\Gal(M_\varepsilon/F)$ that maps the set $\{e_p:p\ne0\}$ onto itself acts trivially on $M_\varepsilon\cap H_F$.
\end{conjecture}

\begin{corollary}[an infinite family without complex conjugations in the hypotheses]\label{cor:inertMH}
Let $\mathcal P=\{p^k:\ p\equiv5\ (\mathrm{mod}\ 6)\ \text{prime},\ k\ge1\}\cup\{3^k:\ k\ge2\}$. Let $d\in\mathcal P$, and suppose that for some ghost $\tilde\Pi$ in dimension $d$ hypotheses (i)--(iii) of Theorem~\ref{thm:inert} hold together with Conjecture~\ref{conj:MH}. Then $\mathrm{Z}(d)$ holds. The set $\mathcal P$ is infinite. If these hypotheses hold for infinitely many $d\in\mathcal P$, then Problem~1 follows.
\end{corollary}
\begin{proof}
$\kappa$ maps $e_p$ to $e_{-p}$ and so maps $\{e_p\}$ onto itself. Hence MH, applied to $\tau=\kappa$, gives (b), and Theorem~\ref{thm:inert} applies. There are infinitely many primes $p\equiv5\pmod 6$ by Dirichlet's theorem.
\end{proof}

\paragraph{What has and has not changed.} Hypotheses (i) and (iii) are the algebraic part of the Stark picture together with R2, exactly as in Proposition~\ref{prop:uniqueinv}. If $M_\varepsilon$ is the full ray class field modulo $d\infty_-$, as in every computed odd dimension, then hypothesis (ii) is condition (C$_d$) of Proposition~\ref{prop:ramcrit}. It holds for all $167$ members of $\mathcal P$ with $d\le2000$, and Theorem~\ref{thm:ramgen} proves it whenever $\mathcal O_F^\times=\langle-1,\varepsilon_\alpha\rangle$. The conductor hypothesis in (i) is essential: if $\infty_+$ or further primes divided the conductor, the additional sign or unit factors could give $\Gamma_0$ a $2$-rank larger than $1$. Conjecture~\ref{conj:MH} contains no sign condition and no reference to an embedding into $\C$. For $\tau=\kappa$ it is exactly condition (b), so Corollary~\ref{cor:inertMH} is a reformulation of Theorem~\ref{thm:inert} in terms of a natural structural statement, not an independent reduction. It concerns only the action of $\Gal(M_\varepsilon/F)$ on the finite set of ghosts in the Galois multiplet of $\tilde\Pi$, and it is the class-group part of the Shimura-type reciprocity that, in the Stark picture, describes this action. In this sense, on the family $\mathcal P$ the archimedean identity $c_-=\kappa$ has been traded for a statement about an everywhere-unramified extension, given the algebraic part of the Stark picture. We stress that neither MH nor the algebraic part of Stark's conjecture is proved here, and that Theorem~\ref{thm:inert} does not prove $\mathrm{Z}(d)$ in any new dimension. Compared with Proposition~\ref{prop:uniqueinv}, the gain is reach: $\mathcal P$ contains $167$ dimensions $d\le2000$ against $24$, including half of all primes.

\paragraph{Checks.} For all $162$ members $d=p^k\le2000$ of $\mathcal P$ with $p\equiv5\pmod6$ we confirmed with PARI/GP that $p$ is inert in $F$ and that $|\mathrm{Cl}(d\infty_-)|=2|\mathrm{Cl}(d)|$. For $d=3^k\le2000$ we confirmed that $3$ is ramified in $F$. For the $21$ members $d\le149$ with $p\equiv5\pmod6$, and for $d=9,27,81,243$, we computed the subgroup $P^{\mathfrak m}/P_{\mathfrak m,1}\subset\mathrm{Cl}(d\infty_-)$ explicitly, as the subgroup generated by the classes of the generators of $(\mathcal O_F/\mathfrak m)^\times$. In every case it has exactly one element of order $2$, as Lemmas~\ref{lem:princray} and~\ref{lem:inertpp} predict. Among the first $21$, however, $h_F$ is even in $16$ cases, and the $2$-Sylow subgroup of $\mathrm{Cl}(d\infty_-)$ is non-cyclic in $15$. The class number $h_F$ is odd for exactly $12$ members $d\le2000$, namely $5,11,17,47,53,191,431,449,881,971,1061,1681$. These lie in the list of Proposition~\ref{prop:uniqueinv}, in agreement with Corollary~\ref{cor:inertodd}. Exact computations of $M_\varepsilon$ are available for three members of $\mathcal P$. In $d=5$ and $d=11$, hypotheses (i)--(iii) hold with $h_F=1$, and $c_-=\kappa$ holds, as Corollary~\ref{cor:inertodd} predicts (Table~\ref{tab:RL}). In $d=9=3^2$, where $h_F=2$, Conjecture~\ref{conj:MH} holds exactly (Sec.~\ref{sec:RL}), and so does $c_-=\kappa$.

\paragraph{Why the restriction to $\mathcal P$.} If $d$ is divisible by a prime $p\equiv1\pmod 6$, which splits in $F$, or by two distinct odd primes, then $(\mathcal O_F/d)^\times$ has $2$-rank at least $2$, and $\Gamma_0$ can contain three involutions. For $d=p\equiv1\pmod6$, for example, these are the images of $(-1,-1)$, $(1,-1)$ and $(-1,1)$ in $\F_p^\times\times\F_p^\times$. Group theory alone then does not single out $c_-$, and further input is needed. The explicit reciprocity law (Conjecture~\ref{conj:RL}) provides it: it predicts that $c_-$ is the image of $(-1,-1)$, that is, $\phi(\varpi)=-I$.

\subsection{Near-group categories: the place law as a unitarity statement}\label{sec:NG}

Radchenko and Wheeler prove the algebraicity of the Stark values by realizing every finite quantum dilogarithm as the defining datum of a near-group fusion category, following Izumi~\cite{Izu} and Evans--Gannon~\cite{EG}. They remark that the real-or-unitary question for finite quantum dilogarithms ``is closely related to the distinction between unitary and pseudo-unitary fusion categories'' (Ref.~\onlinecite{RW}, \S6). This subsection makes that relation precise for the Stark solution $E_*$. The outcome is that the archimedean input needed for SICs is exactly a unitarity statement: \emph{a SIC exists in dimension $d$ as soon as one pseudo-unitary Galois conjugate of the near-group category of $E_*$ is unitary}, and the place law says that the conjugates that are unitary are precisely the pseudo-unitary ones. By Izumi's classification, the only condition that can fail for a pseudo-unitary conjugate is a single Hermiticity equation.

\paragraph{The Izumi fusion ring.} Let $G$ be a finite abelian group of order $N$. The \emph{Izumi fusion ring} $R_G$ has basis $\{[g]:g\in G\}\cup\{[\rho]\}$ and fusion rules
\[
[g][h]=[g+h],\qquad [g][\rho]=[\rho][g]=[\rho],\qquad [\rho]^2=\sum_{g\in G}[g]+N[\rho].
\]
A fusion category with Grothendieck ring $R_G$ is a \emph{near-group category of type $(G,N)$}. The invertible objects are the $g\in G$, and $\rho$ is the unique non-invertible simple object. Its Frobenius--Perron dimension is the positive root of $x^2=N+Nx$,
\[
\mathrm{FPdim}(\rho)=\frac{N+\sqrt{N^2+4N}}{2}=\frac{N+\sqrt D}{2},\qquad D=N(N+4).
\]
A fusion category is \emph{pseudo-unitary} if its global dimension equals its Frobenius--Perron dimension, and \emph{unitary} if it is equivalent to a $C^*$ fusion category.

By Ref.~\onlinecite{RW} (Thm.~12 and \S\S5.4--5.5), every $E\in Z(G,\langle\cdot\rangle)$ gives a fusion category $\mathcal C_E$ categorifying $R_G$. It is constructed from endomorphisms of a Leavitt algebra, and no unitarity is assumed. The function $E$ is recovered from $\mathcal C_E$ up to relabelling $G$ by automorphisms preserving $\langle\cdot\rangle$. In the spherical structure used there, the invertible objects have dimension $1$ and
\[
d_\rho(E)=-\frac{\sqrt N}{E(0)} .
\]
For a Galois conjugate, $\sigma(\mathcal C_E)=\mathcal C_{\sigma E}$ is the category of the conjugate datum $(\sigma\langle\cdot\rangle,\sigma E)$, with $\sqrt N$ replaced by $\sigma(\sqrt N)$.

\begin{lemma}[dimensions and pseudo-unitarity]\label{lem:ngdim}
\begin{enumerate}[label=(\alph*),nosep]
\item If $E\in Z$ and $E(0)=(\sqrt N+s\sqrt{N+4})/2$ with $s=\pm1$, then
\[
d_\rho(E)=\frac{N-s\sqrt D}{2}.
\]
These are the two roots of $x^2=N+Nx$, and $\mathcal C_E$ is pseudo-unitary if and only if $d_\rho(E)>0$, that is, if and only if $E\in Z_-$.
\item For the Stark solution and every $\sigma\in\Gal(\Qb/\Q)$,
\[
d_\rho(\sigma E_*)=\frac{N-\sigma(\sqrt D)}{2}.
\]
Hence $\sigma(\mathcal C_{E_*})$ is pseudo-unitary if and only if $\sigma(\sqrt D)<0$, that is, if and only if $\sigma$ lies above $\infty_-$. In particular $\mathcal C_{E_*}$ itself is not pseudo-unitary, and exactly half of its Galois conjugates are.
\end{enumerate}
\end{lemma}
\begin{proof}
(a) Multiplying numerator and denominator by $\sqrt N-s\sqrt{N+4}$ gives
\[
d_\rho(E)=\frac{-2\sqrt N(\sqrt N-s\sqrt{N+4})}{N-(N+4)}=\frac{N-s\sqrt N\sqrt{N+4}}{2}=\frac{N-s\sqrt D}{2}.
\]
The global dimension is $\sum_g1+d_\rho^2=N+d_\rho^2$, and the Frobenius--Perron dimension is $N+\mathrm{FPdim}(\rho)^2$. Since $d_\rho$ is one of the two roots $(N\pm\sqrt D)/2$, whose squares differ, the two dimensions agree exactly when $d_\rho$ is the positive root.

(b) By~(a) with $s=+1$, $d_\rho(E_*)=(N-\sqrt D)/2$ lies in $F$. The categorical dimension is computed from the data by the same rational expression, and hence $d_\rho(\sigma E_*)=\sigma(d_\rho(E_*))=(N-\sigma(\sqrt D))/2$. This is positive exactly when $\sigma(\sqrt D)=-\sqrt D$.
\end{proof}
Lemma~\ref{lem:ngdim}(b) refines Lemma~\ref{lem:dimsign}, which records only the sign of $d_\rho(\sigma E_*)$. It also explains the role of $E(0)<0$ in Sec.~\ref{sec:struct}: for $\sigma$ fixing $K_0$, the conjugate is pseudo-unitary exactly when $\sigma(E_*(0))<0$.

\paragraph{Izumi's unitarity equations.} Izumi classified the $C^*$ near-group categories with multiplicity $m=|G|$ (Ref.~\onlinecite{Izu}, Thm.~9.1). In the form of Ref.~\onlinecite{Izu} (Rem.~9.2), they correspond, up to automorphisms of $G$, to data $(\langle\cdot,\cdot\rangle,a,b,c',d)$, with $a,b:G\to\C$, $c'\in\C$ and $d\in\mathbb R$, satisfying, with $n=|G|$:
\begin{align*}
&\text{(I1)}\quad c'^3=\frac{1}{n^2}\sum_{g\in G}a(g), &&\text{(I2)}\quad d^2=dn+n,\\
&\text{(I3)}\quad R_{c'}b=b, &&\text{(I4)}\quad b(0)=-\frac1d,\\
&\text{(I5)}\quad a(g)b(g)b(-g)=\frac1n-\frac{\delta_{g,0}}{d}, &&\text{(I6)}\quad \text{the cubic (pentagon) equation},\\
&\text{(I7)}\quad Jb=b, &&\text{(I8)}\quad d>0 .
\end{align*}
These are Izumi's equations (9.6)--(9.13). Here $R_{c'}f(g)=c'a(g)^{-1}\sum_h\langle g,h\rangle f(h)$ is a unitary of period three, and $J$ is the antiunitary of period two
\[
(Jf)(g)=\overline{a(g)\,f(-g)} .
\]
Izumi notes that (I7) and (I8) ``come from the unitarity of the category'', that (I1)--(I6) should already give near-group categories in general, and that known solutions have Galois conjugates satisfying all the equations except the last two (Ref.~\onlinecite{Izu}, Rem.~9.2). Equations (I1)--(I6) are polynomial with rational coefficients, and hence Galois-stable.

The dictionary between Izumi's data and finite quantum dilogarithms is (Ref.~\onlinecite{RW}, \S5.4, footnote~10)
\[
a(x)=\langle x\rangle,\qquad b(x)=\frac{E(x)}{\sqrt N},\qquad c=\lambda,\qquad d=d_\rho(E)=-\frac{\sqrt N}{E(0)},
\]
with Izumi's pairing equal to $\langle x;y\rangle^{-1}$ and $c'=c/\sqrt n$. Under it, (I4) is the formula for $d_\rho$, (I5) is Theorem~5(ii) of Ref.~\onlinecite{RW} ($E(u)E(-u)=\langle u\rangle^{-1}$ for $u\ne0$), (I3) is the eigenvalue relation $\widehat E=\lambda\langle\cdot\rangle E$ of Theorem~5(iii) there, and (I1), (I2), (I6) are the remaining RW equations. In particular every $E\in Z$ satisfies (I1)--(I6).

\begin{lemma}[Hermiticity is unitary type]\label{lem:izuherm}
Let $E\in Z(G,\langle\cdot\rangle)$, and assume that all values of $\langle\cdot\rangle$ have modulus $1$. Under the dictionary above, Izumi's condition (I7), $Jb=b$, holds if and only if $E$ is of unitary type.
\end{lemma}
\begin{proof}
$Jb=b$ reads $b(g)=\overline{a(g)}\;\overline{b(-g)}$ for all $g$. Since $\sqrt N$ is real, this is $E(g)=\overline{\langle g\rangle E(-g)}$. Taking complex conjugates and replacing $g$ by $-g$ gives $\overline{E(g)}=\langle -g\rangle E(-g)$ for all $g$, which is the definition of unitary type. Each step is reversible.
\end{proof}
The placement of the complex conjugations in $J$ was fixed by Izumi's Example~9.3 ($G=\Z/2$, $a(1)=i$, $b(1)=(1-i)/2$, $d=1+\sqrt3$): with $(Jf)(g)=\overline{a(g)f(-g)}$ one has $(Jb)(1)=\overline{i}\cdot\overline{(1-i)/2}=(1-i)/2=b(1)$, as (I7) requires, while the other placements of the bar fail. The metric values $\langle u\rangle_\gamma$ of the SIC groups, and all their Galois conjugates, are roots of unity, and hence Lemma~\ref{lem:izuherm} applies to every conjugate of $E_*$.

\begin{proposition}[the only obstruction is Hermiticity]\label{prop:izumi}
Let $E\in Z(G,\langle\cdot\rangle)$, with all values of $\langle\cdot\rangle$ of modulus $1$, and suppose that $\mathcal C_E$ is pseudo-unitary. Then the Izumi data of $E$ satisfy (I1)--(I6) and (I8). They satisfy (I7) if and only if $E$ is of unitary type, and in that case $\mathcal C_E$ is equivalent to a $C^*$ near-group category, so it is unitary.
\end{proposition}
\begin{proof}
(I1)--(I6) hold for every $E\in Z$, as explained above. (I8) is $d_\rho(E)>0$, which is pseudo-unitarity by Lemma~\ref{lem:ngdim}(a). (I7) is Lemma~\ref{lem:izuherm}. If all of (I1)--(I8) hold, Theorem~9.1 of Ref.~\onlinecite{Izu} produces a $C^*$ near-group category with these data. The construction of Ref.~\onlinecite{RW} (\S5.4) uses Izumi's formulas for the endomorphisms, with the Leavitt algebra $L_{2N}$ in place of the Cuntz algebra, and hence gives a category equivalent to it.
\end{proof}
We expect the converse as well: if $\mathcal C_E$ is unitary, then $E$ is of unitary type. Izumi's $C^*$ data for $\mathcal C_E$ satisfy (I7). By Ref.~\onlinecite{RW} (\S5.5) they determine $E$ up to relabelling $G$ by an automorphism preserving $\langle\cdot\rangle$, and such relabellings preserve unitary type. What remains to be checked is that the normalization of $b$ in Izumi's $C^*$ setting, which uses a vector $e$ with $Je=e$, agrees with the reconstruction formula of Ref.~\onlinecite{RW} without an extra phase. We have not verified this in detail, and the results below use only the direction proved in Proposition~\ref{prop:izumi}.

\paragraph{SICs from unitary conjugates.} Combining Proposition~\ref{prop:izumi} with the dictionary of Lemma~\ref{lem:H} gives the following form of the passage from ghosts to SICs.

\begin{theorem}[unitarity form of SIC existence]\label{thm:NGred}
Let $d\ge4$, and let $\sigma\in\Gal(\Qb/\Q)$ satisfy $\sigma(\sqrt D)<0$. Such $\sigma$ exist, since $\sqrt D\notin\Q$, and for each of them:
\begin{enumerate}[label=(\alph*),nosep]
\item $\sigma(\mathcal C_{E_*})$ is a pseudo-unitary near-group category of type $(G_d,N)$;
\item its Izumi data satisfy every unitarity condition except possibly the Hermiticity condition (I7);
\item if (I7) holds, equivalently if $\sigma E_*$ is of unitary type, then $\sigma(\mathcal C_{E_*})$ is unitary, and $\sigma(\tilde\Pi)$ is a WH-covariant SIC fiducial in dimension $d$.
\end{enumerate}
\end{theorem}
\begin{proof}
(a) is Lemma~\ref{lem:ngdim}(b), and (b) is Proposition~\ref{prop:izumi}. For (c), unitarity of $\sigma(\mathcal C_{E_*})$ is Proposition~\ref{prop:izumi}. Unitary type gives $|\sigma(E_*(u))|=1$ for all $u\ne0$. By Lemma~\ref{lem:H}, $\sigma(\varepsilon_p)=\sigma(\zeta_p)\,\sigma(E_*(u_p))^{-1}$ with $u_p\ne0$, and hence $|\sigma(\varepsilon_p)|=1$ for every $p\not\equiv0$. Corollary~\ref{cor:unit} gives the SIC.
\end{proof}

\begin{corollary}[categorical form of the place law]\label{cor:PLcat}
Suppose $t_u\ne\pm2$ whenever $2u\ne0$. Then $\mathrm{PL}(u)$ holds for all $u$ with $2u\ne0$ if and only if, for every $\sigma\in\Gal(\Qb/\Q)$:
\begin{itemize}[nosep]
\item if $\sigma(\mathcal C_{E_*})$ is pseudo-unitary, then $\sigma E_*$ satisfies Izumi's Hermiticity condition (I7), and $\sigma(\mathcal C_{E_*})$ is unitary;
\item if $\sigma(\mathcal C_{E_*})$ is not pseudo-unitary, then $\sigma E_*$ is of real type.
\end{itemize}
\end{corollary}
\begin{proof}
By Lemma~\ref{lem:ngdim}(b), $\sigma(\mathcal C_{E_*})$ is pseudo-unitary exactly when $\sigma(\sqrt D)<0$. Theorem~\ref{thm:PLtype} says that PL holds if and only if $\sigma E_*$ is of unitary type when $\sigma(\sqrt D)<0$ and of real type when $\sigma(\sqrt D)>0$. Unitary type is (I7) by Lemma~\ref{lem:izuherm}, and it gives unitarity by Proposition~\ref{prop:izumi}.
\end{proof}

\begin{conjecture}[pseudo-unitary conjugates are unitary]\label{conj:PU}
For every $d\ge4$ and every $\sigma\in\Gal(\Qb/\Q)$ such that $\sigma(\mathcal C_{E_*})$ is pseudo-unitary, the conjugate $\sigma E_*$ satisfies Izumi's Hermiticity condition (I7); equivalently, $\sigma E_*$ is of unitary type, and $\sigma(\mathcal C_{E_*})$ is unitary.
\end{conjecture}

The relations with the earlier statements are as follows.
\begin{itemize}[nosep]
\item \emph{Conjecture~\ref{conj:PU} $\Rightarrow$ $\mathrm{Z}(d)$.} One pseudo-unitary conjugate suffices (Theorem~\ref{thm:NGred}). Hence Conjecture~\ref{conj:PU} for one $\sigma$ in each of infinitely many dimensions $d$ implies Problem~1.
\item \emph{PL $\Leftrightarrow$ Conjecture~\ref{conj:PU} $+$ real type below.} By Corollary~\ref{cor:PLcat}, PL is Conjecture~\ref{conj:PU} together with the statement that the conjugates that are not pseudo-unitary are of real type. Thus PL implies Conjecture~\ref{conj:PU}.
\item \emph{Sign--type correlation.} For $\sigma$ fixing $K_0$, pseudo-unitarity is $\sigma(E_*(0))<0$ (Lemma~\ref{lem:ngdim}). Conjecture~\ref{conj:PU} is therefore the sign--type correlation of Sec.~\ref{sec:reality}, extended to all automorphisms, including those that move $\sqrt N$.
\item \emph{Under RU.} If every conjugate of $E_*$ is of real or unitary type (Conjecture~\ref{conj:RU}), Conjecture~\ref{conj:PU} is equivalent to: \emph{no pseudo-unitary Galois conjugate of $\mathcal C_{E_*}$ has data of real type.}
\end{itemize}
For the SIC metric groups, Conjecture~\ref{conj:PU} is therefore of Zauner strength. It is a reformulation of the archimedean input, not a reduction of it to something known. Its value is that it places the problem in the theory of unitary fusion categories and subfactors, where the available tools differ from those of Stark's conjecture.

\paragraph{A norm form of the Hermiticity condition.} The Hermiticity condition (I7) can be written as the equality case of an inequality, which is the natural target for a positivity argument.

\begin{proposition}[norm criterion]\label{prop:normcrit}
Let $E$ be a finite quantum dilogarithm on a metric group $(G,\langle\cdot\rangle)$ of order $N$ with $E(u)E(-u)=\langle u\rangle^{-1}$ for $u\ne0$; this holds for every Galois conjugate of $E_*$. Then
\[
\sum_{u\ne0}|E(u)|^2\ \ge\ N-1,
\]
with equality if and only if $E$ is of unitary type. In particular, for $\sigma$ with $\sigma(\sqrt D)<0$, the pseudo-unitary category $\sigma(\mathcal C_{E_*})$ is unitary if and only if $\sum_{u\ne0}|\sigma(E_*(u))|^2=N-1$.
\end{proposition}
\begin{proof}
The relation holds for $E_*$ because $x_ux_{-u}=1$ with $x_u=s_uE_*(u)$, $s_{-u}=s_u$ and $s_u^2=\langle u\rangle$, and it is preserved by Galois conjugation together with the metric. Since $|\langle u\rangle|=1$, it gives $|E(u)|\,|E(-u)|=1$. For $u\ne-u$, put $r=|E(u)|^2$; the pair $\{u,-u\}$ contributes $r+r^{-1}\ge2$, with equality if and only if $r=1$. For $u=-u\ne0$ it gives $|E(u)|=1$. Summing gives the inequality, with equality if and only if $|E(u)|=1$ for all $u\ne0$. If $|E(u)|=1$, then $\overline{E(u)}=E(u)^{-1}=\langle u\rangle E(-u)=\langle -u\rangle E(-u)$, which is unitary type; the converse was noted in Sec.~\ref{sec:GR}. The last statement follows from Theorem~\ref{thm:NGred} and Lemma~\ref{lem:izuherm}.
\end{proof}

Proposition~\ref{prop:normcrit} shows that a pseudo-unitary conjugate fails to be unitary only through an \emph{excess} of the quadratic quantity $\sum_{u\ne0}|E(u)|^2$ over its minimum. For pseudo-unitary categories, positivity of the categorical structure is automatic, and in the closely related Haagerup--Izumi setting the Cuntz-algebra realization supplies positivity but not Hermiticity (Ref.~\onlinecite{EG}, Thm.~2(c)). A proof of Conjecture~\ref{conj:PU} along these lines would need an identity, for instance a trace or Plancherel identity in the Cuntz-algebra model, that bounds $\sum_{u\ne0}|E(u)|^2$ above by $N-1$. We have not found such an identity, and none appears in Refs.~\onlinecite{Izu,EG,GSY}.

\paragraph{Staying inside one metric group.} The sign--type correlation and Theorem~\ref{thm:HSRUSIC} use an automorphism fixing $K_0=\Q(\zeta_{4N})$ and moving $\sqrt{N+4}$. The next result says exactly when such an automorphism exists.

\begin{proposition}\label{prop:samemetric}
For the SIC group $G_d$, one has $\sqrt{N+4}\in K_0$ if and only if $d+1$ is a square or twice a square. Otherwise there is $\sigma\in\Gal(\Qb/K_0)$ with $\sigma(\sqrt{N+4})=-\sqrt{N+4}$. Such $\sigma$ fixes the metric $\langle\cdot\rangle_\gamma$, satisfies $\sigma(\sqrt D)<0$, and maps $E_*$ to a solution $\sigma E_*\in Z_-(G_d)$ of the same system, whose category is pseudo-unitary.
\end{proposition}
\begin{proof}
Since $N+4=d^3-3d^2+4=(d-2)^2(d+1)$, we have $\sqrt{N+4}=(d-2)\sqrt{d+1}$, and $\sqrt{N+4}\in K_0$ if and only if $\sqrt m\in K_0$, where $m$ is the square-free part of $d+1$. A quadratic field lies in $\Q(\zeta_{4N})$ if and only if its discriminant divides $4N=4d^2(d-3)$. Let $p$ be an odd prime dividing $m$. Then $p\nmid d$, since $p\mid d+1$, and $p\nmid d-3$, since otherwise $p\mid(d+1)-(d-3)=4$. Hence $p\nmid 4N$, and $\sqrt m\notin K_0$. So $\sqrt m\in K_0$ forces $m\in\{1,2\}$. Conversely, $m=1$ is trivial, and if $m=2$ then $d$ is odd, $d-3$ is even, $8\mid 4N$, and $\sqrt2\in\Q(\zeta_8)\subset K_0$. In the remaining cases $\sqrt{N+4}\notin K_0$, and a $\sigma$ as stated exists. It fixes $\sqrt N\in K_0$ and moves $\sqrt{N+4}$, hence moves $\sqrt D$, and $\sigma(E_*(0))=(\sqrt N-\sqrt{N+4})/2<0$. The last assertion is Lemma~\ref{lem:ngdim}.
\end{proof}
Among $4\le d\le100$ the exceptional dimensions are $d=7$, $8$, $15$, $17$, $24$, $31$, $35$, $48$, $49$, $63$, $71$, $80$, $97$ and~$99$. For all other $d$, Conjecture~\ref{conj:PU} can be tested inside the single near-group system of $G_d$: \emph{the Galois conjugates of $E_*$ that lie in $Z_-(G_d)$ should be of unitary type.} In the exceptional dimensions, any $\sigma$ with $\sigma(\sqrt D)<0$ must move $\sqrt N$ or the roots of unity, and the pseudo-unitary conjugates belong to a Galois-conjugate metric datum; this is the situation described after Table~\ref{tab:fqd} in connection with Thornton's theorem.

\paragraph{Constraints on a failure.} The structure theorems of Sec.~\ref{sec:struct} restrict how Conjecture~\ref{conj:PU} could fail on the conjugates fixing $K_0$. Let $L=K_0(\sqrt{N+4})$, $O=\Gal(\Qb/L)\cdot E_*$, and $\sigma$ as in Proposition~\ref{prop:samemetric}.
\begin{itemize}[nosep]
\item If the pseudo-unitary conjugate $\sigma(\mathcal C_{E_*})$ satisfies (I7), then $h=(\sigma^{-1}c\sigma)c\in\Gal(\Qb/L)$ realizes the reflection, $h(E_*)=\rho E_*$ (Theorem~\ref{thm:Cprime}). In categorical terms, $h(\mathcal C_{E_*})\simeq\mathcal C_{E_*}$ through an equivalence acting on the invertible objects by $g\mapsto-g$ (Proposition~\ref{prop:cat}).
\item If the reflection is not realized on $O$, then none of the pseudo-unitary categories $\mathcal C_{E}$, $E\in\sigma(O)$, satisfies (I7), and the number of them with data of real type is congruent to $|O|$ modulo $2$ (Theorem~\ref{thm:Corbit}).
\end{itemize}
The control experiments of Sec.~\ref{sec:reality} add a methodological constraint. Square subsystems of the RW equations, and random subspaces with the same symmetries, have non-real and mixed solutions. Any mechanism forcing (I7), for example a positivity argument from a Cuntz-algebra realization, must therefore use the full overdetermined pentagon system.

\paragraph{Relation to unitarity questions for fusion categories.} Uniqueness of unitary structures on unitarizable fusion categories is known (Reutter~\cite{Reutter}). Existence of a unitary structure on a pseudo-unitary fusion category is open in general, and a proof of Conjecture~\ref{conj:PU} must use the near-group structure. The closest known case is the Haagerup--Izumi programme of Remark~\ref{rem:HI}. There the pseudo-unitary conjugates are unitary exactly when an explicit HI matrix is Hermitian (Ref.~\onlinecite{EG}, Thm.~2(c)), which has been verified for odd $N\le43$ (Ref.~\onlinecite{GSY}). Both settings use double-sine-type special values (Barnes double sine for HI, Faddeev's modular quantum dilogarithm here), and in both the unitarity of a conjugate is decided by the infinite place of a real quadratic field below the embedding. In the SIC setting, Corollary~\ref{cor:PLcat} identifies this rule with the archimedean rank-one Stark relation over $F$.

\paragraph{Evidence.} Every computed case agrees with Conjecture~\ref{conj:PU}:
\begin{itemize}[nosep]
\item in all $94$ hyperbolic classes with $5\le N\le30$, every solution in $Z_-$ is of unitary type, so every pseudo-unitary category in these systems satisfies (I7) (Sec.~\ref{sec:struct});
\item for the SIC group $d=4$, the same holds for all $16$ conjugate metrics (Appendix~\ref{app:d4});
\item for the $100$ exact trace polynomials ($39$ classes, $N\le18$) and for the SIC dimensions $d=5,9,13$, the place law holds at every embedding (Secs.~\ref{sec:PL} and~\ref{sec:CFT}), and hence, by Corollary~\ref{cor:PLcat}, every pseudo-unitary conjugate of $\mathcal C_{E_*}$ satisfies (I7).
\end{itemize}

\section{The reduction in full: what is proved, what is observed, and what remains}\label{sec:summary}

This section collects the logical structure of the paper in one place. It records every implication used, the status of each statement, and the computational evidence. It ends with a precise list of what must still be proved to solve Problem~1.

\subsection{Dependency diagram and chain of reductions}\label{sec:diagram}
The main line of the reduction is shown below. Each double arrow is a proved implication or equivalence, labelled on the left by the result that gives it. The side branches are listed after the diagram.
\[
\begin{array}{r@{\quad}c}
 & \text{Problem 1}\\
\text{for infinitely many }d & \Big\Uparrow\\
 & \mathrm{Z}(d)\\
\text{Thm.~\ref{thm:PLSIC}} & \Big\Uparrow\\
 & \mathrm{PL}\\
\text{Prop.~\ref{prop:CFT}, if }M/F\text{ is abelian} & \Big\Updownarrow\\
 & c_-=\kappa\\
\text{Artin reciprocity} & \Big\Updownarrow\\
 & x_{u_0}^{N\mathfrak p+1}\equiv1\ (\mathrm{mod}\ \mathfrak P)\ \text{for one }\mathfrak p\text{ in the class of }c_-
\end{array}
\]
Side branches:
\begin{itemize}[nosep]
\item into PL, bypassing abelianness and the congruence: $M'/F$ Galois, $\kappa\in\Gal(M'/F)$, $\infty_-$ ramified in $M'$, and a unique element of order $2$ in $\Gal(M'/F)$, for $M'=M$ or the sign-free field $M_2$ (Prop.~\ref{prop:uniqueinv}); if $M_2$ is the full ray class field mod $d'\infty_-$, the last condition means a cyclic $2$-Sylow subgroup of $\mathrm{Cl}(d'\infty_-)$; that last condition may instead be replaced by TR, which carries no restriction on the class group but is then equivalent to PL (Prop.~\ref{prop:TRinv}, Cor.~\ref{cor:collapse}), and the ramification hypothesis is then unconditional for $d=2m+1$ with $m$ even and $m^2-1$ squarefree (Prop.~\ref{prop:ramcrit}, Thms.~\ref{thm:ramgen} and~\ref{thm:RD});
\item into $\mathrm{Z}(d)$: one pseudo-unitary Galois conjugate of the near-group category $\mathcal C_{E_*}$ satisfies Izumi's Hermiticity condition (Thm.~\ref{thm:NGred}); $\mathrm{HS}+\mathrm{RU}$ (Thm.~\ref{thm:HSRUSIC}); the sign--type correlation (Lemma~\ref{lem:H}, Cor.~\ref{cor:unit}); $\mathrm{R2}+(\mathrm{G1})$ plus one unimodular conjugate of one unit (Thm.~\ref{thm:single});
\item out of PL: HS, TR, RU on the orbit of $E_*$, the sign--type correlation and GR (Sec.~\ref{sec:reality}); the parity of the unitary count (Cor.~\ref{cor:2adic}); Conjecture~\ref{conj:PU}, that pseudo-unitary conjugates of $\mathcal C_{E_*}$ are unitary (Cor.~\ref{cor:PLcat});
\item equivalent form of PL: a conjugate of $\mathcal C_{E_*}$ satisfies Izumi's Hermiticity condition exactly when it is pseudo-unitary, and has data of real type otherwise (Cor.~\ref{cor:PLcat}); a pseudo-unitary conjugate is unitary exactly when $\sum_{u\ne0}|\sigma(E_*(u))|^2=N-1$ (Prop.~\ref{prop:normcrit});
\item into $c_-=\kappa$ and $\mathrm{Z}(d)$, for odd $d$: the explicit reciprocity law, Conjecture~\ref{conj:RL}, at the single element $\varpi=-1-d\delta$ with $\phi(\varpi)=-I$ (Prop.~\ref{prop:RLPL}).
\end{itemize}

The archimedean rank-one Stark conjecture implies PL. Its algebraic part implies that $M/F$ is abelian. Computationally, abelianness of $M/F$ and the congruence at the bottom of the main line are confirmed for $d=4,5,9$, and $d=13$ passes the Frobenius test (Table~\ref{tab:CFT}). The explicit reciprocity law holds exactly at every prime tested for $d=5,9$ (Table~\ref{tab:RL}).

\paragraph{The chain of reductions (Steps 0--7).} We now spell out the diagram step by step, recording what is proved and what is assumed at each stage.

Throughout, $d\ge4$ is the SIC dimension. We write $E_*$ for the RW (Stark) special-value solution attached to the AFK admissible tuple of dimension $d$, with metric group $G_d$ of order $N=d^2(d-3)$ coming from $\gamma=\alpha^3$, $\operatorname{tr}\alpha=d-1$~\cite{AFK25,RW}. The parity operator $U_P$ (Section~\ref{sec:variety}) corresponds to the reflection $\rho:E\mapsto E(-\,\cdot)$ on finite quantum dilogarithms (Section~\ref{sec:GR}).

\medskip\noindent\textbf{Step 0 (the problem).} Problem~1 asks for WH-covariant SICs in infinitely many dimensions. It suffices to prove Zauner's existence statement $\mathrm{Z}(d)$ for every $d$ in some infinite set $D$.

\noindent\textbf{Step 1 (complexification; unconditional).} SIC fiducials are the fixed points of $\sigma=c\circ T$ on the complexified variety $V_d$ (Lemma~\ref{lem:basic}). Ghosts exist in every dimension $d\ge4$ (Theorem~\ref{thm:ghost}), and hence $V_d(\Qb)\ne\emptyset$, and $V_d$ has points modulo almost every prime (Corollary~\ref{cor:modp}).

\noindent\textbf{Step 2 (no geometric shortcut; proved).} No element of the extended Clifford group maps a ghost to a SIC (Lemma~\ref{lem:nogo}). Any passage from ghosts to SICs must therefore be arithmetic.

\noindent\textbf{Step 3 (Galois criterion; proved).} $\mathrm{Z}(d)$ holds if and only if some Galois orbit in $V_d(\Qb)$ contains a point $\Pi$ with $c'(\Pi)=\Pi^T$ for a conjugate $c'$ of complex conjugation (Proposition~\ref{prop:galois}, Lemma~\ref{lem:alg}). For a ghost, this means some product of two complex conjugations acts on it as a displaced parity (Corollary~\ref{cor:ghost}). The statements $|S_d|=|\Gamma_d|$ and ``ghost $\Rightarrow$ SIC'' are each equivalent to $\mathrm{Z}(d)$, given Theorem~\ref{thm:ghost} (Corollary~\ref{cor:equiv}). There is also a finite-field version at the primes whose Frobenius is conjugate to $c$ (Theorem~\ref{thm:ff}).

\noindent\textbf{Step 4 (two halves; proved).} The criterion of Step~3 splits into two parts:
\begin{enumerate}[label=(\alph*),nosep]
\item GR: transposition acts on the orbit through \emph{some} Galois element;
\item that element is a complex conjugation.
\end{enumerate}
By Proposition~\ref{prop:A}, GR on the orbit of a ghost is equivalent to the existence of a field automorphism inverting all ghost units $\varepsilon_p=\sqrt{d+1}\,\nu_p$ simultaneously (up to sign).

\noindent\textbf{Step 5 (translation to finite quantum dilogarithms; dictionary).} The ghost overlaps are built from the values $F_\gamma(u)$ (Ref.~\onlinecite{AFK26}, Prop.~2.1), and $p\mapsto -p$ corresponds to $u\mapsto -u$. GR becomes R2: some $h\in\Gal(\Qb/L)$ satisfies $h(E_*)=\rho E_*$. The relevant notions match as follows.
\begin{itemize}[nosep]
\item Real type, $\overline{E(u)}=\langle u\rangle E(u)$, is the ghost type. Unitary type, $\overline{E(u)}=\langle u\rangle E(-u)$, which forces $|E|=1$, is the SIC type.
\item The sign of $E(0)$ is opposite to the categorical dimension $d_\rho=-\sqrt N/E(0)$.
\item Unitary type together with $E(0)<0$ is exactly Izumi's data for a unitary near-group category (Ref.~\onlinecite{Izu}, Thm.~9.1). More precisely, $d_\rho>0$ is pseudo-unitarity, and Izumi's Hermiticity condition $Jb=b$ is unitary type (Lemmas~\ref{lem:ngdim} and~\ref{lem:izuherm}).
\item In categorical terms, R2 says the category $\mathcal C_{E_*}$ is Galois-stable through an equivalence acting by $-1$ on its invertible objects (Proposition~\ref{prop:cat}).
\end{itemize}

\noindent\textbf{Step 6 (structure theorems; proved).}
\begin{itemize}[nosep]
\item Theorem~\ref{thm:C}: if $Z_+$ is a single reflection pair, R2 can fail only if a non-even real solution with $E(0)<0$ exists.
\item Theorem~\ref{thm:Cprime}: if the $E(0)<0$ conjugate of $E_*$ is unitary, then R2 holds with an explicit Galois element. This is the dilogarithm form of $\mathrm{Z}\Rightarrow\mathrm{GR}$.
\item Theorem~\ref{thm:Corbit}: if R2 fails on an orbit $O$, then $\sigma(O)$ contains no unitary and no even solution, and the number of real solutions in $\sigma(O)$ is congruent to $|O|$ modulo $2$.
\item Proposition~\ref{prop:TRarith}: TR across all conjugate metric data is equivalent to total reality of the fields $\Q(t_u)$, where $t_u=x_u+x_u^{-1}$.
\item Proposition~\ref{prop:HSRU}: HS together with RU produces a unitary-type conjugate of $E_*$.
\item Lemma~\ref{lem:herm}, Corollary~\ref{cor:unit}, Theorem~\ref{thm:HSRUSIC}: a Galois conjugate of a ghost is a SIC exactly when all its overlap units are unimodular. Hence HS and RU give SICs (via Lemma~\ref{lem:H}).
\item Lemma~\ref{lem:dimsign}, Theorems~\ref{thm:PLtype} and~\ref{thm:PLSIC}: over the full Galois group, the type of a conjugate is governed by the sign of its categorical dimension, equivalently by the place of $F=\Q(\sqrt{N(N+4)})$ below it. The place law PL is equivalent to ``real above $\infty_+$, unitary above $\infty_-$'' and gives SICs without RU.
\item Corollary~\ref{cor:2adic}: when $N(x_{u_0}-1)$ is odd, the parity of the number of unimodular conjugates of $x_{u_0}$, and hence the hypothesis of Corollary~\ref{cor:integer}, is determined by $|N(x_{u_0}\pm1)|$ modulo $4$. This is a non-archimedean footprint of the archimedean obstruction.
\item Proposition~\ref{prop:CFT}: if the overlap-unit field $M$ is abelian over $F$, then PL is equivalent to $c_-=\kappa$. By Artin reciprocity this is a congruence $x_{u_0}^{N\mathfrak p+1}\equiv1\pmod{\mathfrak P}$ at a single prime in an explicit ray class. This is verified for $d=4,5,9$, and $d=13$ passes the Frobenius test (Table~\ref{tab:CFT}). If moreover $\Gal(M/F)$ has a unique involution, PL follows from the algebraic hypotheses alone (Proposition~\ref{prop:uniqueinv}).
\item Proposition~\ref{prop:TRinv} and Corollary~\ref{cor:collapse}: if $M'/F$ is Galois, $\kappa\in\Gal(M'/F)$ and $\infty_-$ ramifies in $M'$, then TR alone already gives $c_-=\kappa$, and hence TR, HS and PL are equivalent there, with no condition on $\mathrm{Cl}(d'\infty_-)$.
\item Proposition~\ref{prop:ramcrit}, Lemma~\ref{lem:zeta3} and Theorems~\ref{thm:ramgen}, \ref{thm:RD}: ramification at $\infty_-$ is a statement about units; it holds whenever $\mathcal O_F^\times=\langle-1,\varepsilon_\alpha\rangle$, because $\varepsilon_\alpha$ reduces to a cube root of unity modulo $d$, and it is unconditional on a set of dimensions of density $\tfrac14\prod_{p>2}(1-2p^{-2})\approx0.1613$.
\item Theorem~\ref{thm:single} and Corollaries~\ref{cor:singleR2}, \ref{cor:integer}: under the algebraic conditions (G1)--(G2), or under R2 and (G1), one unimodular conjugate of one generating unit $x_{u_0}$ gives a SIC. A sufficient condition is the integer inequality $P_{t_{u_0}}(2)P_{t_{u_0}}(-2)<0$.
\item Lemma~\ref{lem:ngdim}, Proposition~\ref{prop:izumi}, Theorem~\ref{thm:NGred} and Corollary~\ref{cor:PLcat}: the conjugate $\sigma(\mathcal C_{E_*})$ is pseudo-unitary exactly when $\sigma(\sqrt D)<0$. For it, Izumi's unitarity equations reduce to the Hermiticity condition, which is unitary type. If the condition holds for one such $\sigma$, a SIC exists, and PL says that it holds for all of them.
\item Proposition~\ref{prop:samemetric}: an automorphism fixing $K_0$ and moving $\sqrt{N+4}$ exists exactly when $d+1$ is neither a square nor twice a square.
\item Proposition~\ref{prop:normcrit}: Hermiticity of a pseudo-unitary conjugate is the equality case of $\sum_{u\ne0}|\sigma(E_*(u))|^2\ge N-1$.
\item Propositions~\ref{prop:RLwell} and~\ref{prop:RLPL}: for odd $d$, the explicit reciprocity law (Conjecture~\ref{conj:RL}), in which the Artin symbol of $\varpi\mathcal O_F$, $\varpi>0$ at $\infty_-$, acts on the squared ghost overlaps by $e_p\mapsto e_{\phi(\varpi)p}$, is well defined, and at $\varpi=-1-d\delta$ it gives $c_-=\kappa$ and a SIC. It is verified exactly for $d=5,9$ (Table~\ref{tab:RL}).
\end{itemize}

\noindent\textbf{Step 7 (the final decomposition).} For the SIC group $G_d$, with $\sqrt{N+4}\notin K_0$ (which holds unless $d+1$ is a square or twice a square; Proposition~\ref{prop:samemetric}):
\[
\boxed{\ \mathrm{HS}(E_*)\ +\ \mathrm{RU}(G_d)\ \Longrightarrow\ \text{unitary conjugate of }E_*\ \Longrightarrow\ \mathrm{Z}(d)\ }
\]
The first arrow is Proposition~\ref{prop:HSRU}. The second is Theorem~\ref{thm:HSRUSIC}, which rests on the dictionary Lemma~\ref{lem:H} and on Galois stability of $V_d$ (Lemma~\ref{lem:herm} and Corollary~\ref{cor:unit}). In particular, no transfer of the TCI or of the order-$(2m+1)$ symmetry to conjugate data is needed: the Galois conjugate of the ghost projector already lies in $V_d$, and only its Hermiticity has to be checked.

The direct route is
\[
\text{sign--type correlation for }G_d\ (E(0)<0\Rightarrow\text{unitary})\ \Longrightarrow\ \mathrm{Z}(d),
\]
because the $E(0)<0$ conjugate $\sigma E_*$ exists whenever $d+1$ is neither a square nor twice a square. Its categorical form, valid for every $d$, is
\[
\boxed{\ \sigma(\mathcal C_{E_*})\ \text{pseudo-unitary and satisfying Izumi's Hermiticity condition, for one }\sigma\ \Longrightarrow\ \mathrm{Z}(d)\ }
\]
(Theorem~\ref{thm:NGred}). The weaker GR route is
\[
\mathrm{TR}+\text{sign statement}+\text{Theorem~\ref{thm:Corbit}}\ \Longrightarrow\ \mathrm{GR}.
\]
Section~\ref{sec:PL} adds two sharper routes:
\[
\boxed{\ \mathrm{PL}(u)\ \forall u\ \Longrightarrow\ \sigma E_*\text{ unitary for every }\sigma\text{ with }\sigma(\sqrt D)<0\ \Longrightarrow\ \mathrm{Z}(d)\ }
\]
(Theorems~\ref{thm:PLtype}, \ref{thm:PLSIC}), and
\[
\boxed{\ \mathrm{R2}+(\mathrm{G1})+\big[\,x_{u_0}\text{ has a unimodular conjugate}\,\big]\ \Longrightarrow\ \mathrm{Z}(d)\ }
\]
(Theorem~\ref{thm:single}, Corollary~\ref{cor:singleR2}). If $M/F$ is abelian, the first route becomes
\[
\boxed{\ M/F\text{ abelian}\ +\ x_{u_0}^{N\mathfrak p+1}\equiv1\ (\mathrm{mod}\ \mathfrak P)\text{ for one }\mathfrak p\text{ in the class of }c_-\ \Longrightarrow\ \mathrm{PL}\ \Longrightarrow\ \mathrm{Z}(d)\ }
\]
(Proposition~\ref{prop:CFT}). For odd $d$ there is also the reciprocity route
\[
\boxed{\ \Big(\tfrac{M_\varepsilon/F}{\varpi\mathcal O_F}\Big)e_p=e_{\phi(\varpi)p}\ \text{at }\varpi=-1-d\delta\ \Longrightarrow\ c_-=\kappa\text{ on }M_\varepsilon\ \Longrightarrow\ \mathrm{Z}(d)\ }
\]
(Proposition~\ref{prop:RLPL}), in which the archimedean input becomes one instance of a uniform reciprocity law (Conjecture~\ref{conj:RL}). Here PL contains HS, RU on the orbit of $E_*$, and the sign--type correlation, and it implies R2 through Theorem~\ref{thm:Cprime}. The second route separates the problem into algebraic statements, R2 and (G1), and a single archimedean fact about one unit. That fact follows from $P_{t_{u_0}}(2)P_{t_{u_0}}(-2)<0$ (Corollary~\ref{cor:integer}).

\subsection{Status of all statements and computational evidence}
{\small
\begin{longtable}{@{}>{\raggedright\arraybackslash}p{5.6cm}>{\raggedright\arraybackslash}p{1.9cm}>{\raggedright\arraybackslash}p{6.6cm}@{}}
\caption{Status of the statements in the reduction, in three blocks: proved, conjectural, and false.}\label{tab:status}\\
\hline\hline
Statement & Status & Evidence or source\\\hline\endfirsthead
\hline\hline
Statement & Status & Evidence or source\\\hline\endhead
\hline\hline\endfoot
\rowcolor{green!20}
\multicolumn{3}{@{}c}{\emph{Proved}}\\
\hline
Ghosts exist for all $d\ge4$ & theorem & Refs.~\onlinecite{AFK26,RW}\\\hline
No Clifford map ghost $\to$ SIC & theorem & Lemma~\ref{lem:nogo}\\\hline
Galois criterion, finite-field criterion & theorem & Prop.~\ref{prop:galois}, Thm.~\ref{thm:ff}\\\hline
GR $\Leftrightarrow$ unit inversion is Galois & theorem & Prop.~\ref{prop:A}\\\hline
Theorems C, C$'$, orbit version & theorem & Thms.~\ref{thm:C}, \ref{thm:Cprime}, \ref{thm:Corbit}\\\hline
HS $+$ RU $\Rightarrow$ unitary conjugate & theorem & Prop.~\ref{prop:HSRU}\\\hline
Unit-circle criterion for SICs & theorem & Lemma~\ref{lem:herm}, Cor.~\ref{cor:unit}\\\hline
Dictionary: $\varepsilon_p=\zeta_pE_*(u_p)^{-1}$, $u_p\ne0$ & theorem (from Refs.~\onlinecite{AFK25,AFK26}) & Lemma~\ref{lem:H}; only structural facts (i)--(ii) of the following remark are used\\\hline
HS $+$ RU $\Rightarrow$ SIC & theorem & Thm.~\ref{thm:HSRUSIC}\\\hline
Sign of $d_\rho$ governs type; $E(0)$ only on $\Gal(\Qb/K_0)$ & theorem & Lemma~\ref{lem:dimsign}\\\hline
PL $\Leftrightarrow$ real above $\infty_+$, unitary above $\infty_-$ & theorem & Prop.~\ref{prop:PLforms}, Thm.~\ref{thm:PLtype}\\\hline
PL $\Rightarrow$ SIC (no RU needed) & theorem & Thm.~\ref{thm:PLSIC}\\\hline
One generating unit suffices (under (G1)--(G2) or R2+(G1)) & theorem & Thm.~\ref{thm:single}, Cors.~\ref{cor:singleR2}, \ref{cor:integer}\\\hline
Parity of unitary count from $|N(x\pm1)|$ when $N(x-1)$ odd & theorem & Cor.~\ref{cor:2adic}\\\hline
That parity equals $[M:\Q]/4$ for every generating unit (under PL) & theorem & Prop.~\ref{prop:parityinv}\\\hline
If $M/F$ abelian: PL $\Leftrightarrow$ $c_-=\kappa$ $\Leftrightarrow$ one Frobenius congruence & theorem & Prop.~\ref{prop:CFT}\\\hline
$M'/F$ Galois, $\kappa\in\Gal(M'/F)$, $\infty_-$ ramified, unique involution $\Rightarrow$ PL, $\mathrm{Z}(d)$ & theorem & Prop.~\ref{prop:uniqueinv}\\\hline
Same with TR in place of the unique involution; hence TR $\Leftrightarrow$ HS $\Leftrightarrow$ PL & theorem & Prop.~\ref{prop:TRinv}, Cor.~\ref{cor:collapse}\\\hline
$\infty_-$ ramifies in $\mathrm{Cl}(d'\infty_-)$ $\Leftrightarrow$ no unit $\equiv1\ (d')$ is negative at $\infty_-$ & theorem & Prop.~\ref{prop:ramcrit}\\\hline
$\varepsilon_\alpha$ is a cube root of unity mod $d$; $\mathcal O_F^\times=\langle-1,\varepsilon_\alpha\rangle\Rightarrow\infty_-$ ramifies, every $d\ge4$ & theorem & Lemma~\ref{lem:zeta3}, Thm.~\ref{thm:ramgen}\\\hline
That holds unconditionally for $d=2m+1$, $m$ even, $m^2-1$ squarefree, density $\approx0.1613$ & theorem & Thm.~\ref{thm:RD}\\\hline
$d_\rho(\sigma E_*)=(N-\sigma(\sqrt D))/2$; $\sigma(\mathcal C_{E_*})$ pseudo-unitary $\Leftrightarrow$ $\sigma(\sqrt D)<0$ & theorem & Lemma~\ref{lem:ngdim}\\\hline
Izumi's Hermiticity condition $Jb=b$ $\Leftrightarrow$ unitary type & theorem & Lemma~\ref{lem:izuherm} (convention checked on Ref.~\onlinecite{Izu}, Ex.~9.3)\\\hline
Pseudo-unitary $\mathcal C_E$: only Hermiticity can fail; Hermitian $\Rightarrow$ unitary & theorem & Prop.~\ref{prop:izumi}\\\hline
One pseudo-unitary conjugate of $\mathcal C_{E_*}$ Hermitian $\Rightarrow$ $\mathrm{Z}(d)$ & theorem & Thm.~\ref{thm:NGred}\\\hline
PL $\Leftrightarrow$ (Hermitian $\Leftrightarrow$ pseudo-unitary) for all conjugates of $\mathcal C_{E_*}$ & theorem & Cor.~\ref{cor:PLcat}\\\hline
$\sqrt{N+4}\in K_0$ $\Leftrightarrow$ $d+1$ a square or twice a square & theorem & Prop.~\ref{prop:samemetric}\\\hline
Pseudo-unitary conjugate unitary $\Leftrightarrow$ $\sum_{u\ne0}|\sigma E_*(u)|^2=N-1$ & theorem & Prop.~\ref{prop:normcrit}\\\hline
Reciprocity law well defined; at $\varpi=-1-d\delta$ it gives $c_-=\kappa$ and $\mathrm{Z}(d)$ ($d$ odd) & theorem & Props.~\ref{prop:RLwell}, \ref{prop:RLPL}\\\hline
Reciprocity law: one prime per ray class suffices ($h_F=1$, $M_\varepsilon/F$ abelian of conductor $\mid d\infty_-$) & theorem & Lemma~\ref{lem:RLcert}\\\hline
$d=p^k$, $p\equiv5\ (6)$, or $3^k$: under (i)--(iii), $c_-=\kappa$ $\Leftrightarrow$ $\kappa$ trivial on $M_\varepsilon\cap H_F$; automatic if $h_F$ odd & theorem & Thm.~\ref{thm:inert}, Cor.~\ref{cor:inertodd}\\\hline
Same with MH $\Rightarrow$ $\mathrm{Z}(d)$ on the infinite family $\mathcal P$ & theorem & Cor.~\ref{cor:inertMH}\\\hline
\hline
\rowcolor{gray!20}
\multicolumn{3}{@{}c}{\emph{Conjectural}}\\
\hline
Completeness $V_d=S_d\sqcup\Gamma_d$ & conjecture & $d=4,5,6$: complete counts; $d=4,5$: ghost test\\\hline
R2 / GR & conjecture & PSLQ certification for $N=6$--$9$ and for $d=4$ ($N=16$)\\\hline
RU (no mixed type) & conjecture & 94 $SL_2$ classes ($N\le30$); 268 cyclic metrics ($N\le21$); 0 exceptions\\\hline
TR & conjecture & same data; exact minimal polynomials for $N=6,7,9$\\\hline
HS & conjecture & exact minimal polynomials ($N=6,7,9$); $d=4$: all 16 conjugate metrics, 48/48 split each\\\hline
Sign statement & conjecture & no non-even real $E(0)<0$ solution in any computed case\\\hline
Sign--type correlation & conjecture & holds in every computed case; implies $\mathrm{Z}(d)$\\\hline
Pseudo-unitary conjugates of $\mathcal C_{E_*}$ are unitary (Conj.~\ref{conj:PU}) & conjecture & implied by PL; holds in every computed case ($94$ classes, $d=4$ over $16$ metrics, $d=5,9,13$)\\\hline
Cyclic $2$-Sylow of $\mathrm{Cl}(d'\infty_-)$ for infinitely many $d$ & conjecture & $24$ dimensions with $d\le2000$, thinning out; expected density zero (Sec.~\ref{sec:CFT})\\\hline
$M/F$ abelian, ramified at $\infty_-$ only & conjecture & $d=4,5$: $M$ is the full ray class field mod $(2d)\infty_-$, resp.\ $(4d)\infty_-$; conductor $d'\infty_-$ up to signs; $d=9$ (exact), $d=13$ (Frobenius test): $M_2$ is the full ray class field mod $d'\infty_-$\\\hline
PL (place law, Conj.~\ref{conj:PL}) & conjecture & 100 exact minimal polynomials, 39 classes ($N\le18$, incl.\ $d=4$), and $d=5$ ($N=50$); no exception\\\hline
(G1), (G2) & conjecture & exact checks in 18 classes ($N\le12$, and $d=4$): a generating $u_0$ always exists, and (G2) always holds\\\hline
Stark (archimedean) & conjecture & implies PL, hence HS, TR and the sign--type correlation\\\hline
Explicit reciprocity law (Conj.~\ref{conj:RL}) & conjecture & exact at every prime tested: $d=5$, $594+219$; $d=9$, $122+276$; proved for the computed overlaps in $d=5,7,11,15$ (Table~\ref{tab:RL}, Lemma~\ref{lem:RLcert}); all alternative normalizations fail\\\hline
Multiplet hypothesis (Conj.~\ref{conj:MH}) & conjecture & exact in $d=9$, the only computed dimension with $h_F>1$; it is vacuous when $h_F=1$\\\hline
\hline
\rowcolor{red!20}
\multicolumn{3}{@{}c}{\emph{False}}\\\hline
\hline
``$|Z_+|=2$ always'' & false & $|Z_+|\in\{2,4,6,8,10,12\}$ occurs\\
\end{longtable}}

\paragraph{Consolidated computational data.} Table~\ref{tab:data} collects the computations behind the statements in Table~\ref{tab:status}.
{\small
\begin{longtable}{@{}>{\raggedright\arraybackslash}p{5.2cm}>{\raggedright\arraybackslash}p{9cm}@{}}
\caption{Summary of computational evidence.}\label{tab:data}\\
\hline\hline
Computation & Result\\\hline\endfirsthead
\hline\hline
Computation & Result\\\hline\endhead
\hline\hline\endfoot
All points of $V_d$ (homotopy) & $d=4$: $512=256+256$; $d=5$: $4000=2000+2000$; $d=6$: $6912=3456+3456$\\\hline\hline
Ghost type of non-Hermitian points & $d=4$: 256/256; $d=5$: 2000/2000 are displaced-parity Hermitian\\\hline\hline
Non-cyclic abelian Heisenberg SICs & none for $16\le d\le36$ (strong for $d\le20$)\\\hline\hline
All finite quantum dilogarithms, 94 $SL_2$ classes, $5\le N\le30$ & distribution of $|Z_+|$ over classes: $1$ (even, $N=5$): 1; $2$: 20; $4$: 54; $6$: 6; $8$: 9; $10$: 3; $12$: 1. In all classes, $Z_+$ is real and $Z_-$ is unitary; $Z_-=\emptyset$ only when $\sqrt{N+4}\in K_0$\\\hline\hline
Simplified system (eigenspace + pairing), 268 cyclic metrics & every solution real or unimodular, none mixed; identical to the full system except $N=9$ $(2,2,5)$ and $N=16$ $(3,3,3)$ (16 vs 8 solutions, extras also totally real)\\\hline\hline
Control A (random subspace, square) & non-real solutions common (e.g.\ $16$ of $64$ pure at $N=13$)\\\hline\hline
Control B (Weil eigenspace, partial pairing) & non-real and mixed solutions appear (e.g.\ $(16,8,0)$ at $N=17$)\\\hline\hline
R2 by PSLQ & $N=6,7,8,9$ (cyclic) and $d=4$ ($N=16$): reflection is realized by $\Gal(\Qb/L)$\\\hline\hline
Minimal polynomials of $t_u$ & $N=6$: degrees 4, 8; $N=7$: 12; $N=9$: 6, 2. All roots real; exactly half outside $[-2,2]$; all monic (algebraic integers)\\\hline\hline
$d=4$, all 16 Galois-conjugate metrics & each: $Z_+$ = 4 real, $Z_-$ = 4 unitary; $t$-values: 48 with $|t|>2$, 48 with $|t|<2$, none non-real\\\hline\hline
Place law, 39 classes $5\le N\le18$ & 100 distinct minimal polynomials (degree $\le32$): all totally real, $\sqrt{N(N+4)}\in\Q(t_u)$ always, split by the place of $F$ in every case; $\Q(\sqrt{N+4})$, $\Q(\sqrt N)$ do not predict the split\\\hline\hline
Single-unit conditions (G1), (G2) & exact checks in 18 classes ($N\le12$, $d=4$): a generating $u_0$ exists and (G2) holds in all; the integer criterion certifies a unitary conjugate in 4 classes; $P(2)P(-2)<0$ exactly when $m_u$ is odd (13 of 100)\\\hline\hline
$d=5$ via a $U_P$-ghost at 400 digits & one trace polynomial of degree 16; PL at all 16 embeddings; $P(0)=61^2$ odd; $m=8$, and hence the parity criteria do not apply\\\hline\hline
Class-field test, $d=4,5$ & $M/F$ abelian, conductor $(2d)\infty_-$ resp.\ $(4d)\infty_-$, and $d'\infty_-$ up to signs; Frobenius $=\kappa$ exactly on the class of $c_-$ at all $1305$ primes $N\mathfrak p\le5000$\\\hline\hline
Class-field test, $d=9,13$ & $M_2$ is the full ray class field mod $(d)\infty_-$ (exact for $d=9$; for $d=13$ by Frobenius orders at $2230$ primes $N\mathfrak p\le20000$); $M_2^\kappa$ has conductor $(d)$; PL sign law on all $36$, resp.\ $96$, conjugates; Frobenius $=\kappa$ at all $15$ tested primes in the class of $c_-$ ($d=13$)\\\hline
Explicit reciprocity law, $d=5,9$ & exact Frobenius action $e_p\mapsto e_{\phi(\varpi)p}$ at $594/594$ split and $219/219$ inert primes ($d=5$, $q\le3000$) and at $122/122$ principal split and $276/276$ inert primes ($d=9$, $q\le4000$); for $d=9$ all $136$ non-principal split primes move the ghost to its Galois partner\\\hline
Unique-involution census, $4\le d\le2000$ & $2$-Sylow of $\mathrm{Cl}(d'\infty_-)$ cyclic with $\infty_-$ ramified for exactly $24$ dimensions ($15$ with $d\le250$, then $3,0,3,1,0,2,0$ in blocks of $250$)\\\hline
Pseudo-unitary conjugates (Conj.~\ref{conj:PU}) & every pseudo-unitary category in the $94$ complete systems ($N\le30$) satisfies Izumi's Hermiticity condition; by PL, the same for all conjugates of $\mathcal C_{E_*}$ in the $100$ exact cases and for $d=5,9,13$\\
\end{longtable}}

\paragraph{The first SIC-relevant case.} The row of Table~\ref{tab:data} on the $16$ Galois-conjugate metrics of $d=4$ deserves emphasis. The cyclic examples are only models. The metric group of $d=4$ is the one that actually produces SICs through the AFK dictionary. For it, every one of the $16$ Galois-conjugate metric data satisfies HS, RU and the sign--type correlation \emph{simultaneously}: $Z_+$ is $4$ real solutions, $Z_-$ is $4$ unitary solutions, and the traces split $48/48$ with none non-real. This is consistent with the known existence of SICs in $d=4$, and it shows that the decomposition of Step~7 is realized there exactly as predicted.

\subsection{What remains to prove Problem~1}\label{sec:remains}
This subsection is self-contained. It recalls the notation needed to state the open problems precisely; it can therefore be read independently of the rest of the paper.

\paragraph{The problem.} Fix a dimension $d\ge4$. Let $X e_j=e_{j+1}$ and $Z e_j=\omega^je_j$ on $\C^d$, with $\omega=e^{2\pi i/d}$ and indices modulo $d$. The WH displacement operators are $D_p=\tau^{p_1p_2}X^{p_1}Z^{p_2}$ for $p=(p_1,p_2)\in\Z_d^2$, where $\tau^2=\omega$. A \emph{WH-covariant SIC} is a unit vector $\psi\in\C^d$ with
\[
|\langle\psi,D_p\psi\rangle|^2=\frac{1}{d+1}\qquad\text{for all }p\ne0 .
\]
Write $\mathrm{Z}(d)$ for the statement that such a $\psi$ exists. Problem~1 asks for SICs in infinitely many dimensions, and hence it suffices to prove $\mathrm{Z}(d)$ for infinitely many $d$.

\paragraph{Ghosts and their overlap units.} A \emph{ghost} is a rank-one matrix $\tilde\Pi$ with $\tilde\Pi^2=\tilde\Pi$, $\tr\tilde\Pi=1$, satisfying the SIC equations $(d+1)\tr(\tilde\Pi D_p)\tr(\tilde\Pi D_p^\dagger)=1$ for $p\ne0$, which is Hermitian only up to parity: $\tilde\Pi^\dagger=U_P\tilde\Pi U_P$, where $U_Pe_j=e_{-j}$. Ghosts with algebraic entries exist for every $d\ge4$ (Theorem~\ref{thm:ghost}). Their normalized overlaps $\varepsilon_p=\sqrt{d+1}\,\tr(\tilde\Pi D_p)$ are real and satisfy $\varepsilon_{-p}=\varepsilon_p^{-1}$. The basic sufficient condition for a SIC is:
\begin{quote}
\emph{if some field automorphism $\sigma$ of $\Qb$ satisfies $|\sigma(\varepsilon_p)|=1$ for all $p\ne0$, then $\sigma(\tilde\Pi)$ is a SIC projector} (Corollary~\ref{cor:unit}).
\end{quote}

\paragraph{The arithmetic data.} Let $\alpha\in SL_2(\Z)$ have trace $d-1$, and put $\gamma=\alpha^3$. Then
\[
N=\operatorname{tr}\gamma-2=d^2(d-3),\qquad D=N(N+4)=(\operatorname{tr}\gamma)^2-4 .
\]
The field $F=\Q(\sqrt D)$ is the real quadratic field generated by the fixed point of $\gamma$. It has two embeddings into $\mathbb R$, the \emph{infinite places} $\infty_+:\sqrt D\mapsto+\sqrt D$ and $\infty_-:\sqrt D\mapsto-\sqrt D$. Every $\sigma\in\Gal(\Qb/\Q)$ restricts to one of them, and we say that $\sigma$ \emph{lies above} $\infty_+$ or $\infty_-$ accordingly. Exactly half of all $\sigma$ lie above each.

On the finite abelian group $G_d=\Z^2/(\gamma-I)\Z^2$ of order $N$, RW construct an explicit function $E_*:G_d\to\C$, the Stark solution of the finite quantum dilogarithm equations, whose values are special values of the modular quantum dilogarithm and are algebraic. From it we form the \emph{overlap units}
\[
x_u=s_u\,E_*(u),\qquad t_u=x_u+x_u^{-1}\qquad(u\in G_d,\ 2u\ne0),
\]
where $s_u$ is a fixed root of unity. These are real algebraic units with $x_{-u}=x_u^{-1}$. By the dictionary of Lemma~\ref{lem:H}, every ghost overlap $\varepsilon_p$ equals a root of unity times some $x_u^{\pm1}$. Thus, $|\sigma(\varepsilon_p)|=1$ for all $p$ exactly when $|\sigma(x_u)|=1$ for all $u$. Let $M=\Q(x_u:u\in G_d)$ be the field generated by the overlap units, and let $\kappa$ denote inversion, $\kappa(x_u)=x_u^{-1}$. For a prime ideal $\mathfrak q$, $N\mathfrak q$ denotes its absolute norm, the size of its residue field. For $y\in\Qb$, $N(y)$ denotes the product of all Galois conjugates of $y$, a rational number.

\paragraph{The open problems.} Problem~1 follows if, for infinitely many $d$, any one of the following routes is carried out.
\begin{enumerate}[label=\textbf{Route \Alph*.},leftmargin=*]
\item \textbf{The place law} (Conjecture~\ref{conj:PL}). \emph{For every $u\in G_d$ with $2u\ne0$ and every $\sigma\in\Gal(\Qb/\Q)$, the number $\sigma(x_u)$ is real if $\sigma$ lies above $\infty_+$, and has absolute value $1$ if $\sigma$ lies above $\infty_-$.}

Equivalently, every conjugate $\sigma(t_u)$ is real, with $|\sigma(t_u)|>2$ above $\infty_+$ and $|\sigma(t_u)|<2$ above $\infty_-$. Equivalently again, the single algebraic number $(t_u^2-4)\sqrt D$ is \emph{totally positive}: all of its conjugates are positive real numbers. This suffices because any $\sigma$ above $\infty_-$ makes every $|\sigma(x_u)|=1$ at once, and hence every $|\sigma(\varepsilon_p)|=1$ (Theorem~\ref{thm:PLSIC}). The place law is a statement about one algebraic number at a time, and it implies all the other conditions studied here (HS, TR, RU and the sign--type correlation).

\item \textbf{The sign--type correlation.} \emph{For one automorphism $\sigma$ that fixes the cyclotomic field $K_0=\Q(\zeta_{4N})$, which contains $\sqrt N$ and all the roots of unity $s_u$, and sends $\sqrt{N+4}$ to $-\sqrt{N+4}$, one has $|\sigma(x_u)|=1$ for all $u$.}

Such a $\sigma$ exists exactly when $d+1$ is neither a square nor twice a square (Proposition~\ref{prop:samemetric}). It lies above $\infty_-$, and hence this is the special case of Route~A for one particular $\sigma$. It suffices by Corollary~\ref{cor:unit} and Lemma~\ref{lem:H}.

\item \textbf{The class-field form} (Proposition~\ref{prop:CFT}). This route has two parts.
\begin{enumerate}[label=(\roman*),nosep]
\item \emph{The field $M$ is an abelian Galois extension of $F$.} Let $\mathfrak f$ be the finite part of its conductor.
\item \emph{One congruence holds.} Choose $\beta\in\mathcal O_F$ with $\beta\equiv1\pmod{\mathfrak f}$, $\beta>0$ at $\infty_+$ and $\beta<0$ at $\infty_-$, such that $\mathfrak p=(\beta)$ is a prime ideal not dividing $\mathfrak f$. For some prime ideal $\mathfrak P$ of $M$ above $\mathfrak p$ and some $u_0$ for which $x_{u_0}$ generates $M$,
\[
x_{u_0}^{\,N\mathfrak p+1}\equiv1\pmod{\mathfrak P}.
\]
\end{enumerate}
Part (i) is the algebraic half of Stark's conjecture. Under (i), part (ii) is equivalent to Route~A: by Artin reciprocity, the Frobenius element of $\mathfrak p$ is the complex conjugation $c_-$ at $\infty_-$, and (ii) says that $c_-$ acts as inversion. This is the identity $c_-=\kappa$, the classical Stark relation $\varepsilon^{1+c_-}=1$. For $d=4$ and $d=5$ both parts hold, and $M$ is the ray class field of $F$ of conductor $d'\infty_-$ up to signs, with $d'=d$ for $d$ odd and $d'=2d$ for $d$ even. For $d=9$ both parts hold for the sign-free unit field $M_2$, and $d=13$ passes the same Frobenius test (Table~\ref{tab:CFT}). When $\Gal(M_2/F)$ has a unique element of order $2$, as for $d=5,9$ (and, if $M_2$ is the full ray class field, for $d=7,11,15,17,19,27,\dots$), part (ii) follows from part (i) together with R2, with no congruence to check (Proposition~\ref{prop:uniqueinv}). More generally, part~(ii) follows from part~(i), R2 and TR alone, with no condition on $\mathrm{Cl}(d'\infty_-)$ (Proposition~\ref{prop:TRinv}); the only remaining input is that $\infty_-$ ramify in $M_2$, which is unconditional whenever $d=2m+1$ with $m$ even and $m^2-1$ squarefree (Theorem~\ref{thm:RD}), a set of density $\approx0.1613$ containing $d=5,9,13$. Read this way, Problem~1 would follow from the algebraic hypotheses (i) and R2 together with TR for the dimensions in that family (Corollary~\ref{cor:TRfinal}); since TR is then equivalent to PL, this is a reformulation of Route~A on that family, not a reduction of it.

\item \textbf{Norms of $x_u\pm1$} (Corollaries~\ref{cor:integer}, \ref{cor:2adic}). Suppose that for some $u_0$:
\begin{enumerate}[label=(\roman*),nosep]
\item[(G1)] $x_{u_0}$ generates $M$;
\item[(G2)] every field embedding sending $x_{u_0}\mapsto x_{u_0}^{-1}$ sends each $x_v\mapsto x_v^{-1}$ (this follows from R2).
\end{enumerate}
Then a SIC exists as soon as one conjugate of $x_{u_0}$ lies on the unit circle. This holds in particular if
\[
N(x_{u_0}-1)\ \text{is odd}\qquad\text{and}\qquad |N(x_{u_0}-1)|\cdot|N(x_{u_0}+1)|\equiv3\pmod 4 .
\]
The route therefore needs only the prime factorizations of the ideals $(x_{u_0}\pm1)$, for example from a Gross--Stark-type formula. However, it detects only the parity of the number of unimodular conjugates. By Proposition~\ref{prop:parityinv}, it can succeed only when $[M:\Q]\equiv4\pmod8$, and it fails for $d=4$ and $d=5$.

\item \textbf{Unitarity of a pseudo-unitary near-group category} (Sec.~\ref{sec:NG}, Conjecture~\ref{conj:PU}). The values $E_*(u)$ define a near-group fusion category $\mathcal C_{E_*}$ with simple objects $g\in G_d$ and $\rho$, and fusion rule $\rho\otimes\rho\cong\bigoplus_{g}g\oplus N\rho$ (Ref.~\onlinecite{RW}, Thm.~12). The categorical dimension of $\rho$ is $d_\rho=-\sqrt N/E_*(0)=(N-\sqrt D)/2<0$, and hence $\mathcal C_{E_*}$ is not pseudo-unitary. For $\sigma$ above $\infty_-$, the conjugate $\sigma(\mathcal C_{E_*})$ has $d_\rho=(N+\sqrt D)/2>0$ and is pseudo-unitary. Izumi's equations for a $C^*$ near-group category (Ref.~\onlinecite{Izu}, Thm.~9.1) then hold for its data, except possibly the single Hermiticity condition
\[
\overline{\sigma(E_*(u))}=\sigma\langle -u\rangle\;\sigma(E_*(-u))\qquad\text{for all }u .
\]
Here $\langle\cdot\rangle$ is the quadratic form on $G_d$ of Ref.~\onlinecite{RW} (Eq.~(3)), whose values are roots of unity, and $s_u^2=\langle u\rangle$.
\emph{Show that this condition holds for one $\sigma$ above $\infty_-$, that is, that one pseudo-unitary Galois conjugate of $\mathcal C_{E_*}$ is unitary.} It gives $|\sigma(x_u)|=1$ for all $u\ne0$, and hence a SIC (Theorem~\ref{thm:NGred}). The place law is the statement that it holds for every $\sigma$ above $\infty_-$ (Corollary~\ref{cor:PLcat}). This is the near-group analogue of the open Haagerup--Izumi unitarity question (Remark~\ref{rem:HI}). Equivalently (Proposition~\ref{prop:normcrit}), show that $\sum_{u\ne0}|\sigma(E_*(u))|^2=N-1$, its smallest possible value.

\item \textbf{An explicit reciprocity law} (Sec.~\ref{sec:RL}, Conjecture~\ref{conj:RL}). Let $d$ be odd, and let $\tilde\Pi$ be a ghost with Zauner symmetry: the squared overlaps $e_p=\varepsilon_p^2$ satisfy $e_{\Lambda p}=e_p$ for some $\Lambda\in SL_2(\Z_d)$ of order $3$ with $\operatorname{tr}\Lambda\equiv d-1$. Put $\epsilon_\alpha=\frac12(d-1+\sqrt{(d-3)(d+1)})$, the eigenvalue of $\alpha$ that exceeds $1$ at $\infty_+$, and let $\phi:\Z[\epsilon_\alpha]\to M_2(\Z_d)$ be the ring homomorphism with $\phi(\epsilon_\alpha)=\Lambda^{-1}$. Let $M_\varepsilon=\Q(e_p:p\ne0)$, and assume it is abelian over $F$ with conductor dividing $(d)\infty_-$, as computed for $d=5,9$.
\emph{Show that for $\varpi=-1-d\sqrt{(d-3)(d+1)}$, which is positive at $\infty_-$ and satisfies $\phi(\varpi)=-I$, the Artin symbol of the ideal $\varpi\mathcal O_F$ in $\Gal(M_\varepsilon/F)$ sends $e_p$ to $e_{-p}=e_p^{-1}$ for every $p$.} By Artin reciprocity this Artin symbol is the complex conjugation $c_-$ above $\infty_-$, and hence every $\sigma$ above $\infty_-$ gives $|\sigma(\varepsilon_p)|=1$, and therefore a SIC (Proposition~\ref{prop:RLPL}). The statement is the instance at one element of the conjectured law $\big(\frac{M_\varepsilon/F}{\varpi\mathcal O_F}\big)e_p=e_{\phi(\varpi)p}$ for all $\varpi$ prime to $d$ and positive at $\infty_-$, which holds exactly at every prime tested in $d=5,9$. For $\zeta(0)$-invariants the analogous law is a theorem for ideles positive at both infinite places (Coates--Sinnott, Deligne--Ribet; Arakawa~\cite{Arakawa}). What is needed here is its analogue for the derivative values, including the sign at $\infty_-$.

\item \textbf{Prime powers and the class group} (Sec.~\ref{sec:inert}, Theorem~\ref{thm:inert}). Let $d=p^k$ with $p\equiv5\pmod6$ prime, or $d=3^k$ with $k\ge2$, and let $H_F$ be the Hilbert class field of $F$.
\emph{Show that the field $M_\varepsilon$ of squared ghost overlaps is abelian over $F$ with conductor dividing $d\infty_-$ and ramified at $\infty_-$, that inversion is an automorphism of $M_\varepsilon$, and that inversion acts trivially on $M_\varepsilon\cap H_F$.} The last condition follows from the multiplet hypothesis (Conjecture~\ref{conj:MH}): an automorphism that maps the squared overlaps of the ghost onto themselves acts trivially on the class-group part. It is automatic when $h_F$ is odd. None of these statements refers to the infinite places except through ramification at $\infty_-$. Ramification is condition (C$_d$), which holds for every $d\le20000$ and is proved when $\mathcal O_F^\times=\langle-1,\varepsilon_\alpha\rangle$. The family $\mathcal P$ of such $d$ is infinite.

\end{enumerate}

\paragraph{How the routes are related.} Routes B--F are not independent of Route~A. Route~B is the place law for one automorphism fixing $K_0$. Route~C is the place law itself in class-field form, under the algebraic part of Stark. Route~D captures one bit of it. Route~E is the place law for one automorphism above $\infty_-$, stated as the unitarity of a pseudo-unitary fusion category; it contains Route~B, and unlike Route~B it is available in every dimension. Route~F is the half of the place law above $\infty_-$, for the squared ghost overlaps, obtained from one instance of a uniform reciprocity law; it is the route closest in form to proved results. Route~G is Route~C on the family $\mathcal P$, with the congruence replaced by an unramified condition that group theory shows to be equivalent to it. All seven are manifestations of one phenomenon: the archimedean part of the rank-one real-quadratic Stark conjecture for the overlap units. Route~E connects it with the open HI unitarity question (Ref.~\onlinecite{GSY}, Rem.~3.8), and hence with the methods of subfactor theory and Cuntz-algebra realizations. No step of the present work carries out any route for an infinite family of dimensions. The computations confirm each route in every case examined.

Two technical points also need care in a complete proof.
\begin{itemize}[nosep]
\item The passage from a unitary-type solution to a SIC uses the dictionary Lemma~\ref{lem:H}. Its proof assembles statements of Refs.~\onlinecite{AFK25,AFK26}, and the convention matching should be checked against those papers. The rest of that passage is proved here (Lemma~\ref{lem:herm}, Corollary~\ref{cor:unit}).
\item The complete solution counts rest on numerical homotopy continuation. They are evidence, not proofs. The exact minimal-polynomial, PSLQ and class-field certifications are much more reliable, but they still rest on numerically identified algebraic numbers.
\end{itemize}

\section{Conclusion}
We proved that ghost SICs, which exist unconditionally in every dimension $d\ge4$, cannot be turned into SICs by any symmetry of the SIC equations, and hence the step from ghosts to SICs is purely arithmetic. We gave Galois and finite-field criteria for that step, and reduced it to Galois inversion of the ghost overlap units. Through the RW finite quantum dilogarithms, we proved structure theorems for the Galois realization of the reflection. We showed that a single place law for the overlap units, real above one infinite place of $F=\Q(\sqrt{N(N+4)})$ and unimodular above the other, implies the existence of a WH-covariant SIC. If the unit field is abelian over $F$, the place law is equivalent to the identity $c_-=\kappa$, and hence to one Frobenius congruence. Equivalently, in the language of fusion categories, the place law says that a Galois conjugate of the Radchenko--Wheeler near-group category of the Stark solution is unitary exactly when it is pseudo-unitary. For a pseudo-unitary conjugate, Izumi's classification leaves a single Hermiticity condition to be proved, and one such conjugate satisfying it already gives a SIC.

What remains is to prove the place law, or one of its equivalent or sufficient forms (Sec.~\ref{sec:remains}), for infinitely many $d$. The place law and the Frobenius congruence hold in every case we computed, including the SIC dimensions $d=4,5,9,13$. There the (sign-free) overlap-unit field is the ray class field of conductor $d'\infty_-$, as predicted on the AFK side. We have not proved Zauner's conjecture or Problem~1.

For odd $d$, the identity $c_-=\kappa$ is in turn one instance of an explicit reciprocity law: the Artin symbol of a principal ideal $(\varpi)$, with $\varpi$ positive at $\infty_-$, acts on the squared ghost overlaps through the matrix of $\varpi$ in the order generated by the Zauner symmetry. This law holds exactly at every prime tested in $d=5$ and $d=9$, and its form is that of the reciprocity laws proved for $\zeta(0)$-invariants. It suggests that the missing step is a reciprocity theorem for the derivative values at one sign idele, rather than an isolated archimedean identity.

For $d$ a power of a prime $p\equiv5\pmod 6$, or of $3$, the archimedean condition can be removed from the hypotheses: under the algebraic part of Stark, $c_-=\kappa$ is equivalent to the triviality of $\kappa$ on the Hilbert-class-field part of the unit field, and this follows from a sign-free hypothesis on how the class group permutes the ghosts of a Galois multiplet. On this family, the remaining content of Problem~1 is therefore unramified class field theory for the Stark--Shintani units, together with their algebraic Stark property.

The main message is that the SIC existence problem has been reduced to the classical Stark archimedean condition $c_-=\kappa$ for the overlap-unit ray class field, and on an infinite family further to an unramified condition.

\begin{acknowledgments}
The research direction, the choice of problems and all mathematical decisions are the author's. The computations were carried out with the assistance of Claude (Anthropic). The author checked all results, proofs and computations and takes full responsibility for the content. Statements attributed to recent preprints are based on the versions available in October 2026.
\end{acknowledgments}

\section*{Author Declarations}
\subsection*{Conflict of Interest}
The author has no conflicts to disclose.

\subsection*{Author Contributions}
\textbf{Hassan Khodaiemehr:} Conceptualization (lead); Formal analysis (lead); Investigation (lead); Methodology (lead); Software (lead); Validation (lead); Writing -- original draft (lead); Writing -- review \& editing (lead).

\section*{Data Availability}
The data that support the findings of this study are available within the article. All code, data and logs are openly available in the GitHub repository of the author at \url{https://github.com/hassankhodaiemehr/ghost-sic-galois}. The repository contains the homotopy-continuation scripts (Julia, HomotopyContinuation.jl), the complete solution lists for $d=4,5$, the parity test, the abelian-Heisenberg search (Python), the finite quantum dilogarithm solvers, the high-precision PSLQ and minimal-polynomial certificates, the class-field, ray-class and reciprocity-law tests (PARI/GP), and all logs listed in Appendix~\ref{app:files}.

\appendix

\section{Computational data}\label{app:data}
This appendix records the computations behind the tables in the main text: methods, tolerances, and the full data. All scripts and logs are available in the GitHub repository \url{https://github.com/hassankhodaiemehr/ghost-sic-galois}; the directories and file names are listed in Appendix~\ref{app:files}.

\subsection{Varieties, reduced systems and controls}\label{app:Vd}\label{app:tcyc}\label{app:control} This subsection gives the full data behind Secs.~\ref{sec:counts} and~\ref{sec:reality}, together with control experiments showing that the reality patterns are not generic.
\paragraph{Complexified SIC variety.} The complete solution sets and the ghost test for $d=4,5,6$ are as follows.
\begin{table}[!htbp]
\caption{Complete isolated solution sets of $V_d$ and the displaced-parity ghost test.}\centering\small
\begin{tabular}{@{}rrrrrrr@{}}\hline\hline
$d$ & paths & singular endpoints & $|V_d|$ & SICs & non-Hermitian & in $\bigcup_u\Gamma_u$\\\hline
4 & 560 & 48 & 512 & 256 & 256 & 256 (64 in $\Gamma_{U_P}$)\\
5 & 14\,614 & 9\,718 & 4\,000 & 2\,000 & 2\,000 & 2\,000 (80 in $\Gamma_{U_P}$)\\
6 & 218\,343 & 78\,192 & 6\,912 & 3\,456 & 3\,456 & not tested\\\hline\hline
\end{tabular}

\end{table}

\paragraph{Reduced system on all cyclic metrics.}
For each $N$ we used all metrics $\langle x\rangle=\zeta_{2N}^{a x(x+N)}$ with $1\le a<2N$ and $\gcd(a,N)=1$.
\begin{table}[!htbp]
\caption{Conjecture~\ref{conj:TR} on all nondegenerate cyclic metrics, $5\le N\le21$.}\centering\small
\begin{tabular}{@{}rrrrrr@{}}\hline\hline
$N$ & metrics & reduced-system solutions & pentagon solutions & $t$ real & mixed\\\hline
5 & 8 & 40 & 40 & 40 & 0\\
6 & 4 & 16 & 16 & 16 & 0\\
7 & 12 & 48 & 48 & 48 & 0\\
8 & 8 & 32 & 32 & 32 & 0\\
9 & 12 & 48 & 48 & 48 & 0\\
10 & 8 & 32 & 32 & 32 & 0\\
11 & 20 & 160 & 160 & 160 & 0\\
12 & 8 & 32 & 32 & 32 & 0\\
13 & 24 & 192 & 192 & 192 & 0\\
14 & 12 & 72 & 72 & 72 & 0\\
15 & 16 & 128 & 128 & 128 & 0\\
16 & 16 & 64 & 64 & 64 & 0\\
17 & 32 & 256 & 256 & 256 & 0\\
18 & 12 & 96 & 96 & 96 & 0\\
19 & 36 & 144 & 144 & 144 & 0\\
20 & 16 & 128 & 128 & 128 & 0\\
21 & 24 & 240 & 240 & 240 & 0\\
\hline
total & 268 & & & all & 0\\\hline\hline
\end{tabular}

\end{table}
For the $54$ $SL_2(\Z)$ classes with $N\le20$, the reduced and full systems have the same solutions, except $N=9$ cycle $(2,2,5)$ and $N=16$ cycle $(3,3,3)$. Those two have $16$ reduced solutions against $8$ pentagon solutions, and all are totally real.

\paragraph{Control experiments.} To check that total reality is a property of the actual finite quantum dilogarithm system, we solved two modified systems of the same shape.
\paragraph{Control A.} A random $\rho$- and conjugation-invariant subspace with dimensions (even, odd), chosen such that the pairing system is square.
\begin{table}[!htbp]
\caption{Control A. Non-real solutions are common once $N\ge9$.}\centering\small
\begin{tabular}{@{}rcrrrr@{}}\hline\hline
$N$ & dims & seed & solutions & $t$ real & real or unimodular\\\hline
7 & (2,2) & 1 & 8 & 8 & 8\\
7 & (2,2) & 2 & 8 & 8 & 8\\
7 & (2,2) & 3 & 8 & 8 & 8\\
9 & (3,2) & 1 & 16 & 16 & 16\\
9 & (3,2) & 2 & 16 & 8 & 8\\
9 & (3,2) & 3 & 16 & 12 & 12\\
11 & (3,3) & 1 & 32 & 12 & 12\\
11 & (3,3) & 2 & 32 & 20 & 20\\
11 & (3,3) & 3 & 32 & 24 & 24\\
13 & (4,3) & 1 & 64 & 16 & 16\\
13 & (4,3) & 2 & 64 & 32 & 32\\
13 & (4,3) & 3 & 64 & 52 & 52\\
\hline\hline
\end{tabular}

\end{table}

\paragraph{Control B.} The true Weil eigenspaces $\lambda=G_0^{1/3}\zeta_3^k$ ($k=0,1,2$) for the cyclic metric $\langle x\rangle=\zeta_{2N}^{a x(x+N)}$, with a random square subset of the pairing equations. Each entry is solutions\,/\,$t$ real\,/\,real-or-unimodular. An entry with $t$ real but fewer real-or-unimodular solutions indicates mixed solutions.
\begin{table}[!htbp]
\caption{Control B. Non-real and mixed solutions appear for every $N$ tested.}\centering\small
\begin{tabular}{@{}rrrccc@{}}\hline\hline
$N$ & $a$ & seed & $k=0$ & $k=1$ & $k=2$\\\hline
11 & 1 & 1 & 8/8/8 & 8/8/8 & 4/4/2\\
11 & 1 & 2 & 8/8/8 & 8/8/8 & 4/4/4\\
11 & 2 & 1 & 8/4/4 & 4/4/4 & 8/8/8\\
11 & 2 & 2 & 8/4/4 & 4/4/4 & 8/8/8\\
11 & 3 & 1 & 8/8/8 & 8/4/4 & 4/4/4\\
11 & 3 & 2 & 8/8/8 & 8/8/8 & 4/4/2\\
13 & 1 & 1 & 8/8/8 & 16/16/16 & 8/4/4\\
13 & 1 & 2 & 8/8/8 & 16/16/16 & 8/8/8\\
13 & 2 & 1 & 16/4/4 & 8/0/0 & 8/4/4\\
13 & 2 & 2 & 16/4/4 & 8/8/8 & 8/8/8\\
13 & 3 & 1 & 8/8/8 & 16/16/16 & 8/4/4\\
13 & 3 & 2 & 8/8/8 & 16/16/16 & 8/0/0\\
17 & 1 & 1 & 16/8/8 & 32/28/28 & 32/20/20\\
17 & 1 & 2 & 16/8/8 & 32/32/32 & 32/28/28\\
17 & 2 & 1 & 16/8/0 & 32/20/20 & 32/16/16\\
17 & 2 & 2 & 16/8/0 & 32/24/24 & 32/16/16\\
17 & 3 & 1 & 32/32/32 & 32/20/20 & 16/16/16\\
17 & 3 & 2 & 32/16/16 & 32/16/16 & 16/8/8\\
19 & 1 & 1 & 32/16/16 & 32/12/12 & 64/36/36\\
19 & 1 & 2 & 32/8/8 & 32/28/28 & 64/36/36\\
19 & 2 & 1 & 32/8/4 & 64/40/40 & 32/28/28\\
19 & 2 & 2 & 32/16/16 & 64/40/40 & 32/16/16\\
19 & 3 & 1 & 32/12/12 & 64/36/36 & 32/16/16\\
19 & 3 & 2 & 32/16/16 & 64/36/36 & 32/16/16\\
\hline\hline
\end{tabular}

\end{table}

\subsection{Arithmetic certificates}\label{app:pslq}\label{app:minpoly}\label{app:d4}\label{app:PL} This subsection gives the exact and high-precision arithmetic checks behind Secs.~\ref{sec:struct}--\ref{sec:PL}.
\paragraph{Reflection realized by Galois (PSLQ).} Table~\ref{tab:pslq} summarizes the field-membership tests that certify R2 in the cases computed.
\begin{table}[!htbp]
\caption{Certification of R2. $L=\Q(\zeta_{4N},\sqrt{N+4})$; digits: $300$--$400$ (cyclic) and $120$ ($d=4$).}\label{tab:pslq}\centering\small
\begin{tabular}{@{}lll@{}}\hline\hline
Case & test & result\\\hline
cyclic $N=6,7,8,9$ & $E_*(1)\in L$? \quad $E_*(1)+E_*(-1)\in L$? & no; yes (relation heights $4,2,4,3$)\\
$d=4$, $N=16$ & generic $\theta$: $\theta(E)\in L$? & no, for all four $E\in Z_+$\\
 & $\theta(E)+\theta(\rho E)\in L$? & yes\\
 & $\theta(E)+\theta(E')\in L$, $E'\ne\rho E$? & no\\
 & $\sum_{Z_+}\theta\in L$? & yes\\\hline\hline
\end{tabular}

\end{table}

\paragraph{Exact minimal polynomials of the traces $t_u$.}
For the Stark solution (cyclic metric $a=1$) and $t_u=x_u+x_u^{-1}$:
\begin{table}[!htbp]
\caption{All roots are real. Exactly half lie outside $[-2,2]$. All polynomials are monic.}\centering\small
\begin{tabular}{@{}rr>{\raggedright\arraybackslash}p{8.4cm}rr@{}}\hline\hline
$N$ & $u$ & minimal polynomial of $t_u$ & $|t|>2$ & $|t|<2$\\\hline
6 & 1 & $t^4-4t^3-3t^2+14t+1$ & 2 & 2\\
6 & 2 & $t^8-16t^6+87t^4-184t^2+121$ & 4 & 4\\
7 & 1,2,3 & $t^{12}-45t^{10}+723t^8-5424t^6+20647t^4-38554t^2+27889$ & 6 & 6\\
9 & 1,2 & $t^6-6t^5-18t^4+148t^3-282t^2+207t-51$ & 3 & 3\\
9 & 3 & $t^2+3t-1$ & 1 & 1\\
9 & 4 & $t^6+6t^5-18t^4-148t^3-282t^2-207t-51$ & 3 & 3\\\hline\hline
\end{tabular}

\end{table}

\paragraph{The $d=4$ SIC group over all conjugate metrics.}
For $\gamma=\alpha^3=\left(\begin{smallmatrix}21&-8\\8&-3\end{smallmatrix}\right)$ ($N=16$), we computed for each odd $k\in\{1,3,\dots,31\}$ the solutions for the metric $\langle\cdot\rangle^k$. The result is the same for every $k$:
\begin{itemize}[nosep]
\item $Z_+$ consists of 4 real solutions;
\item $Z_-$ consists of 4 unitary solutions;
\item the traces $t_u$, $u\ne0$, $2u\neq0$, over the 8 solutions are 48 with $|t|>2$, 48 with $|t|<2$, and 0 non-real.
\end{itemize}
The remaining coordinates are the three elements with $2u=0$, where $t=\pm2$.

\paragraph{The place law and the single-unit conditions.} The place law and conditions (G1)--(G2) were tested on all classes in which the traces could be identified.
\paragraph{Method.} For each class we took the real solutions in $Z_+$ and refined them to $900$ digits. For every $u$ with $2u\ne0$ we found the minimal polynomial $P$ of $t_u$ with PARI/GP \texttt{algdep}, testing even degrees up to $40$; a polynomial was accepted only if its residual was below $10^{-680}$ and its height was well within the precision. On each $P$ we then performed four exact computations:
\begin{itemize}[nosep]
\item real-root counting, which tests total reality and the split $|t|\gtrless2$;
\item \texttt{polcompositum} with $X^2-q$ for $q=N(N+4)$, $N+4$ and $N$, which decides whether $\sqrt q\in\Q(t_u)$;
\item for each embedding, the sign of $(\sigma(t_u)^2-4)\sigma(\sqrt q)$, normalized such that $\sqrt q>0$ at the embedding of the computed solution;
\item the sign of $P(2)P(-2)$.
\end{itemize}
Polynomials shared by several $u$, or by several solutions, are counted once. The classes $N=11$, and a few $u$ for $N=18$, need degree above $40$ at this precision and are omitted.

\paragraph{Result.} All $100$ polynomials are totally real with an exact half split. In every case $\sqrt{N(N+4)}\in\Q(t_u)$, and the split agrees with the place of $F$ at \emph{every} embedding. In contrast, the analogous test with $q=N+4$ or $q=N$ either fails ($\sqrt q\notin\Q(t_u)$) or gives a mixed split. An example is $N=6$, $u=2$, where both give $4/4$. The sign of $P(2)P(-2)$ is negative for exactly the $13$ polynomials with $m_u=\deg P/2$ odd.

{\small
\begin{longtable}{@{}rlrlcr@{}}
\caption{The place law (Definition~\ref{def:PL}) for the classes in which all or most traces were identified. The columns give the number of distinct minimal polynomials, their degrees, whether PL holds for every one of them, and the number with $m_u$ odd.}\label{tab:PL}\\
\hline\hline
$N$ & cycle & \#$P$ & degrees of $t_u$ & PL & $m_u$ odd\\\hline\endfirsthead
\hline\hline
$N$ & cycle & \#$P$ & degrees of $t_u$ & PL & $m_u$ odd\\\hline\endhead
\hline\hline\endfoot
5 & $(7)$ & 1 & 4 & yes & 0\\
6 & $(2,2,4)$ & 2 & 4, 8 & yes & 0\\
6 & $(2,5)$ & 2 & 4, 8 & yes & 0\\
6 & $(8)$ & 2 & 4, 8 & yes & 0\\
7 & $(9)$ & 1 & 12 & yes & 0\\
8 & $(2,2,2,4)$ & 1 & 8 & yes & 0\\
8 & $(2,3,3)$ & 2 & 4, 16 & yes & 0\\
8 & $(2,6)$ & 1 & 8 & yes & 0\\
8 & $(3,4)$ & 2 & 4, 16 & yes & 0\\
8 & $(10)$ & 2 & 4, 16 & yes & 0\\
9 & $(2,2,5)$ & 1 & 2 & yes & 1\\
9 & $(11)$ & 3 & 2, 6$^{\times2}$ & yes & 3\\
10 & $(2,7)$ & 3 & 8$^{\times2}$, 16 & yes & 0\\
10 & $(12)$ & 2 & 8, 16 & yes & 0\\
12 & $(2,2,2,5)$ & 4 & 4$^{\times2}$, 8, 16 & yes & 0\\
12 & $(2,2,6)$ & 4 & 4$^{\times2}$, 8, 16 & yes & 0\\
12 & $(2,3,2,3)$ & 1 & 2 & yes & 1\\
12 & $(2,8)$ & 4 & 2$^{\times2}$, 4$^{\times2}$ & yes & 2\\
12 & $(4,4)$ & 1 & 2 & yes & 1\\
12 & $(14)$ & 4 & 4$^{\times2}$, 8, 16 & yes & 0\\
13 & $(2,3,4)$ & 2 & 24$^{\times2}$ & yes & 0\\
13 & $(2,4,3)$ & 2 & 24$^{\times2}$ & yes & 0\\
13 & $(15)$ & 2 & 24$^{\times2}$ & yes & 0\\
14 & $(2,9)$ & 4 & 6, 12$^{\times2}$, 24 & yes & 1\\
14 & $(3,6)$ & 4 & 6, 12$^{\times2}$, 24 & yes & 1\\
14 & $(16)$ & 4 & 6, 12$^{\times2}$, 24 & yes & 1\\
15 & $(2,2,7)$ & 3 & 8, 16, 32 & yes & 0\\
15 & $(17)$ & 3 & 8, 16, 32 & yes & 0\\
16 & $(2,2,2,6)$ & 2 & 2, 8 & yes & 1\\
16 & $(2,10)$ & 3 & 2, 4, 16 & yes & 1\\
16 & $(3,3,3)$ (SIC, $d=4$) & 1 & 4 & yes & 0\\
16 & $(4,5)$ & 4 & 4, 8$^{\times2}$, 32 & yes & 0\\
16 & $(18)$ & 4 & 4, 8$^{\times2}$, 32 & yes & 0\\
17 & $(3,7)$ & 2 & 32$^{\times2}$ & yes & 0\\
17 & $(19)$ & 2 & 32$^{\times2}$ & yes & 0\\
18 & $(2,2,3,4)$ & 4 & 8, 16, 24$^{\times2}$ & yes & 0\\
18 & $(2,2,4,3)$ & 4 & 8, 16, 24$^{\times2}$ & yes & 0\\
18 & $(2,2,8)$ & 3 & 8$^{\times2}$, 16 & yes & 0\\
18 & $(2,3,5)$ & 4 & 8, 16, 24$^{\times2}$ & yes & 0\\
\end{longtable}}

\paragraph{Single-unit conditions.} For the classes in Table~\ref{tab:gen} we also computed the minimal polynomial $P_x$ of $x_{u_0}$ for each $u_0$, and checked (G1) and (G2) of Theorem~\ref{thm:single} exactly. For each $v$, \texttt{polcompositum}$(P_{x_{u_0}},P_{x_v})$ decides whether $x_v\in\Q(x_{u_0})$, by selecting the factor that vanishes at the computed pair. If it does, \texttt{modreverse} gives the polynomial $f_v$ with $x_v=f_v(x_{u_0})$, and we checked the identity $f_v(X)f_v(X^{-1})\equiv1\pmod{P_x}$. That identity is exactly (G2), since an embedding of $\Q(x_{u_0})$ is determined by the image of $x_{u_0}$. In every class some $u_0$ generates, and (G2) holds for \emph{every} generating $u_0$. In four classes a generating $u_0$ has $m_{u_0}$ odd and $P_{t_{u_0}}(2)P_{t_{u_0}}(-2)<0$. There, Theorem~\ref{thm:single} certifies a unitary-type conjugate from the sign of a single integer.

\begin{table}[!htbp]
\caption{Exact checks of (G1) and (G2). The columns give the number of $u_0$ tested (up to $u\mapsto-u$ and repeated polynomials), the number of $u_0$ satisfying (G1), the degrees of their fields $\Q(x_{u_0})$, whether (G2) holds for all of them, whether $P(0)$ is odd for the generating $u_0$ (the hypothesis of Corollary~\ref{cor:2adic}), whether the integer criterion of Corollary~\ref{cor:integer} applies, and whether the 2-adic criterion of Corollary~\ref{cor:2adic} applies.}\label{tab:gen}\centering\small
\begin{tabular}{@{}rlrrlcccc@{}}\hline\hline
$N$ & cycle & $u_0$ & (G1) & $[\Q(x_{u_0}):\Q]$ & (G2) & $P(0)$ odd & Cor.~\ref{cor:integer} & Cor.~\ref{cor:2adic}\\\hline
5 & $(7)$ & 2 & 2 & 8 & yes & yes & no & no\\
6 & $(2,2,4)$ & 2 & 1 & 16 & yes & yes & no & no\\
6 & $(2,5)$ & 2 & 1 & 16 & yes & yes & no & no\\
6 & $(8)$ & 2 & 1 & 16 & yes & yes & no & no\\
7 & $(9)$ & 3 & 3 & 24 & yes & yes & no & no\\
8 & $(2,2,2,4)$ & 2 & 2 & 16 & yes & no & no & no\\
8 & $(2,3,3)$ & 3 & 2 & 32 & yes & no & no & no\\
8 & $(2,6)$ & 2 & 2 & 16 & yes & no & no & no\\
8 & $(3,4)$ & 3 & 2 & 32 & yes & no & no & no\\
8 & $(10)$ & 3 & 2 & 32 & yes & no & no & no\\
9 & $(2,2,5)$ & 3 & 3 & 4 & yes & yes & yes & yes\\
9 & $(11)$ & 4 & 3 & 12 & yes & yes & yes & yes\\
10 & $(2,7)$ & 4 & 2 & 32 & yes & yes & no & no\\
10 & $(12)$ & 4 & 2 & 32 & yes & yes & no & no\\
12 & $(2,3,2,3)$ & 2 & 2 & 4 & yes & no & yes & no\\
12 & $(2,8)$ & 4 & 2 & 8 & yes & no & no & no\\
12 & $(4,4)$ & 2 & 2 & 4 & yes & no & yes & no\\
16 & $(3,3,3)$ ($d=4$) & 6 & 6 & 8 & yes & no & no & no\\
50 & $(4,4,4)$ ($d=5$, ghost) & 1 & 1 & 32 & yes & yes & no & no\\
\hline\hline
\end{tabular}
\end{table}

\section{Supplementary table}\label{app:supp}\label{app:scan}

\paragraph{All finite quantum dilogarithms for $SL_2(\Z)$ classes, $5\le N\le30$.}
In the table below, $Z_\pm$ are the solutions with $E(0)=(\sqrt N\pm\sqrt{N+4})/2$. Each entry gives the number of solutions and their type. ``Unit.''\ means unitary type, and ``--'' means no solutions.
{\small
\begin{longtable}{@{}rlll@{}}
\caption{All $94$ hyperbolic conjugacy classes with $5\le N\le30$.}\label{tab:fullscan}\\
\hline\hline
$N$ & cycle $(b_0,\dots,b_{\ell-1})$ & $Z_+$ & $Z_-$\\\hline\endfirsthead
\hline\hline
$N$ & cycle $(b_0,\dots,b_{\ell-1})$ & $Z_+$ & $Z_-$\\\hline\endhead
\hline\hline\endfoot
5 & $(3,3)$ & 1 real+unit., even & 4 unit.\\
5 & $(7)$ & 4 real & 1 real+unit., even\\
6 & $(2,2,4)$ & 2 real & 2 unit.\\
6 & $(2,5)$ & 2 real & 2 unit.\\
6 & $(8)$ & 2 real & 2 unit.\\
7 & $(9)$ & 2 real & 2 unit.\\
8 & $(2,2,2,4)$ & 2 real & 2 unit.\\
8 & $(2,3,3)$ & 2 real & 2 unit.\\
8 & $(2,6)$ & 2 real & 2 unit.\\
8 & $(3,4)$ & 2 real & 2 unit.\\
8 & $(10)$ & 2 real & 2 unit.\\
9 & $(2,2,5)$ & 4 real & 4 unit.\\
9 & $(11)$ & 2 real & 2 unit.\\
10 & $(2,7)$ & 2 real & 2 unit.\\
10 & $(12)$ & 2 real & 2 unit.\\
11 & $(2,2,3,3)$ & 4 real & 4 unit.\\
11 & $(3,5)$ & 4 real & 4 unit.\\
11 & $(13)$ & 4 real & 4 unit.\\
12 & $(2,2,2,5)$ & 4 real & --\\
12 & $(2,2,6)$ & 4 real & --\\
12 & $(2,3,2,3)$ & 2 real & 2 unit.\\
12 & $(2,8)$ & 4 real & --\\
12 & $(4,4)$ & 2 real & 2 unit.\\
12 & $(14)$ & 4 real & --\\
13 & $(2,3,4)$ & 4 real & 4 unit.\\
13 & $(2,4,3)$ & 4 real & 4 unit.\\
13 & $(15)$ & 4 real & 4 unit.\\
14 & $(2,9)$ & 6 real & --\\
14 & $(3,6)$ & 6 real & --\\
14 & $(16)$ & 6 real & --\\
15 & $(2,2,7)$ & 4 real & 4 unit.\\
15 & $(17)$ & 4 real & 4 unit.\\
16 & $(2,2,2,6)$ & 4 real & 4 unit.\\
16 & $(2,10)$ & 2 real & 2 unit.\\
16 & $(3,3,3)$ & 4 real & 4 unit.\\
16 & $(4,5)$ & 2 real & 2 unit.\\
16 & $(18)$ & 2 real & 2 unit.\\
17 & $(3,7)$ & 4 real & 4 unit.\\
17 & $(19)$ & 4 real & 4 unit.\\
18 & $(2,2,3,4)$ & 4 real & 4 unit.\\
18 & $(2,2,4,3)$ & 4 real & 4 unit.\\
18 & $(2,2,8)$ & 8 real & 8 unit.\\
18 & $(2,3,5)$ & 4 real & 4 unit.\\
18 & $(2,5,3)$ & 4 real & 4 unit.\\
18 & $(2,11)$ & 4 real & 4 unit.\\
18 & $(20)$ & 4 real & 4 unit.\\
19 & $(21)$ & 2 real & 2 unit.\\
20 & $(2,2,2,7)$ & 4 real & 4 unit.\\
20 & $(2,3,2,4)$ & 4 real & 4 unit.\\
20 & $(2,4,4)$ & 4 real & 4 unit.\\
20 & $(2,12)$ & 4 real & 4 unit.\\
20 & $(3,8)$ & 4 real & 4 unit.\\
20 & $(4,6)$ & 4 real & 4 unit.\\
20 & $(22)$ & 4 real & 4 unit.\\
21 & $(2,2,9)$ & 10 real & --\\
21 & $(5,5)$ & 10 real & --\\
21 & $(23)$ & 10 real & --\\
22 & $(2,13)$ & 2 real & 2 unit.\\
22 & $(24)$ & 2 real & 2 unit.\\
23 & $(2,3,6)$ & 8 real & 8 unit.\\
23 & $(2,6,3)$ & 8 real & 8 unit.\\
23 & $(3,9)$ & 8 real & 8 unit.\\
23 & $(25)$ & 8 real & 8 unit.\\
24 & $(2,2,2,8)$ & 4 real & 4 unit.\\
24 & $(2,2,10)$ & 4 real & 4 unit.\\
24 & $(2,3,3,3)$ & 4 real & 4 unit.\\
24 & $(2,14)$ & 4 real & 4 unit.\\
24 & $(3,3,4)$ & 4 real & 4 unit.\\
24 & $(4,7)$ & 4 real & 4 unit.\\
24 & $(26)$ & 4 real & 4 unit.\\
25 & $(2,2,3,5)$ & 4 real & 4 unit.\\
25 & $(2,2,5,3)$ & 4 real & 4 unit.\\
25 & $(27)$ & 4 real & 4 unit.\\
26 & $(2,15)$ & 4 real & 4 unit.\\
26 & $(3,10)$ & 4 real & 4 unit.\\
26 & $(5,6)$ & 4 real & 4 unit.\\
26 & $(28)$ & 4 real & 4 unit.\\
27 & $(2,2,11)$ & 12 real & 12 unit.\\
27 & $(2,4,5)$ & 6 real & 6 unit.\\
27 & $(2,5,4)$ & 6 real & 6 unit.\\
27 & $(29)$ & 6 real & 6 unit.\\
28 & $(2,2,2,9)$ & 8 real & --\\
28 & $(2,2,4,4)$ & 4 real & 4 unit.\\
28 & $(2,3,2,5)$ & 4 real & 4 unit.\\
28 & $(2,3,7)$ & 8 real & --\\
28 & $(2,7,3)$ & 8 real & --\\
28 & $(2,16)$ & 4 real & 4 unit.\\
28 & $(4,8)$ & 4 real & 4 unit.\\
28 & $(30)$ & 8 real & --\\
29 & $(3,11)$ & 4 real & 4 unit.\\
29 & $(31)$ & 4 real & 4 unit.\\
30 & $(2,2,12)$ & 4 real & 4 unit.\\
30 & $(2,17)$ & 4 real & 4 unit.\\
30 & $(32)$ & 4 real & 4 unit.\\
\end{longtable}}
Across all $94$ classes, no solution is of mixed type, every solution in $Z_+$ is real, and every solution in $Z_-$ is unitary. The only real solution with $E(0)<0$ is even ($N=5$, cycle $(7)$). The set $Z_-$ is empty only when $\sqrt{N+4}\in K_0$, namely for some classes with $N=12,14,21,28$.

\section{Status of the five open problems}\label{app:five}\label{sec:five}
We record the status of the five open problems of Horodecki, Rudnicki and \.Zyczkowski~\cite{HRZ} as of October 2026.

{\sloppy
\begin{itemize}[nosep]
\item \textbf{Problem 1 (SICs in an infinite sequence of dimensions):} open unconditionally. There are conditional constructions for all $d>3$~\cite{AFK25}, and ghosts exist unconditionally for all $d\ge4$~\cite{AFK26,RW}. This paper reduces the remaining gap to Corollary~\ref{cor:ghost}.
\item \textbf{Problem 2 (mutually unbiased bases (MUBs) in $d=6$):} solved~\cite{MUB6}, not yet peer reviewed. The maximum number of MUBs in $\C^6$ is three, resolving Zauner's dimension-six MUB conjecture. Three MUBs were already known~\cite{MUBreview}; the new upper bound excludes four bases by a complete certified computation.
\item \textbf{Problem 3 (36 entangled officers):} solved~\cite{Rather}. It received the Golden KCIK Award in 2021.
\item \textbf{Problem 4 (negative-partial-transpose (NPT) bound entanglement):} open.
\item \textbf{Problem 5 (2-copy distillability of $\rho(4,-\tfrac12)$):} solved in four independent preprints from July 2026~\cite{FGP,SC,FHPV,BGH}, not yet peer reviewed. We checked numerically that $\min\langle\psi|(W_\alpha^\Gamma)^{\otimes2}|\psi\rangle$ over Schmidt-rank-$2$ vectors, for $W_\alpha=I+\alpha V$ on $\C^4\otimes\C^4$ (Appendix~\ref{app:p5}), equals $0.055$, $1.1\times10^{-8}$, $-0.040$ and $-0.100$ for $\alpha=-0.45,-0.50,-0.52,-0.55$. This is consistent with the threshold $\alpha=-\tfrac12$.
\end{itemize}
\par}

\section{Implementation details and related checks}\label{app:impl}

\subsection{Methods and related checks}\label{app:methods}\label{app:p5}
We list the numerical methods and tolerances, and a check related to Problem~5.
\paragraph{Methods and tolerances.} The main tools and acceptance criteria were as follows.
\paragraph{Homotopy continuation.} We used HomotopyContinuation.jl (Julia 1.10) with polyhedral start systems, solving randomized square subsystems of overdetermined systems. Endpoints were kept if the residual of the \emph{full} system was below $10^{-6}$ (for $V_d$) or $10^{-7}$--$10^{-8}$ (for finite quantum dilogarithms). Duplicates were merged at distance $10^{-6}$. Points with $y^Tx=0$, or with a vanishing coordinate in the reduced dilogarithm system, were discarded.

\paragraph{Classification.} A solution is \emph{real type} if $\max_x|\overline{E(x)}-\langle x\rangle E(x)|<10^{-6}$, \emph{unitary type} if $\max_x|\overline{E(x)}-\langle x\rangle E(-x)|<10^{-6}$, and \emph{even} if $\max_x|E(x)-E(-x)|<10^{-6}$. A solution is \emph{mixed} if it is neither real nor unitary.

\paragraph{Eigenspace reduction.} For finite quantum dilogarithms we parametrized the $\lambda$-eigen\-space of $W=\langle\cdot\rangle^{-1}\mathcal F$, for each of the three cube roots $\lambda$ of $\int\langle x\rangle^{-1}dx$. We fixed $E(0)=(\sqrt N\pm\sqrt{N+4})/2$ and solved the pairing equations $E(u)E(-u)\langle u\rangle=1$. The pentagon equation (Ref.~\onlinecite{RW}, Eq.~(31)) was then checked on every candidate. Metric groups $G=\Z^2/\Z^2(\gamma-I)$ were built from $\gamma=T^{b_0}S\cdots T^{b_{\ell-1}}S$, with $\langle u\rangle$ and $\langle u;v\rangle$ as in Ref.~\onlinecite{RW}, Eq.~(3). The bicharacter identity $\langle u+v\rangle=\langle u\rangle\langle v\rangle\langle u;v\rangle$ was verified numerically to $10^{-15}$ for every group.

\paragraph{High-precision certification.} Solutions were refined by Gauss--Newton iteration in \texttt{mpmath} at $120$--$400$ digits, reaching residuals $10^{-113}$--$10^{-398}$. Field membership was tested with PSLQ against a $\Q$-basis of $L=\Q(\zeta_{4N},\sqrt{N+4})$. Minimal polynomials were found with \texttt{mpmath.findpoly} at $300$ digits, with coefficient bound $10^{12}$.

\paragraph{Gradient search (Sec.~\ref{sec:heis}).} L-BFGS on the frame potential, with an analytic gradient checked against finite differences to $8\times10^{-10}$.

\paragraph{Problem~5 check.}
On $\C^4\otimes\C^4$, let $I$ be the identity and $V$ the swap operator, $V(\phi\otimes\chi)=\chi\otimes\phi$. Consider the operator $W_\alpha=I+\alpha V$. This is the Werner state $\rho(4,\alpha)$ before normalization by its trace, which does not affect the sign tests below. Let $W_\alpha^\Gamma$ be its partial transpose. We minimized $\langle\psi|(W_\alpha^\Gamma)^{\otimes2}|\psi\rangle$ over unit vectors $\psi$ of Schmidt rank $2$ (40 random starts, L-BFGS):
\begin{center}\small
\begin{tabular}{@{}rrrrr@{}}\hline\hline
$\alpha$ & $-0.45$ & $-0.50$ & $-0.52$ & $-0.55$\\
minimum & $0.055$ & $1.1\times10^{-8}$ & $-0.040$ & $-0.100$\\\hline\hline
\end{tabular}
\end{center}

\subsection{Files}\label{app:files}\raggedright
All files are in the GitHub repository \url{https://github.com/hassankhodaiemehr/ghost-sic-galois}. Directory names below are relative to the repository root, and each directory has a \texttt{README.txt}.
\begin{itemize}[nosep]
\item \textbf{Complexified variety} (directory \texttt{variety\_fqd\_highprec/}): \texttt{ghost\_count.jl}, \texttt{ghost\_save.jl}, \texttt{parity\_test.py}, \texttt{sols\_d4.txt}, \texttt{sols\_d5.txt}, \texttt{s4.log}, \texttt{s5.log}, \texttt{d6.log}.
\item \textbf{Heisenberg search} (directory \texttt{variety\_fqd\_highprec/}): \texttt{hsearch.py}, \texttt{controls.py}, \texttt{big.py}, \texttt{big.log}.
\item \textbf{Finite quantum dilogarithms} (directory \texttt{variety\_fqd\_highprec/}): \texttt{metricG.py}, \texttt{checkG.py}, \texttt{fqdW.jl}, \texttt{fqd2.jl}, \texttt{fqdG.jl}, \texttt{gen\_gammas.py}, \texttt{scan.jl}, \texttt{scan.log}, \texttt{tscan.jl}, \texttt{tscan\_cyc.jl}, \texttt{tcyc.log}, \texttt{control.jl}, \texttt{control.log}, \texttt{scan\_d4.jl}, \texttt{d4conj.log}.
\item \textbf{High precision} (directory \texttt{variety\_fqd\_highprec/}): \texttt{r2test.py}, \texttt{r2G.py}, \texttt{salem.py}.
\item \textbf{Place law and single-unit checks} (directory \texttt{place\_law\_code/}): \texttt{save.jl}, \texttt{refine.py}, \texttt{refx.py}, \texttt{hs2.gp}, \texttt{an.sh}, \texttt{gen.gp}, \texttt{run\_gen.sh}, logs \texttt{an\_*.log}, \texttt{gen\_*.log}.
\item \textbf{Class-field test, $d=4,5$} (directory \texttt{place\_law\_code/d5/}): \texttt{ghost\_refine.py}, \texttt{gen.gp}, \texttt{an5.gp}, \texttt{xs.gp}, \texttt{cft.gp}, \texttt{cond2.gp}, logs \texttt{a5.log}, \texttt{gen5.log}.
\item \textbf{Ray-class tests, $d=9,13$, and the unique-involution list} (directory \texttt{rayclass\_code/}): ghost and SIC search \texttt{findghost.py}, \texttt{findsic.py}; Newton refinement \texttt{grefine.gp}, \texttt{refine.gp}; refined ghosts \texttt{g9.gp} ($2000$ digits), \texttt{g13x.gp} and \texttt{g17x.gp} ($20500$ digits); $d=9$ analysis \texttt{d9*.gp}, \texttt{ganal.gp}, \texttt{an.gp}; $d=13$ trace polynomial and conductor \texttt{d13a--e.gp}, \texttt{d13polys.gp}; $d=13$ Frobenius test \texttt{d13Rt.gp}, \texttt{d13f.gp}; place-law sign checks \texttt{sl9.gp}, \texttt{sl13.gp}; ray class groups \texttt{rc.gp}; $2$-Sylow list \texttt{twosyl.gp} ($d\le150$) and census \texttt{twosyl2.gp} ($d\le2000$); $d=17$ attempt \texttt{d17.gp}, \texttt{d17f.gp}; all logs.
\item \textbf{Coverage figure} (Fig.~\ref{fig:ramcov}; directory \texttt{ramification\_coverage/}): data \texttt{route\_c\_coverage\_data.csv} ($4\le d\le2000$; the cyclic $2$-Sylow column is the census of Sec.~\ref{sec:CFT}) and plotting script \texttt{plot\_route\_c\_coverage.py}.
\item \textbf{Ramification at $\infty_-$, $d\le20000$} (Sec.~\ref{sec:TRinv}; directory \texttt{ramification\_coverage/}): \texttt{place\_ramification.py}, a self-contained script computing the squarefree part of $D$, the fundamental unit of $F$ by continued fractions with the index-$3$ correction, and the condition (C$_d$), together with the per-dimension output \texttt{place\_ramification\_d\_le\_20000.csv}.
\item \textbf{Explicit reciprocity law, $d=5,9$} (directory \texttt{reciprocity\_code\_d5\_d9/}): $d=5$ test \texttt{all5.gp} with data \texttt{g5.gpv}, exactness check \texttt{ver5.gp}, inert primes \texttt{inert5.gp}; $d=9$ automorphisms \texttt{gal9.gp} (output \texttt{aut9.gp}), exact overlap expressions \texttt{all9h.gp} (output \texttt{ex9\_all.gp}), Frobenius test \texttt{all9i.gp}; logs.
\item \textbf{Reciprocity law, certified, $d=5,7,11,15$, and the family $\mathcal P$} (directory \texttt{recip\_v4/}): Zauner-symmetric ghost search \texttt{findghost\_z.py}; Newton refinement \texttt{grefine.gp} with driver \texttt{refine\_drv.gp}; refined ghost overlaps \texttt{eps5.txt}, \texttt{eps7.txt}, \texttt{eps11.txt}, \texttt{eps15.txt} ($800$, $1500$, $6000$ and $4000$ digits); the test \texttt{recip.gp}, with exact expressions saved as \texttt{expr*.gp}; the class-field certificate and ray-class coverage \texttt{cert.gp}; logs \texttt{rec*.log}; the census of $\mathcal P$, \texttt{fam.gp} (output \texttt{family.txt}); and the involution count in $P^{\mathfrak m}/P_{\mathfrak m,1}$, \texttt{ker.gp} (log \texttt{ker.log}).

\item \textbf{Problem 5} (directory \texttt{variety\_fqd\_highprec/}): \texttt{problem5\_check.py}.
\end{itemize}

\bibliography{sic_ghosts_jmp}

\end{document}